\documentclass{aa}  

\usepackage{graphicx}
\usepackage{tablefootnote} 
\usepackage{txfonts}
\usepackage{hyperref} 
\usepackage{lscape}
\usepackage{longtable}
\usepackage{adjustbox}
\usepackage{ragged2e}
\usepackage{natbib}
\usepackage{balance}
\usepackage{afterpage}
\bibpunct{(}{)}{;}{a}{}{,} 

\newcommand{\teff}{\ensuremath{\mathrm{T_{eff}}}}
\newcommand{\kms}{\ensuremath{\mathrm{km\,s^{-1}}}}
\newcommand{\logg}{\ensuremath{\mathrm{\log g}}}
\newcommand{\vt}{\ensuremath{\mathrm{v_{turb}}}}

\newcommand{\Msol}{$\rm M_\odot$}

\usepackage{hyperref}
\hypersetup{colorlinks=true,linkcolor=blue,citecolor=blue,filecolor=blue,urlcolor=blue}

\begin{document}

   \title{Chemical Evolution of R-process Elements in Stars (CERES)}

   \subtitle{III. Chemical abundances of neutron capture elements from Ba to Eu}

   \author{L. Lombardo
          \inst{1}
\and C. J. Hansen \inst{1}
\and F. Rizzuti \inst{2}
\and G. Cescutti \inst{2,3,4}
\and L. I. Mashonkina \inst{5}
\and P. Fran\c{c}ois \inst{6,7}
\and P. Bonifacio \inst{6}
\and E. Caffau \inst{6}
\and A. Alencastro Puls \inst{1}
\and R. Fernandes de Melo \inst{1}
\and A. J. Gallagher \inst{8}
\and \'A. Sk\'ulad\'ottir \inst{9}
\and A. J. Koch-Hansen \inst{10}
\and L. Sbordone \inst{11}
          }

   \institute{Goethe University Frankfurt, Institute for Applied Physics (IAP), Max-von-Laue-Str. 12, 60438, Frankfurt am Main \\
              \email{Lombardo@iap.uni-frankfurt.de}
        \and Dipartimento di Fisica, Sezione di Astronomia, Università di Trieste, Via G. B. Tiepolo 11, 34143 Trieste, Italy
        \and  INAF, Osservatorio Astronomico di Trieste, Via Tiepolo 11, 34143 Trieste, Italy 
        \and INFN, Sezione di Trieste, Via A. Valerio 2, 34127 Trieste, Italy 
         \and  Institute of Astronomy, Russian Academy of Sciences, Pyatnitskaya 48, 119017, Moscow, Russia 
         \and GEPI, Observatoire de Paris, Universit\'{e} PSL, CNRS, 5 place Jules Janssen, 92195 Meudon, France   
         \and UPJV, Université de Picardie Jules Verne, Pôle Scientifique, 33 rue St Leu, 80039 Amiens, France 
         \and Leibniz-Institut f\"{u}r Astrophysik Potsdam, An der Sternwarte 16, 14482 Potsdam, Germany
         \and Dipartimento di Fisica e Astronomia, Universitá degli Studi di Firenze, Via G. Sansone 1, I-50019 Sesto Fiorentino, Italy
         \and Zentrum f\"ur Astronomie der Universit\"at Heidelberg, Astronomisches Rechen-Institut, M\"onchhofstr. 12, 69120 Heidelberg, Germany
         \and ESO-European Southern Observatory, Alonso de Cordova 3107, Vitacura, Santiago, Chile
}

   \date{Received September 15, 1996; accepted March 16, 1997}

 
  \abstract
   {The chemical abundances of elements such as barium and the lanthanides are essential to understand the nucleosynthesis of heavy elements in the early Universe as well as the contribution of different neutron capture processes (for example slow versus rapid) at different epochs.}
   {The Chemical Evolution of R-process Elements in Stars (CERES) project aims to provide a homogeneous analysis of a sample of metal-poor stars ([Fe/H]$<$$-1.5$) to improve our understanding of the nucleosynthesis of neutron capture elements, in particular the $r$-process elements, in the early Galaxy.}
   {Our data consist of a sample of high resolution and high signal-to-noise ratio UVES spectra. The chemical abundances were derived through spectrum synthesis, using the same model atmospheres and stellar parameters as derived in the first paper of the CERES series.}
   {We measured chemical abundances or upper limits of seven heavy neutron capture elements (Ba, La, Ce, Pr, Nd, Sm, and Eu) for a sample of 52 metal-poor giant stars. We estimated through the mean shift clustering algorithm that at [Ba/H]=$-2.4$ and [Fe/H]=$-2.4$ a variation in the trend of [X/Ba], with X=La,Nd,Sm,Eu, versus [Ba/H] occurs. This result suggests that, for [Ba/H]$<$$-2.4$, Ba nucleosynthesis in the Milky Way halo is primarily due to the $r$-process, while for [Ba/H]$>$$-2.4$ the effect of the $s$-process contribution begins to be visible. In our sample, stars with [Ba/Eu] compatible with a Solar System pure $r$-process value (hereafter, $r$-pure) do not show any particular trend compared to other stars, suggesting $r$-pure stars may form in similar environments to stars with less pure $r$-process enrichments.
   }
   {
   Homogeneous investigations of high resolution and signal-to-noise ratio spectra are crucial for studying the heavy elements formation, as they provide abundances that can be used to test nucleosynthesis models as well as Galactic chemical evolution models. 
   }


   \maketitle
%

\section{Introduction}

Since the first observations of very metal-poor stars ([Fe/H]$<$$-2$) enriched in rapid ($r$-) neutron capture (n-capture) process elements \citep{Pagel1965Natur.206..282P, SpiteSpite1978A&A....67...23S}, it has become  clear that the heavy element (Z$>$30) production in the early Galaxy followed a different path from that followed in the Galactic disc at solar metallicity.
The observed trends with [Fe/H] for [Ba/Fe] and [Eu/Fe] at low metallicities as well as the chemical abundance patterns of very metal-poor stars seem to suggest that heavy elements in the early Universe are mostly produced by the $r$-process, as first proposed by \citet{Truran1981A&A....97..391T}. 
This scenario is supported by the fact that the primary source of slow neutron capture ($s$-) process at solar metallicity, namely low- and intermediate-mass (1-8 \Msol) asymptotic giant branch (AGB) stars, have lifetimes too long to significantly enrich the interstellar medium at very low metallicities \citep[see e.g.][]{Busso1999ARA&A..37..239B,Kappler2011RvMP...83..157K,Karakas2014PASA...31...30K}.

Observations at high resolution show that very metal-poor stars are characterised by a broad range of heavy elements' abundances, from stars with [Eu/Fe]$<$$-0.3$ 
such as HD~122563
\citep{Butcher1975ApJ...199..710B,Honda2006ApJ...643.1180H} and
HD~88609 \citep{Honda2007ApJ...666.1189H}
to stars with [Eu/Fe]=+1.6 such as CS 22892-052 \citep{Sneden2003ApJ...591..936S}, and even  
[Eu/Fe]$>$2 such as the recently discovered 2MASS J22132050–5137385 \citep{Roederer2024ApJ...971..158R}, but seemingly robust chemical patterns in the lanthanide region (56$<$Z$<$72) \citep[see e.g.][]{Sneden2008ARA&A..46..241S} are found.
This so-called ``robustness'' or ``universality'' of the $r$-process is, however, not observed for the light n-capture elements (30$<$Z$<$50).  
Such abundance pattern variations can only occur if the physical conditions vary during the $r$-process
event, or if multiple different formation sites contribute \citep[see e.g.][]{Travaglio2004ApJ...601..864T, Qian2008ApJ...687..272Q, Hansen2014ApJ...797..123H, Frischknecht2016MNRAS.456.1803F, Spite2018A&A...611A..30S}.

The large neutron densities required to sustain the $r$-process can be reached in several astrophysical environments, such as magneto-rotational driven supernovae (MRD SNe), collapsars, and compact mergers of two neutron stars or of a neutron star and a black hole \citep[see e.g. the review from ][and references therein]{Cowan2021RvMP...93a5002C}.
The recent discovery of the $r$-process signature in the kilonova following the neutron star merger (NSM) event GW170817 strongly supports NSM as $r$-process site \citep[see e.g.][and references therein]{AbbottPhysRevLett.119.161101,Watson2019Natur.574..497W}. 
However, the scenario in which NSM are the only $r$-process source cannot explain the entire production of heavy elements in the Galaxy, as it faces problems in reproducing the chemical abundances observed in metal-poor stars in both Milky Way and its satellite dwarf galaxies \citep[see e.g.][and references therein]{Roederer_ncap2014ApJ...784..158R,Cote2019ApJ...875..106C,SkuladottirSalvadori2020A&A...634L...2S}.
Future direct follow-up of kilonovae are infrequent, and still face uncertainties in modelling the $r$-process element lines under the proper conditions \citep[see e.g.][and references therein]{Watson2019Natur.574..497W,Gillanders2022MNRAS.515..631G,Domoto2022ApJ...939....8D,Perego2022ApJ...925...22P,Gillanders2024MNRAS.529.2918G}.
Hence, to make further progress in understanding the exact physical conditions of the $r$-process including possible delay times, our best option is still to use indirect observations of old low-mass stars, as their abundances reflect the chemical composition of the gas they were born from.

The Chemical Evolution of R-process Elements in Stars (CERES) project aims to measure the abundances of as many n-capture elements as possible in a sample of metal-poor ([Fe/H]$<$$-1.5$) giant stars. 
The final goal of the project is to improve our understanding of the n-capture  processes, in particular the $r$-process, through the availability of numerous chemical abundances, that can be used to test the prediction of nucleosynthesis models (yields) as well as of Galactic chemical evolution (GCE) models. For this reason the abundances need to be derived consistently and homogeneously.
In \citealt{Lombardo2022A&A...665A..10L} (hereafter Paper I), 
we presented our sample which constitutes of 52 giant stars.
We performed a homogeneous analysis on this set of spectra and provided the stellar parameters and chemical abundances of 18 elements, from Na to Zr, for the stars in our sample.
In Fernandes de Melo et al. (accepted, hereafter Paper II), we completed the chemical abundance analysis of light elements, deriving the abundances of C, N, O, and Li. 
In this paper, we extend the analysis to the heavy n-capture elements, presenting the abundances of Ba and the rare earth elements La, Ce, Pr, Nd, Sm, and Eu for our sample stars. 
We also compare our results for Sr, Ba and Eu with the prediction of GCE models from \citet{Cescutti2014A&A...565A..51C},  \citet{Cescutti2015A&A...577A.139C}, and \citet{Rizzuti2021MNRAS.502.2495R}.

\section{Data set}\label{Sect:dataset}
As described in Paper I, we selected the stars in our sample with the aim of having complete abundance patterns, especially in the heavy elements' region. 
The stars were selected 
to be metal-poor ([Fe/H]$<$$-1.5$) and with less than five heavy elements (Z$>$$30$)  measured in the literature. 
Coordinates, CERES names  and one other designation for each of our target stars can be found in Table A.1 of Paper\,I.
We here use the CERES name to refer to any star in our sample.
The targets were observed with the Ultraviolet and Visual Echelle Spectrograph (UVES) of the Very Large Telescope (VLT) at the European Southern Observatory \citep[ESO;][]{Dekker2000SPIE.4008..534D} during two runs in November 2019 and March 2020 (PI: C.J.Hansen, Proposal ID: 0104.D-0059). Our observations were complemented with UVES archival data of comparable quality. The spectra were observed with different setups, using the BLUE346 and/or BLUE390 arms, and the RED564 and/or RED580 arms. The ranges of wavelength covered by different arms are: 303$<$$\lambda$$<$388 nm (BLUE346), 326$<$$\lambda$$<$454 nm (BLUE390), 458$<$$\lambda$$<$668 nm (RED564), and 476$<$$\lambda$$<$684 nm (RED580).
The details of the observations and the complete data set are presented in Table A.1. of Paper I. 

\section{Analysis}\label{Sect:analysis}
\subsection{Stellar parameters}

The stellar parameters for our sample of stars were derived in Paper I using photometry and parallaxes from the \textit{Gaia} Early Data Release 3 \citep[EDR3;][]{Gaia2016A&A...595A...1G, Gaia2021A&A...649A...1G}, and reddening maps from \citet{SchalflyFinkbeiner2011ApJ...737..103S}. Effective temperatures (\teff) and surface gravities (\logg) were derived iteratively until the variation between the parameters of consecutive iterations was $< 50$ K in \teff\ and $< 0.05$ dex in \logg. The macroturbulence velocities (\vt) were estimated using the calibration in \citet{Mashonkina2017A&A...604A.129M}. 
The [Fe/H] abundances were obtained using MyGIsFOS \citep{Sbordone2014A&A...564A.109S}, an automatic pipeline that measures abundances by comparing the observed spectral lines with a grid of synthetic spectra computed with the SYNTHE code \citep[see][]{Sbordone2004MSAIS...5...93S,Kurucz2005MSAIS...8...14K} and based on one-dimensional (1D), plane-parallel ATLAS12 model atmospheres \citep{Kurucz2005MSAIS...8...14K}, in the approximation of local thermodynamic equilibrium (LTE).
The stellar parameters and [Fe/H] derived in Paper I and adopted in this study are listed in Table A.1 in appendix A\footnote{\href{https://doi.org/10.5281/zenodo.14218032}{https://doi.org/10.5281/zenodo.14218032}}. 
The uncertainties on stellar parameters are $\Delta$\teff\,=\,100\,K, $\Delta$\logg\,=\,0.04\,dex, $\Delta$\vt\,=\,0.5\,\kms, and $\Delta$[Fe/H]\,=\,0.13\,dex (Paper I). 

\subsection{Abundances}
The chemical abundances for the target elements were derived from spectrum synthesis, by matching observed spectra with synthetic ones computed with the code MOOG \citep[version 2019]{Sneden2012MOOG}. The computed synthetic spectra are based on ATLAS12 \citep{Kurucz2005MSAIS...8...14K} model atmospheres assuming LTE. 
When computing the synthetic spectra, we assumed for the other elements the abundances derived in Paper I  and Paper II. 
Line lists for atomic and molecular species were generated with Linemake\footnote{\url{https://github.com/vmplacco/linemake}}\ \citep{Linemake,2021ascl.soft04027P}. The list of adopted lines for the studied elements is shown in Table A.2 in appendix A\footnote{\href{https://doi.org/10.5281/zenodo.14218032}{https://doi.org/10.5281/zenodo.14218032}}. 

\paragraph{\it Barium.}
Barium abundances were derived from three \ion{Ba}{II} lines at 5853.67 \AA, 6141.71 \AA, and 6496.90 \AA. For all lines we took into account the hyperfine and isotopic structure, as well as the $\log gf$, provided by \citet{Gallagher2020A&A...634A..55G}, and we assumed a solar isotopic ratio. The detailed list of \ion{Ba}{II} lines is shown in Table A.3\footnote{\href{https://doi.org/10.5281/zenodo.14218032}{https://doi.org/10.5281/zenodo.14218032}}. 
For nine stars in our sample, it was not possible to obtain Ba abundances because the available spectra for these stars only cover the wavelength range in the blue (BLUE346 or BLUE390), and not in the red, where the Ba lines are located. 

\paragraph{\it Lanthanum.}
We determined lanthanum abundances for our sample of stars using four \ion{La}{II} lines at 3949.10 \AA, 4086.71 \AA, 4123.22 \AA, and 4920.98 \AA. We adopted $\log gf$ values and hyperfine structures provided by \citet{Lawler2001ApJ...556..452L}. The adopted list of \ion{La}{II} lines is shown in Table A.4\footnote{\href{https://doi.org/10.5281/zenodo.14218032}{https://doi.org/10.5281/zenodo.14218032}}.
No hyperfine structure is available for the 4920 \AA\ line.

\paragraph{\it Cerium.}
Cerium abundances were derived using nine \ion{Ce}{II} lines at 3577.46 \AA, 3999.24 \AA, 4073.47 \AA, 4083.22 \AA, 4118.14 \AA, 4120.83 \AA, 4137.65 \AA, 4165.60 \AA, and 5274.23 \AA. The adopted $\log gf$ values are taken from \citet{Lawler2009ApJS..182...51L}.

\paragraph{\it Praseodymium.}
We derived praseodymium abundances using the \ion{Pr}{II} lines at 4408.81 \AA, 5259.73 \AA, and 5322.77 \AA. We adopted $\log gf$ values and hyperfine structures provided by \citet{Li2007PhyS...76..577L} and \citet{Ivarsson2001PhyS...64..455I}. The detailed list of Pr lines is shown in Table A.5\footnote{\href{https://doi.org/10.5281/zenodo.14218032}{https://doi.org/10.5281/zenodo.14218032}}.

\paragraph{\it Neodymium.}
Neodymium abundances were derived using eight \ion{Nd}{II} lines, at 3784.24 \AA, 3826.41 \AA, 4021.33 \AA, 4446.38 \AA, 4959.12 \AA, 5255.51 \AA, 5293.16 \AA, and 5319.81 \AA. The adopted $\log gf$ values are taken from \citet{DenHartog2003ApJS..148..543D}, while the isotopic and hyperfine components of the line at 4446.38 \AA\ are taken from \citet[][Table A.6\footnote{\href{https://doi.org/10.5281/zenodo.14218032}{https://doi.org/10.5281/zenodo.14218032}}]{Roederer2008ApJ...675..723R}.

\paragraph{\it Samarium.}
We derived samarium abundances from the \ion{Sm}{II} lines at 4434.32 \AA\ and 4704.40 \AA.
The adopted $\log gf$ values for these lines are from \citet{Lawler2006ApJS..162..227L}. 

\paragraph{\it Europium.}
Europium abundances were derived using the \ion{Eu}{II} lines at 3819.67 \AA, 4129.72 \AA, and 6645.06 \AA. 
We adopted $\log gf$ values, hyperfine and isotopic structures provided by \citet{Lawler2001ApJ...563.1075L}, and we assumed a solar isotopic ratio. The detailed list of Eu lines is shown in Table A.7\footnote{\href{https://doi.org/10.5281/zenodo.14218032}{https://doi.org/10.5281/zenodo.14218032}}.


\begin{table*}[ht]
\centering
 \caption[]{\label{tab:abundances}Chemical abundances with uncertainties for our sample of stars. 
 }
\begin{tabular}{lrrrrrrr}
 \hline \hline
  Star &  [FeII/H] &  nl(Ba) &  A(Ba) &  $\sigma$(Ba) &  [Ba/H] &  [Ba/Fe] &  ...\\
 \hline
  CES0031$-$1647 & $-$2.31 & 3 & $-$0.33 & 0.06 & $-$2.51 & $-$0.20  & ...\\
  CES0045$-$0932 & $-$2.80 & 3 & $-$1.42 & 0.03 & $-$3.60 & $-$0.80    & ...\\
  CES0048$-$1041 & $-$2.33 & 3 & 0.15 & 0.09 & $-$2.03 & 0.30    & ...\\
  CES0055$-$3345 & $-$2.24 & 3 & 0.10 & 0.06 & $-$2.08 & 0.16    & ...\\
  CES0059$-$4524 & $-$2.26 & 0 &  &  &  &                     & ...\\
  CES0102$-$6143 & $-$2.84 & 3 & $-$0.29 & 0.18 & $-$2.47 & 0.37  & ...\\
  ... & ... & ... & ... & ... & ... & ...  & ...\\
\hline
\end{tabular}
\tablefoot{ The value "nl(X)" corresponds to the number of measured lines for a given element X. 
The uncertainty "$\sigma$(X)" represents the line-to-line scatter when more than a single 
line of a given element X is available. The complete table is available in machine readable format at the CDS.}
\end{table*}

\section{Results}\label{Sect:results}

\begin{figure*}[h]
   \centering
   \includegraphics[width=0.33\hsize]{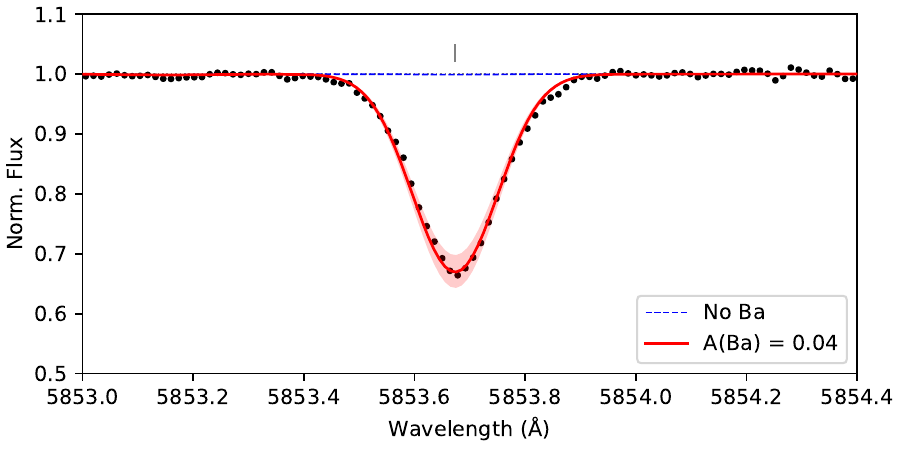}
   \includegraphics[width=0.33\hsize]{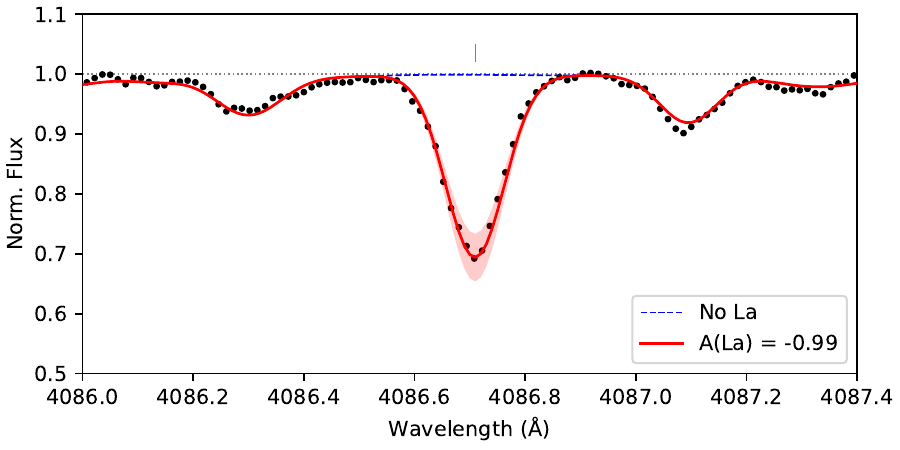}
   \includegraphics[width=0.33\hsize]{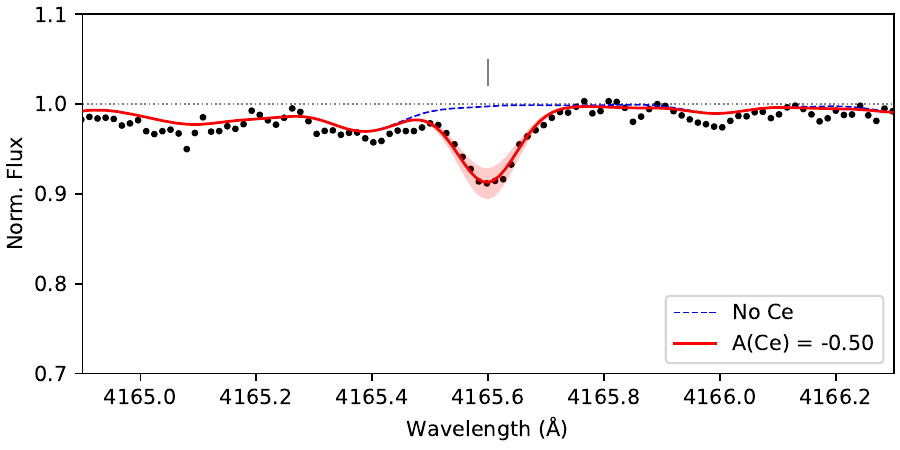}
   \includegraphics[width=0.33\hsize]{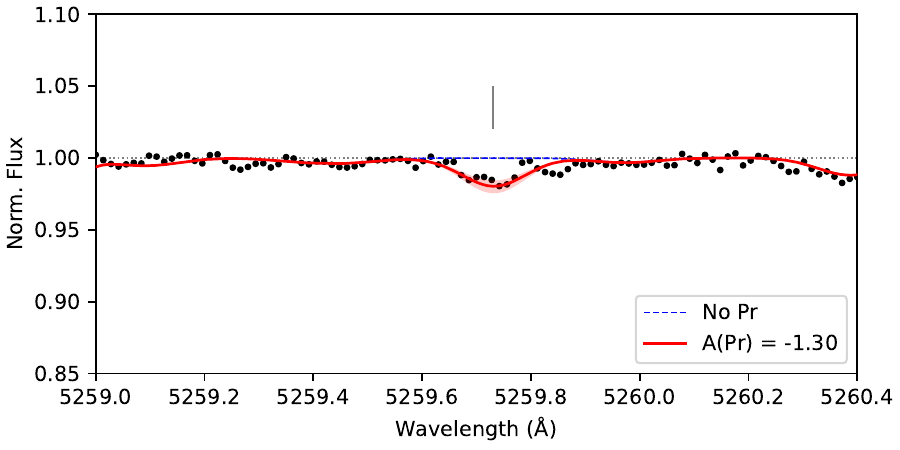}
   \includegraphics[width=0.33\hsize]{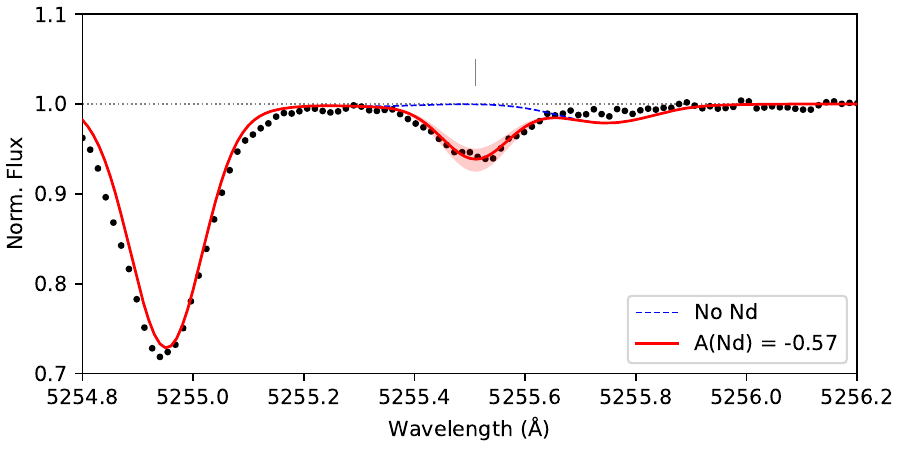}
   \includegraphics[width=0.33\hsize]{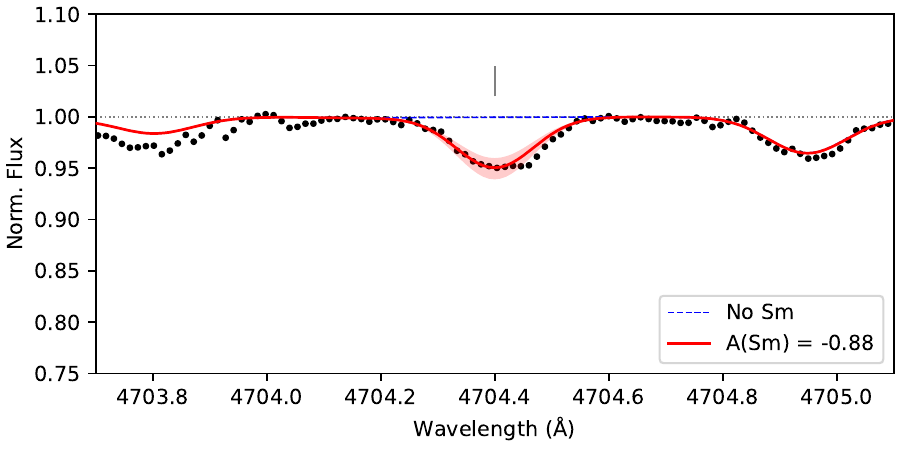}
   \includegraphics[width=0.33\hsize]{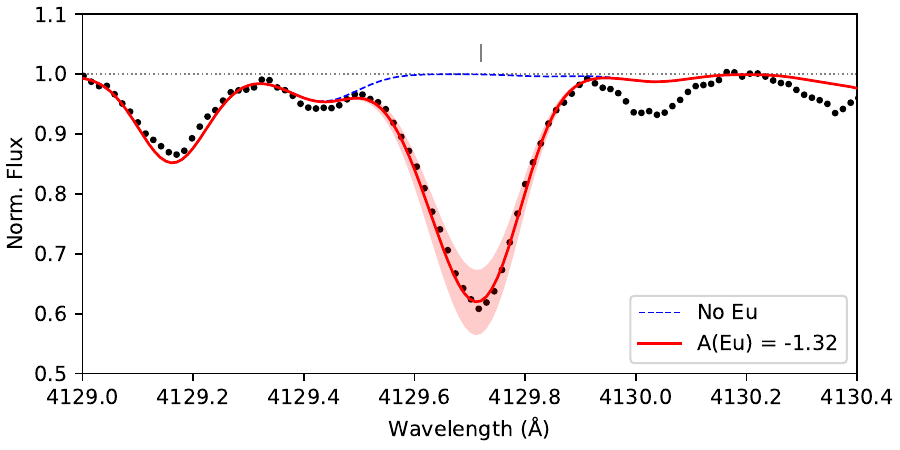}
     \caption{
     Portion of the spectra of the star CES0048$-$1041 around a selection of studied lines:  \ion{Ba}{II} at 5853.67 \AA, \ion{La}{II} at 4086.71 \AA, \ion{Ce}{II} at 4165.60 \AA, \ion{Pr}{II} at 5259.73 \AA, \ion{Nd}{II} at 5255.51 \AA, \ion{Sm}{II} at 4704.40 \AA, and \ion{Eu}{II} at 4129.72 \AA. The black dots represent the observed spectra. The red solid lines represent the best fit, with the respective A(X) shown in the label on the lower right corner of each panel. The red shaded areas show $\pm0.1$ dex interval around the best fit. The blue dashed lines show synthetic spectra without the studied element. 
              }
         \label{Fig:spec_synthesis}
   \end{figure*}

The derived chemical abundances with uncertainties or upper limits of our sample of stars are provided in machine readable format at the Centre de Donn\'ees astronomiques de Strasbourg  (CDS). An example is shown in Table~\ref{tab:abundances}.
The chemical abundances are expressed in the form A(X) and [X/H], where $\mathrm{A(X) = \log_{10}(X/H) + 12}$, and $\mathrm{[X/H] = \log_{10}(X/H) - \log_{10}(X/H)_\odot }$.
The adopted solar abundances $\mathrm{A(X)_\odot}$ are taken from \citet{Asplund2009}, except for Fe, for which we adopted the value from \citet{CaffauSun}, in order to be consistent with Paper I. 
Since we measured only singly ionised species, we refer to abundance ratios [X/Fe] as [\ion{X}{II}/\ion{Fe}{II}]\,=\,[\ion{X}{II}/H]--[\ion{Fe}{II}/H].
The uncertainty $\mathrm{\sigma(X)}$ represents the line-to-line scatter when more than a single line of a given element X is available. 
The chemical abundance errors due to uncertainties in stellar parameters are listed in Table~\ref{tab:sensitivities}.
These errors were estimated by varying the stellar parameters according to the uncertainties in the model atmosphere of the star CES0031--1647.
We selected this star as representative of the sample because its stellar parameters roughly coincide with the average stellar parameters of the sample.
An example of spectrosynthesis for a selection of lines of the studied elements in the star CES0048$-$1041 is shown in Fig~\ref{Fig:spec_synthesis}.
The derived abundances are generally in good agreement with those derived in previous studies, taking into account any differences in atmospheric parameters and atomic data. The detailed comparison is presented in appendix~\ref{App:comparison}.
In this study we also present for the first time the abundances of heavy elements in the star HE0428-1340 (CES0430-1334). In fact, only iron and carbon abundances for this star are present in literature. For six other stars in the sample, however, the only abundances in the literature are those provided by the 3rd data release of the GALAH survey \citep[GALAH DR3][]{DeSilva2015MNRAS.449.2604D,Buder2021MNRAS.506..150B}: TYC\,5922-517-1 (CES0547-1739), TYC\,4840-159-1 (CES0747-0405), TYC\,8931-1111-1 (CES0900-6222), TYC\,8939-2532-1 (CES0908-6607), TYC\,9200-2292-1 (CES0919-6958), and TYC\,9427-1414-1 (CES1413-7609). 
Thus, in our study, we present for the first time abundances obtained at high resolution and high signal-to-noise (S/N) ratio.

\begin{table}[h!]
\centering
 \caption[]{\label{tab:sensitivities}Sensitivities of chemical abundances with respect to the stellar parameters for a typical star in our sample (CES0031--1647).}
\begin{tabular}{lcccc}
 \hline \hline
  [X/H] & $\Delta$\teff & $\Delta$\logg & $\Delta$\vt & $\Delta$[Fe/H]\\
   & +100 K & +0.04 dex & +0.50 \kms & +0.13 dex\\
 \hline
  \ion{Ba}{II} & +0.08 & +0.01 & --0.14 & +0.03\\
  \ion{La}{II} & +0.08 & +0.01 & --0.01 & --0.01\\
  \ion{Ce}{II} & +0.10 & +0.04 & --0.01 & +0.00\\
  \ion{Pr}{II} & +0.09 & +0.01 & +0.01 & +0.00\\
  \ion{Nd}{II} & +0.07 & +0.00 & +0.01 & --0.01\\
  \ion{Sm}{II} & +0.07 & +0.01 & --0.02 & --0.02\\
  \ion{Eu}{II} & +0.07 & +0.01 & +0.01 & +0.01\\
\hline
\end{tabular}
\end{table}

\subsection{[X/Fe] versus [Fe/H]}

\begin{figure*}[h!]
  \centering
  \includegraphics[width=0.33\hsize]{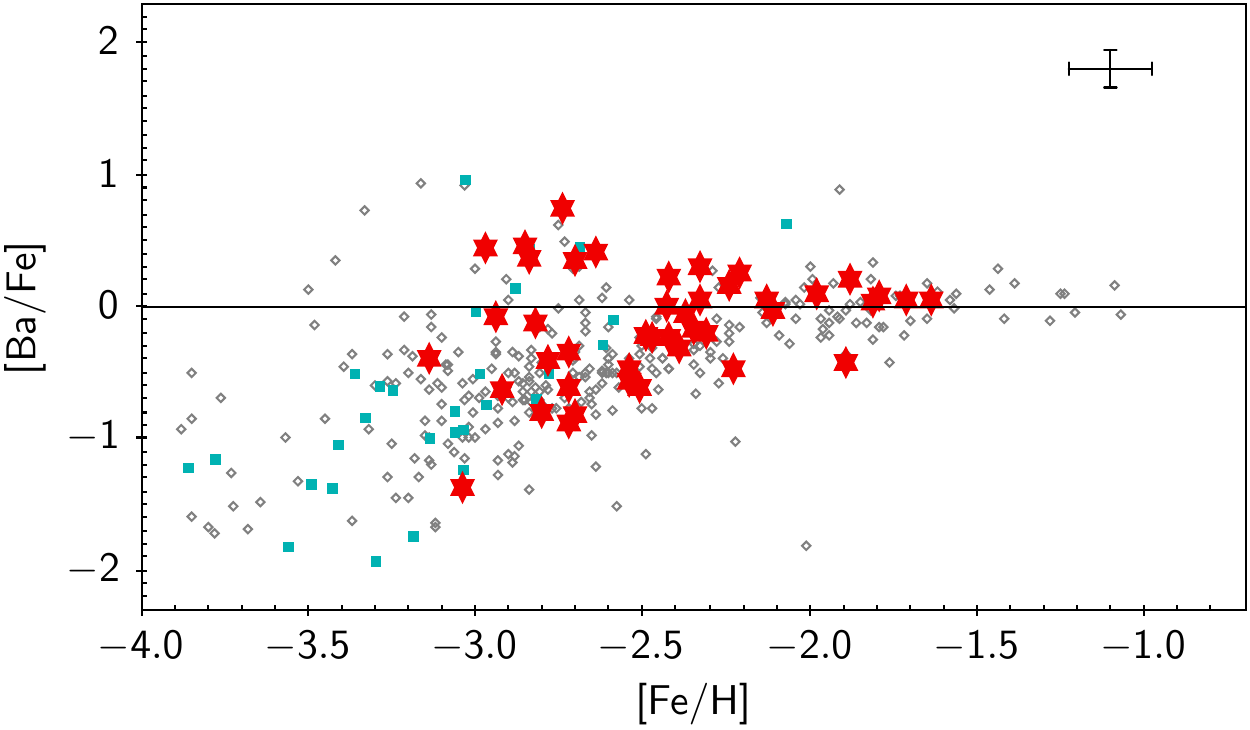}
  \includegraphics[width=0.33\hsize]{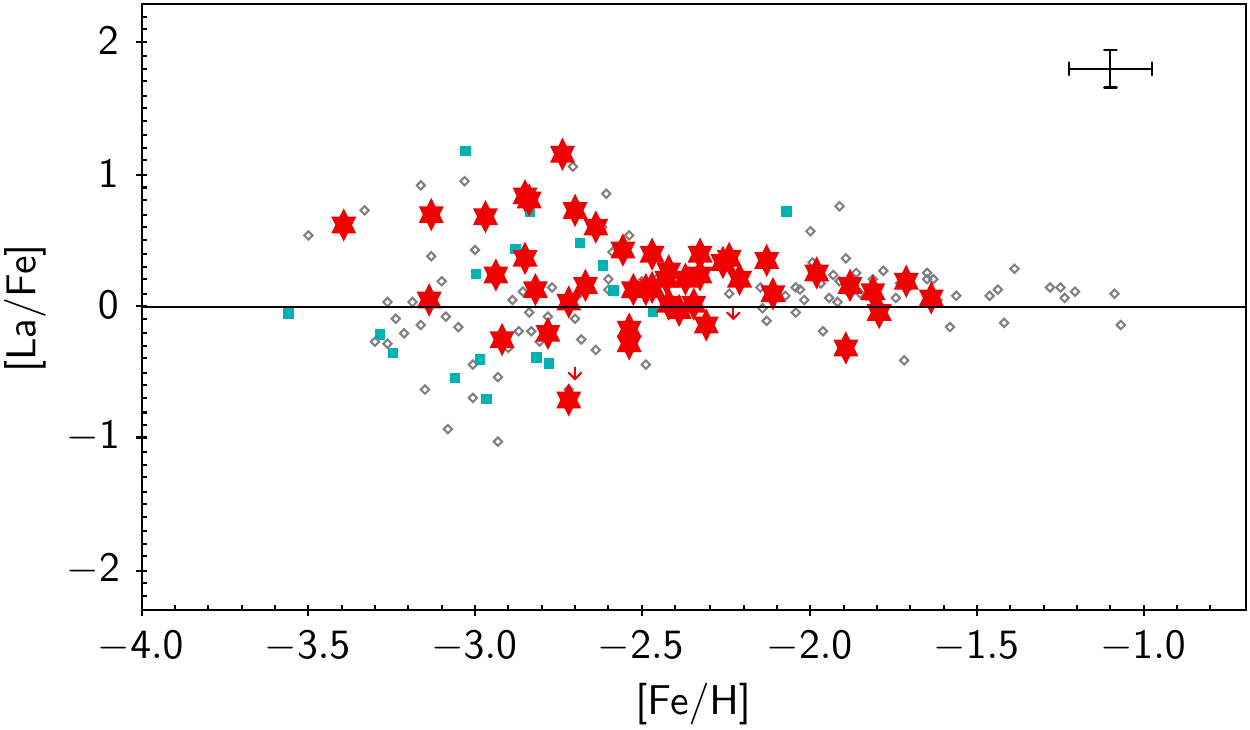}
  \includegraphics[width=0.33\hsize]{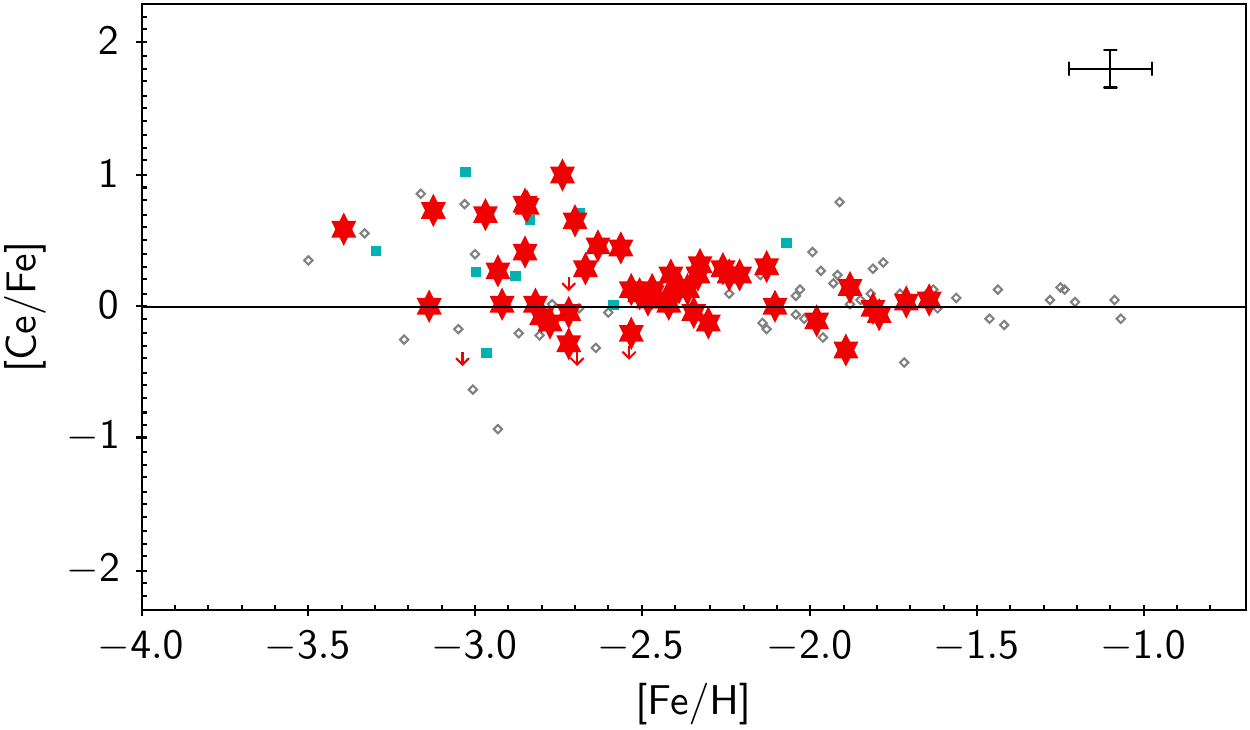}
  \includegraphics[width=0.33\hsize]{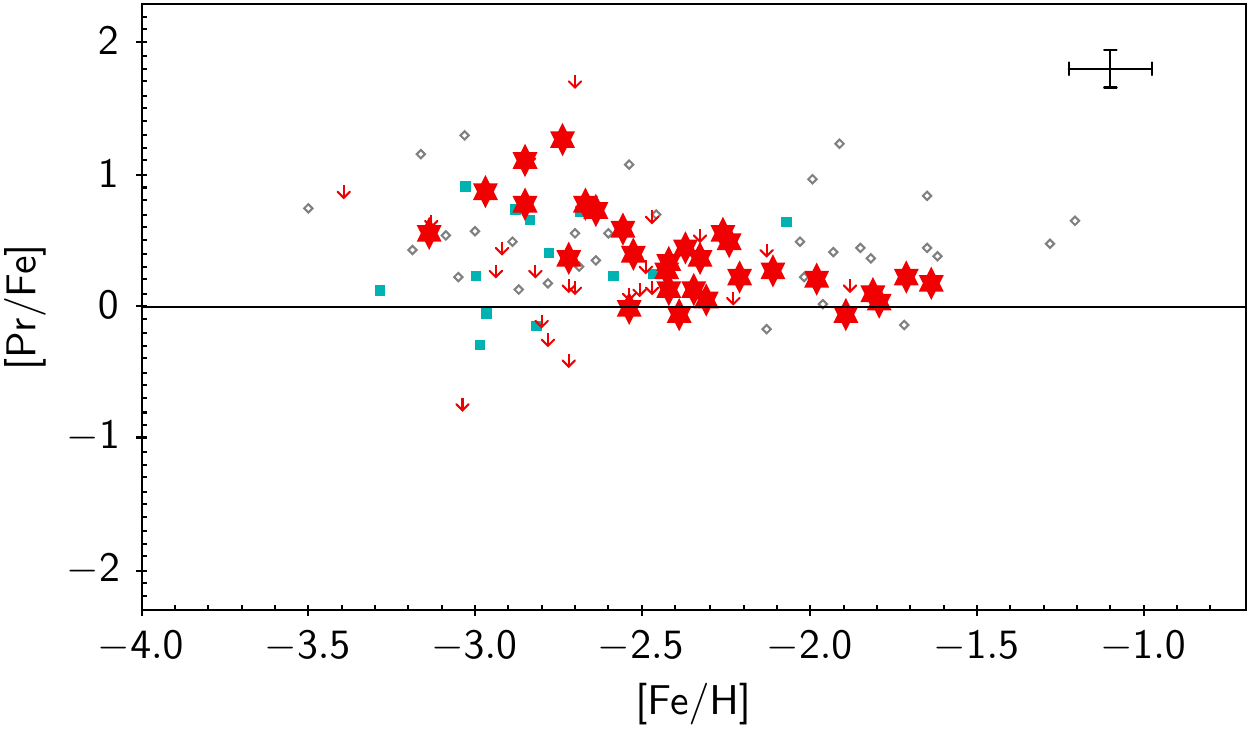}
  \includegraphics[width=0.33\hsize]{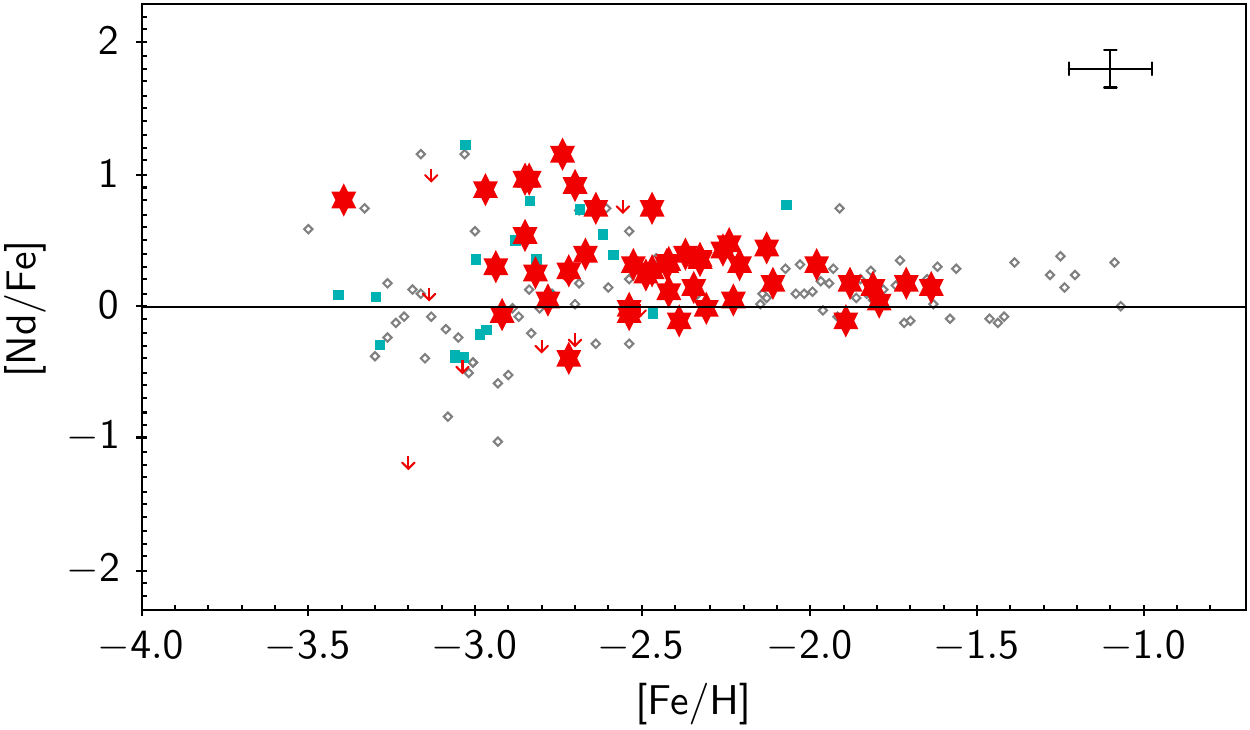}
  \includegraphics[width=0.33\hsize]{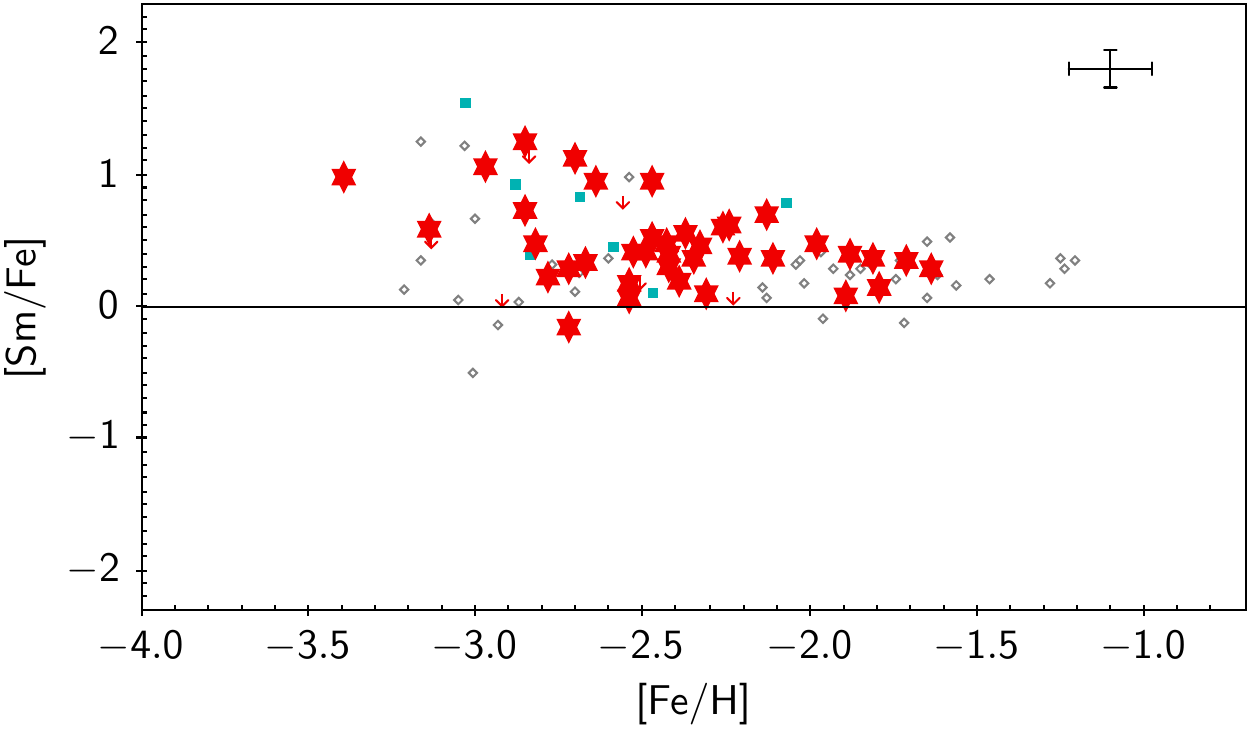}
  \includegraphics[width=0.33\hsize]{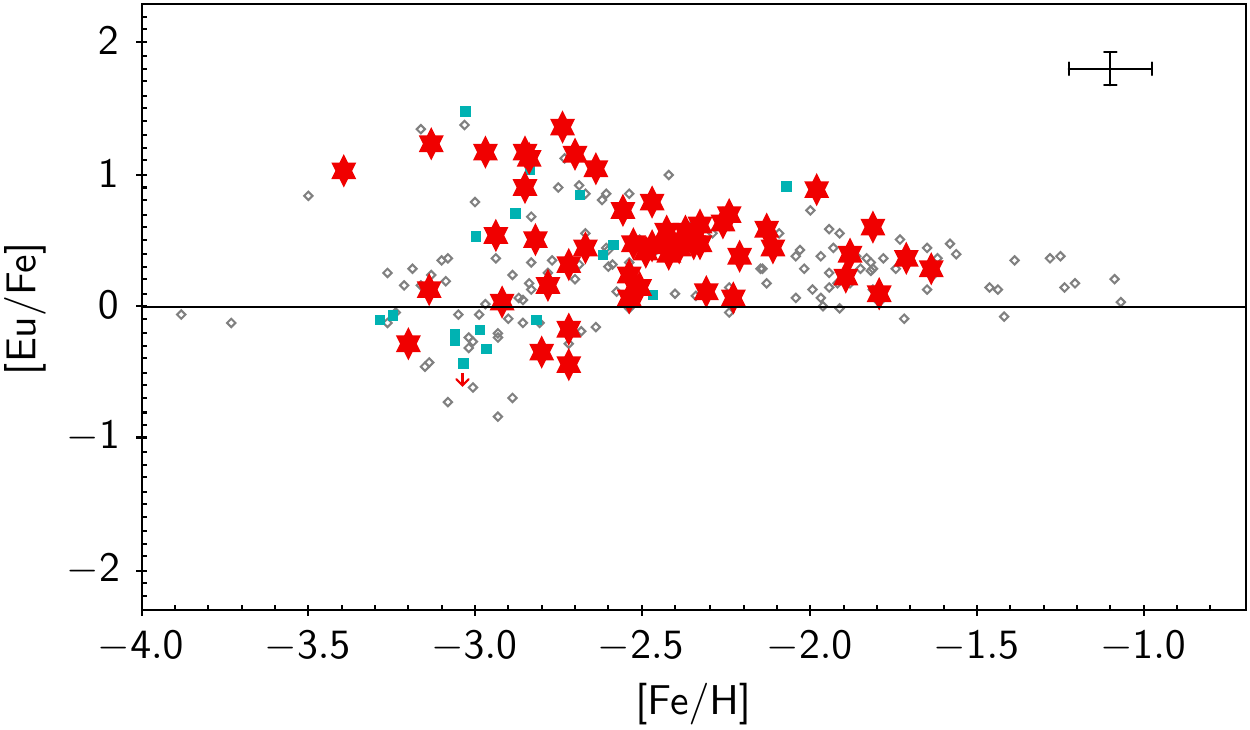}
  \caption{[Ba/Fe], [La/Fe], [Ce/Fe], [Pr/Fe], [Nd/Fe], [Sm/Fe], and [Eu/Fe] abundance ratios as a function of [Fe/H] for our sample of stars (red star symbols) compared to results obtained by \citealt{Francois2007A&A...476..935F} (cyan squares) and \citealt{Roederer2014AJ....147..136R} (grey open diamonds). 
             }
        \label{Fig:BaFe_LaFe_CeFe_PrFe_NdFe_SmFe_EuFe_FeH}
  \end{figure*}

Figure~\ref{Fig:BaFe_LaFe_CeFe_PrFe_NdFe_SmFe_EuFe_FeH} shows the abundance ratios [X/Fe] as a function of the metallicity ([Fe/H]=[\ion{Fe}{II}/H]) for our stellar sample, compared to the results obtained by \citet{Francois2007A&A...476..935F} and \citet{Roederer2014AJ....147..136R}. 
Our results are in good agreement with the literature data, and do not show any systematic offset with previous analyses. 
Similarly to other previous studies, we observe a large dispersion in [Ba/Fe] ratios, which increases at [Fe/H]$<$$-2.5$. 
For [Fe/H]$<$$-3.0$, the literature samples show a decreasing trend with metallicity, with many stars having [Ba/Fe] lower than solar yet with a fair fraction of stars having enhanced Ba \citep[especially in carbon enhanced metal-poor stars, see][]{Hansen2016A&A...588A..37H, Hansen2019A&A...623A.128H}. 
Our results seem to agree with this general trend, although we were only able to derive Ba abundances for two stars at these metallicities.
For [La/Fe] and [Ce/Fe] abundance ratios, at lower metallicities, we observe a large spread at [Fe/H]$<$$-2.5$,  
similarly to the one found for [Ba/Fe], but we do not observe the same declining trend with decreasing metallicity.
In our opinion, this mismatch is more likely due to an observational bias rather than to a different production site and/or mechanism 
as other studies show La abundance behaviour similar to that of Ba \citep{Hansen2012A&A...545A..31H}.
In fact, for metallicities below $-3$, it is usually only possible to derive upper limits on the abundances of La and Ce, unless the stars are enriched in these elements, while the Ba lines are strong enough to allow an actual measurement.
A high dispersion is also observed for [Pr/Fe] and [Nd/Fe] at [Fe/H]\,$<$\,$-2.5$, which increases towards lower metallicities. 
Similarly to La and Ce, the trend at the lowest metallicities is still not clear, as usually only upper limits can be obtained. 
The [Sm/Fe] and [Eu/Fe] abundance ratios behave similar to the other heavy n-capture elements, displaying an increasing scatter for metallicities below $-2.5$ dex.

\subsection{[X/H] versus [C/N]}
In Paper~II, we investigated the effect of extra mixing that occurs after the red giant branch (RGB) bump, which alters the surface abundances of C and N \citep{Gratton2000A&A...354..169G}, and whether it could have altered the abundance ratios of light n-capture elements, that is,  Sr, Y, and Zr.
The results indicate that there is no clear correlation between their abundances and [C/N] ratio, which suggests that the abundances are unaffected by mixing. 
In the present work, we also found no clear correlation for n-capture elements investigated here, that is, Ba, La, Ce, Pr, Nd, Sm, and Eu, as shown in Fig. C.1\footnote{\href{https://doi.org/10.5281/zenodo.14218032}{https://doi.org/10.5281/zenodo.14218032}} in appendix C.
This suggests that even these heavy elements are unaffected by the extra mixing.

\subsection{NLTE corrections}

Barium and europium lines are known to be sensitive to departures from LTE, also referred to as non-LTE (NLTE) effects. The NLTE effects on these lines can be strong in metal-poor stars, so it is important to understand how this effect affects the derived Ba and Eu abundances in order to investigate the nucleosynthesis of these elements.
To study the NLTE effect on our derived abundances, we applied 1D NLTE corrections for Ba and Eu lines computed by L.I.Mashonkina\footnote{\url{https://spectrum.inasan.ru/nLTE2/}} database \citep{Mashonkina2023MNRAS.524.3526M}. 
The available grids include 1D NLTE abundance corrections for five \ion{Ba}{II} lines (4554.0298 \AA, 4934.0801 \AA, 5853.7002 \AA, 6141.6099 \AA, and 6496.8999 \AA) and for 11 Eu lines (3724.9299 \AA, 3819.6699 \AA, 3907.1101 \AA, 3930.5 \AA, 3971.97 \AA, 4129.7002 \AA, 4205.02 \AA, 4435.5801 \AA, 4522.5801 \AA, 6437.6401 \AA, and 6645.0601 \AA)
in the range of stellar parameters covered by metal-poor stars. 
The details for the computation of NLTE corrections are described in \citet{Mashonkina2019AstL...45..341M} for \ion{Ba}{II} and \citet{Mashonkina2000A&A...364..249M} for \ion{Eu}{II}. 

   \begin{figure*}[h!]
   \centering
   \includegraphics[width=0.49\hsize]{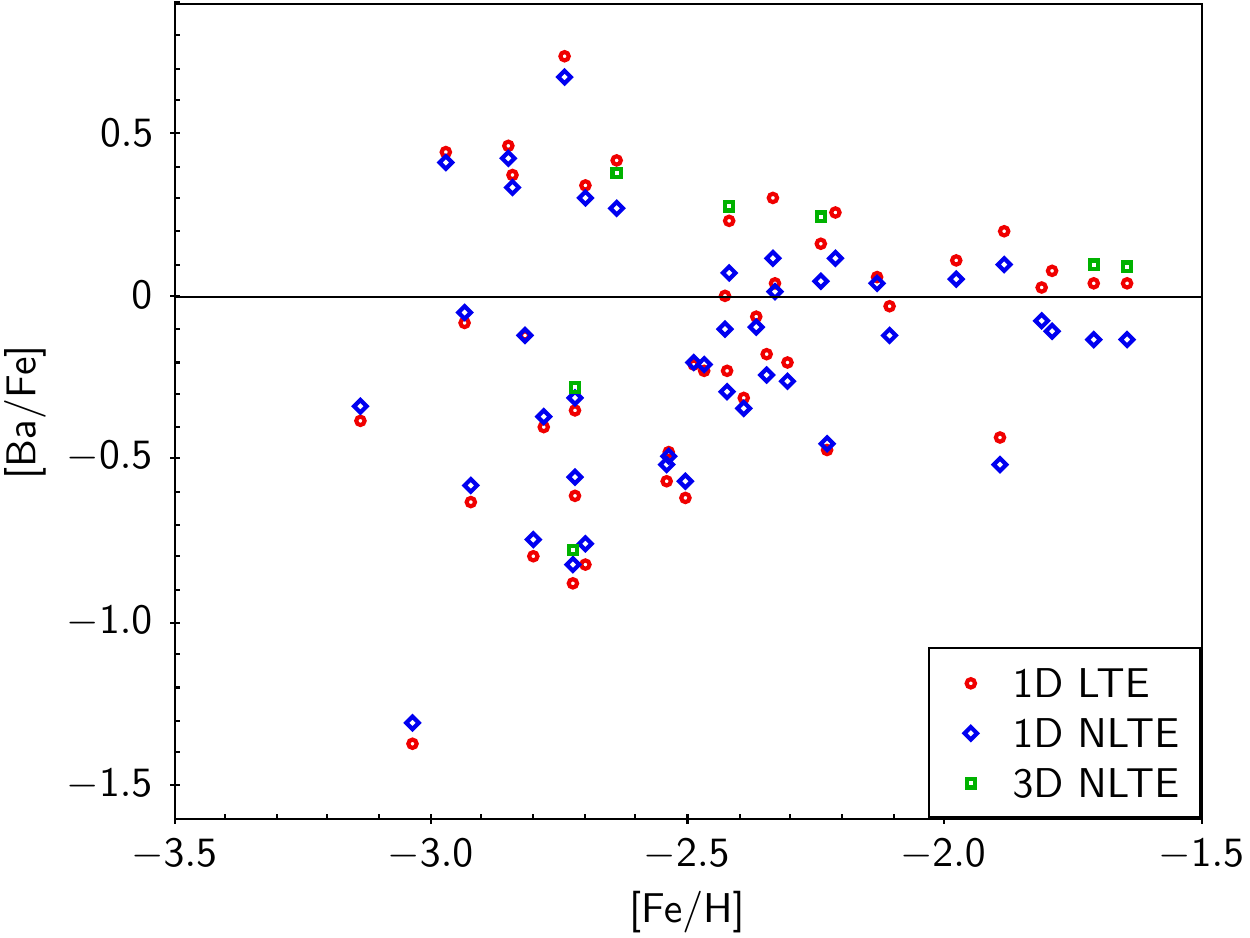}
   \includegraphics[width=0.49\hsize]{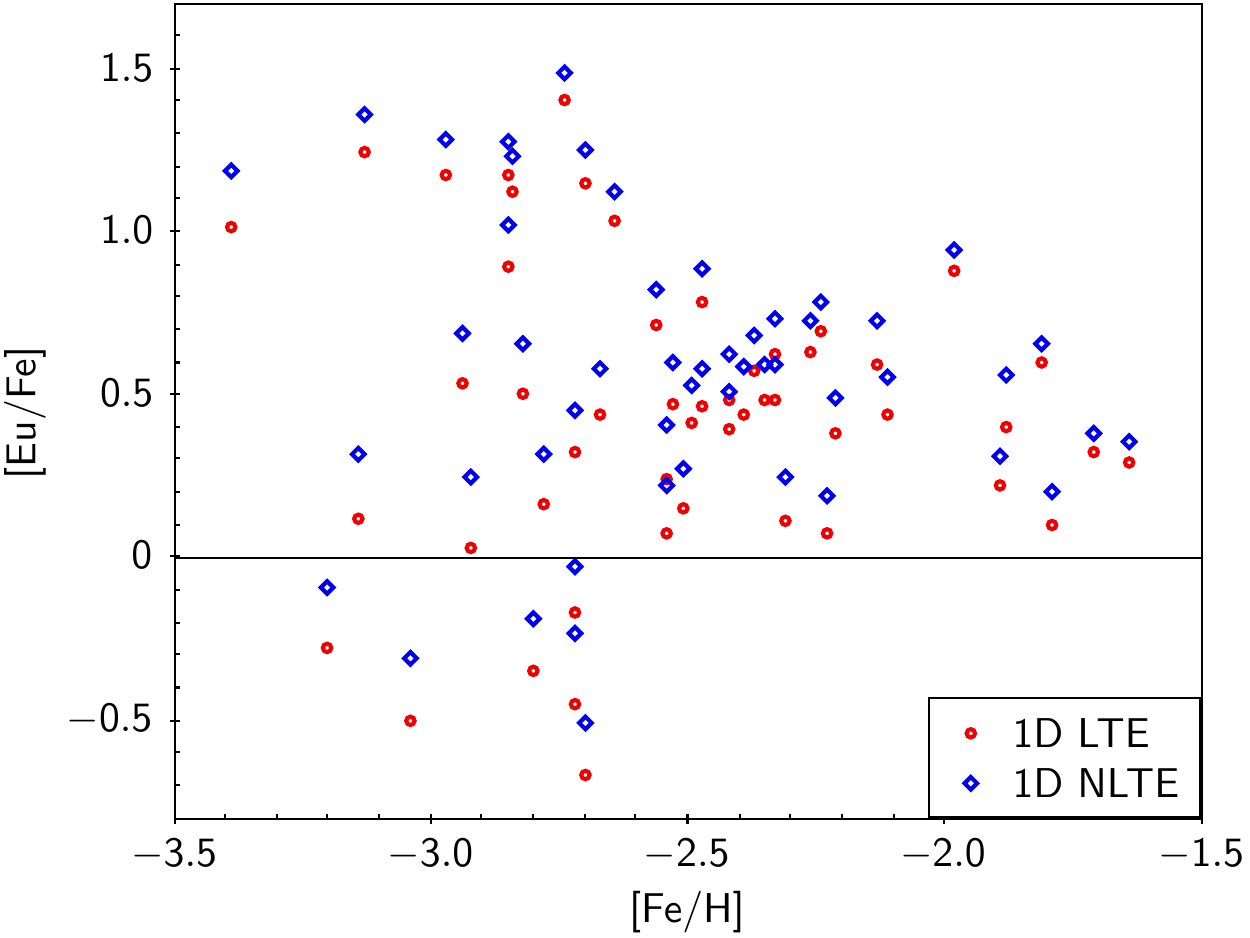}
      \caption{ [Ba/Fe] and [Eu/Fe] abundance ratios as a function of [Fe/H]. Red open circles represent 1D LTE abundances. Other coloured symbols are the computed 1D NLTE (blue open diamonds) and 3D NLTE (green open squares) abundance corrections.
              }
         \label{Fig:BaFe_EuFe_FeH_nlte}
   \end{figure*}

Figure \ref{Fig:BaFe_EuFe_FeH_nlte} shows [Ba/Fe] and [Eu/Fe] abundance ratios as a function of [Fe/H] before and after applying NLTE corrections.
The 1D NLTE corrections for Ba tend to be larger for stars with [Fe/H]$>$$-2.5$ and [Ba/Fe]$>$0 and to decrease the Ba abundances. 
They also seem to slightly reduce the dispersion at metallicities below $-2.5$, but the large scatter is still present. 
On the other hand, the 1D NLTE corrections for Eu are all positive, and on average they increase the Eu abundances by 0.13 dex. 
For consistency, \ion{Fe}{II} abundances should also be corrected for NLTE effects. 
Contrary to \ion{Ba}{II} and \ion{Eu}{II} lines, the NLTE effects on \ion{Fe}{II} lines seem to be negligible, even in very metal-poor stars \citep{Mashonkina2011A&A...528A..87M,Bergemann2012MNRAS.427...27B,Lind2012MNRAS.427...50L}. 
Nevertheless, we checked if the NLTE effects on \ion{Fe}{II} lines in our stars were actually negligible using the 1D NLTE corrections for \ion{Fe}{II} lines provided by \citet{Bergemann2012MNRAS.427...27B}\footnote{\url{https://nlte.mpia.de/index.php}}. 
Since the NLTE corrections were available only for a sub-sample of the adopted lines, we compared the LTE \ion{Fe}{II} abundances derived from the lines in the sub-sample to the corresponding abundances corrected for NLTE effects. 
We found that only for the \ion{Fe}{II} line at 4923.932 \AA\ the NLTE corrections are above 0.01 in all stars. 
It varies from +0.02 to +0.08 dex for this line, however, even when the correction is as large as +0.08, the resulting difference between the overall Fe II LTE and NLTE abundance does not exceed 0.005 dex.
For this reason, Fig.~\ref{Fig:BaFe_EuFe_FeH_nlte} shows [Fe/H] in LTE.

Additionally, we investigated the three-dimensional (3D) NLTE effects on Ba lines using 3D NLTE - 1D LTE abundance corrections grids computed by Gallagher et al. (in prep). 
The grids are available on the ChETEC-INFRA site\footnote{\url{https://www.chetec-infra.eu/3dnlte/}}, which provides the details of the grid computation and the instructions on how to derive and apply the 3D NLTE corrections. 
Seven stars in our sample have stellar parameters inside the grid range, and the [Ba/Fe] abundances ratios corrected for 3D NLTE corrections are shown as green open squares in Fig.~\ref{Fig:BaFe_EuFe_FeH_nlte}. 
We note that the 3D NLTE corrections are, on average, smaller than the 1D NLTE corrections and positive in sign, that is,  they tend to increase Ba abundances.

   \begin{figure}[h!]
   \centering
   \includegraphics[width=\hsize]{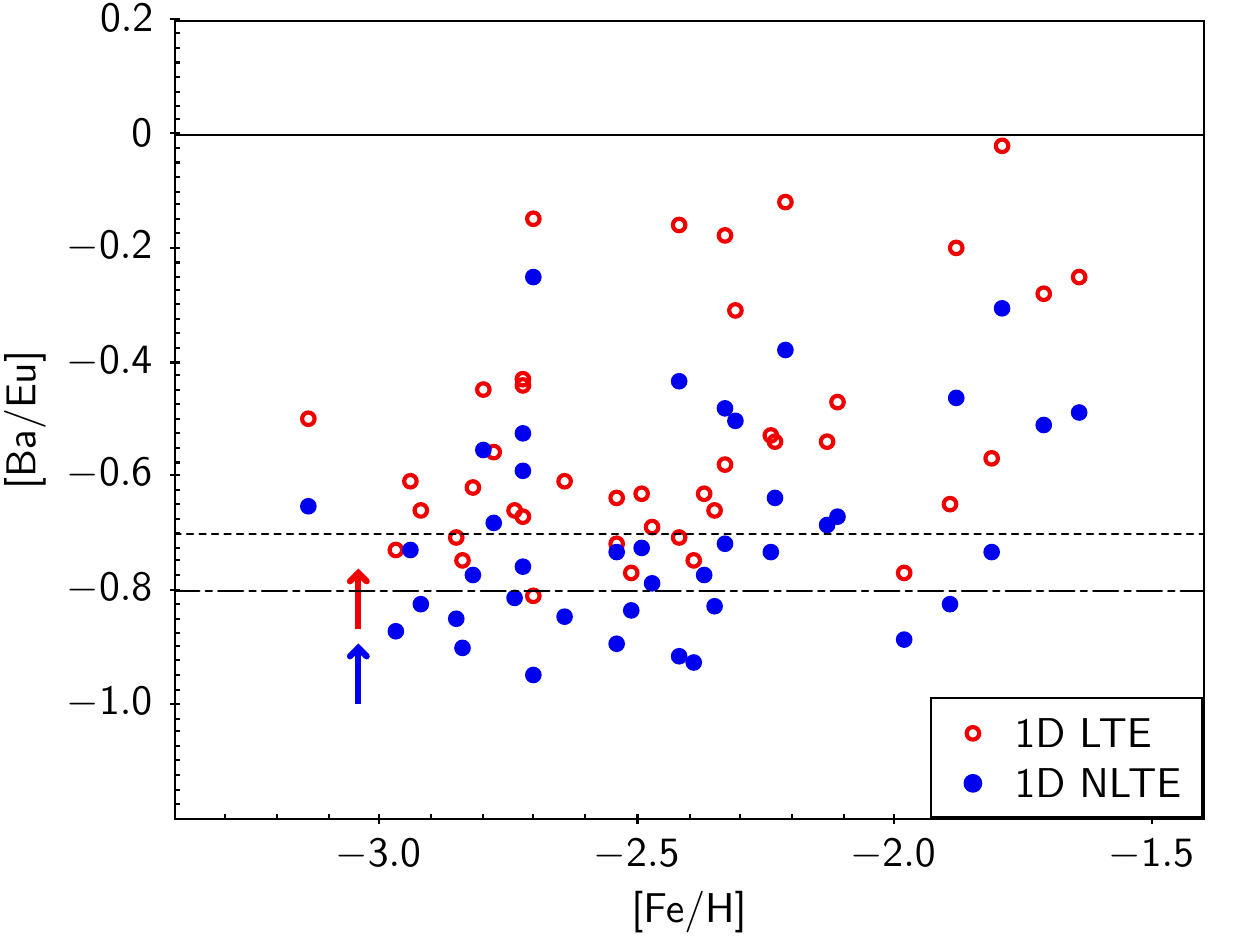}
   \caption{Comparison between 1D LTE (red open circles) and 1D NLTE (blue circles) [Ba/Eu] ratios as a function of [Fe/H] for our sample of stars. The dashed black line represent the Solar System pure $r$-process value according to \citealt{Arlandini1999ApJ...525..886A}, while the dashed-dotted line represent the same value according to \citealt{Bisterzo2014ApJ...787...10B}.}
   \label{Fig:baeu_feh_lte_nlte}
   \end{figure}

\citet{Mashonkina2014A&A...565A.123M} evaluated the empirical $r$-process log(Ba/Eu)\footnote{With log(Ba/Eu)=A(Ba)$-$A(Eu)} ratio using homogeneously derived 1D NLTE abundances in a sample of very metal-poor $r$-process enhanced stars with [Eu/Fe]$>$1. 
In their study, the authors found that, for their sample of stars, NLTE corrections lead to lower Ba, but higher Eu abundances. 
This is in agreement with what we observe in our sample.
They also found an average log(Ba/Eu)=0.78$\pm$0.06 in 1D NLTE for the $r$-process enhanced stars. 
For the six $r$-process enhanced stars with [Eu/Fe]$>$1 in our sample we find a mean log(Ba/Eu)=0.79$\pm$0.04 in 1D NLTE, which corresponds to [Ba/Eu]=$-0.87$. 
This result is in excellent agreement with the previous findings.

Figure~\ref{Fig:baeu_feh_lte_nlte} shows [Ba/Eu] ratios as a function of [Fe/H] in 1D LTE and 1D NLTE. 
In 1D LTE, 16 out of 43 stars have [Ba/Eu]$\leq$$-0.7$ within uncertainties, where [Ba/Eu]=$-0.7$ is the Solar System pure $r$-process value according to \citet[][hereafter A99]{Arlandini1999ApJ...525..886A}. 
If we instead take [Ba/Eu]=$-0.8$ as the Solar System pure $r$-process value \citep[][hereafter B14]{Bisterzo2014ApJ...787...10B}, we note that the number of stars compatible with solar pure $r$-process [Ba/Eu] is reduced to 7 out of 43. 
When the 1D NLTE corrections are applied, we see that 30 out of 43 stars have [Ba/Eu]$\leq$$-$0.7 within errors, and 24 have also [Ba/Eu]$\leq$$-$0.8. 
Since the 1D NLTE corrections are available only for Ba and Eu, the results presented in this study are 1D LTE abundances. 
As a reference for the contributions of $s$- and $r$- processes in the Solar System, we refer to the values presented in \citet{Arlandini1999ApJ...525..886A}.

\begin{figure*}[h!]
   \centering
   \includegraphics[width=0.40\hsize]{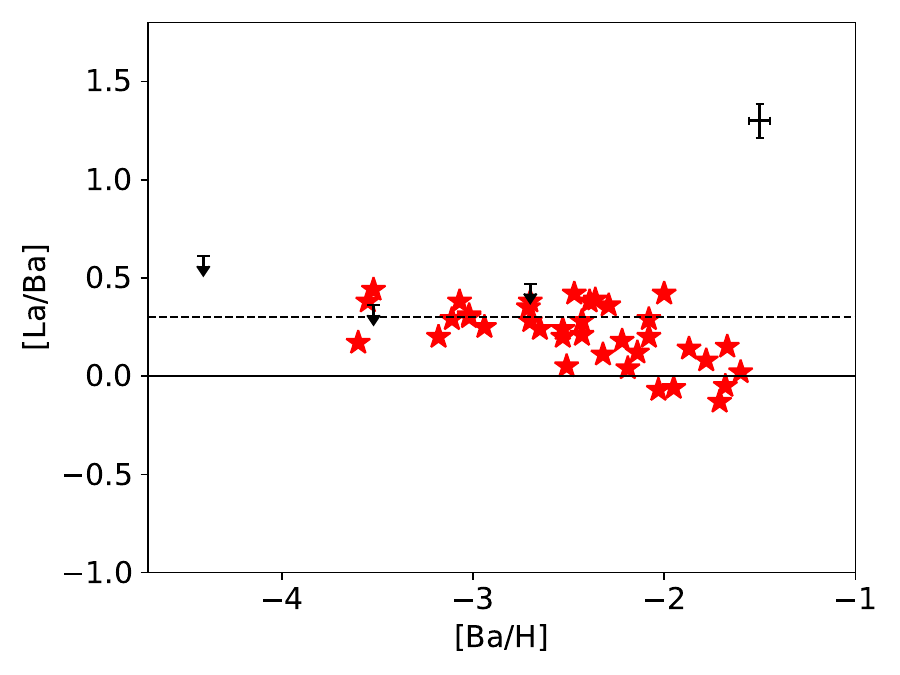}
   \includegraphics[width=0.40\hsize]{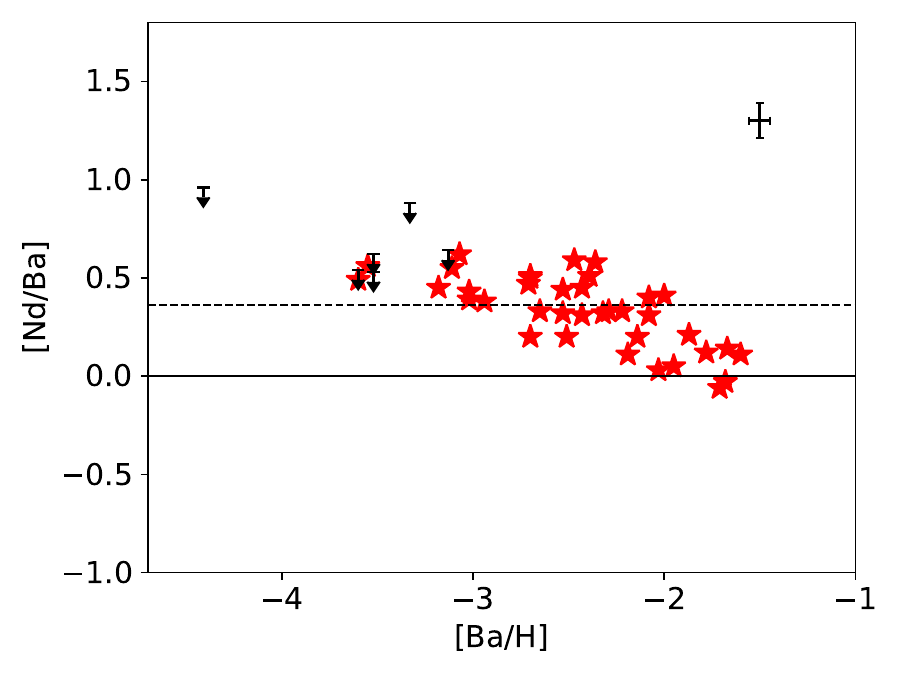}
   \includegraphics[width=0.40\hsize]{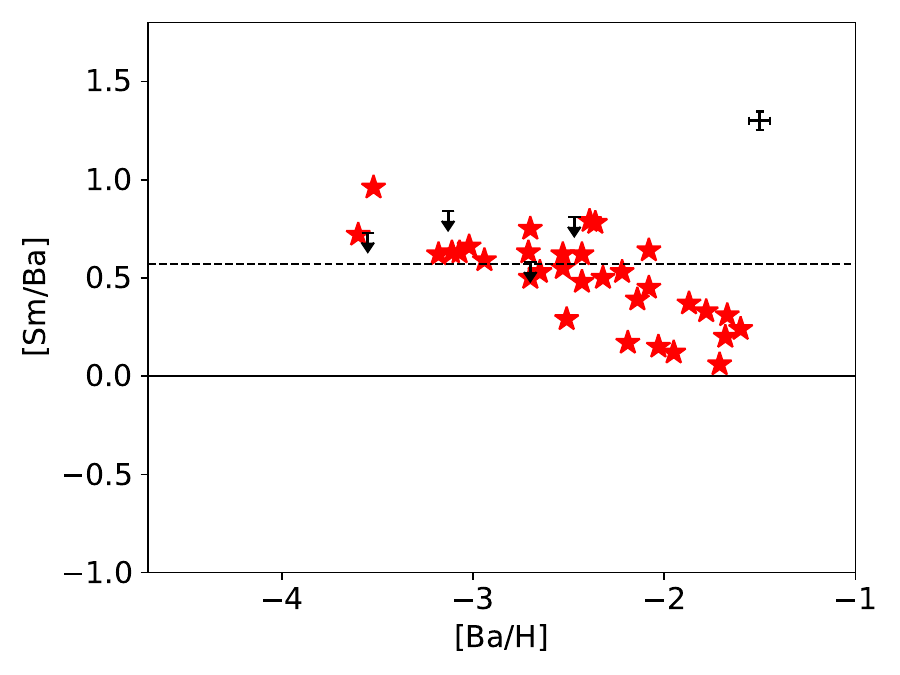}
   \includegraphics[width=0.40\hsize]{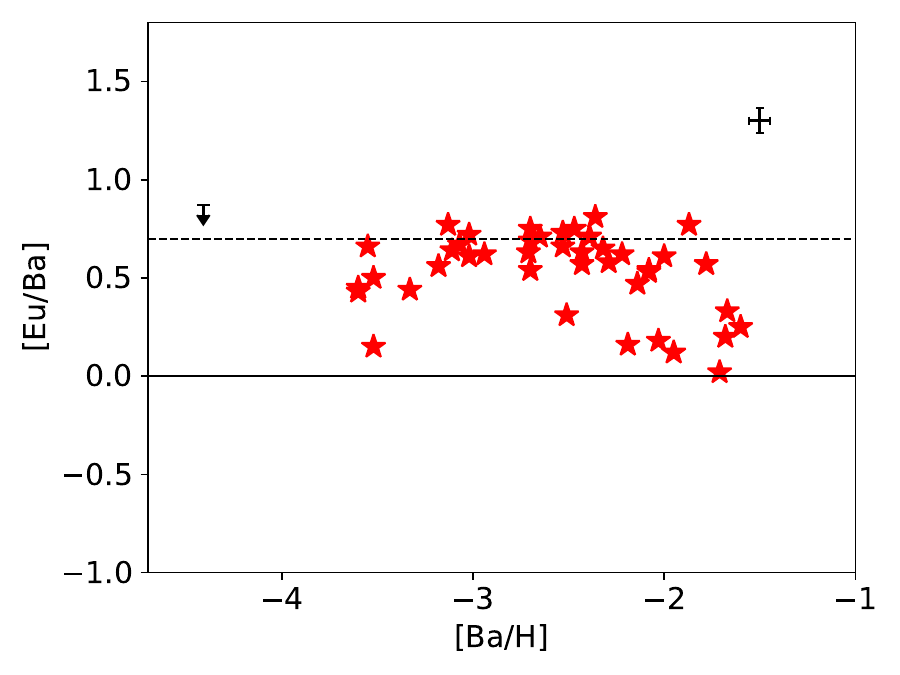}
      \caption{[La/Ba], [Nd/Ba], [Sm/Ba] and [Eu/Ba] as a function of [Ba/H] for our sample of stars. The black arrows represent upper limits of abundance ratios. 
      A representative error bar is shown in the upper-right corner of each panel. 
      The dashed black lines represent the Solar System pure $r$-process values according to \citet{Arlandini1999ApJ...525..886A}.
              }
         \label{Fig:correlations_bah}
   \end{figure*}


\section{Discussion}\label{Sect:discussion}

\subsection{$s$-process versus $r$-process}

The observed increasing scatter of [Ba/Fe] ratios towards low metallicities shown in Fig.~\ref{Fig:BaFe_LaFe_CeFe_PrFe_NdFe_SmFe_EuFe_FeH}  
is thought to be caused by a variation of the Ba nucleosynthesis at [Fe/H]\,$\approx$\,$-2.5$ and below \citep[see e.g.][]{GrattonSneden1994A&A...287..927G,McWilliam1998AJ....115.1640M,Burris2000ApJ...544..302B,Honda2004ApJ...607..474H,Simmerer2004ApJ...617.1091S,Francois2007A&A...476..935F, Hansen2012A&A...545A..31H}. 
At later times and higher metallicities, the $s$-process elements such as Ba are dominantly produced by low- and intermediate-mass AGB stars 
\citep[see e.g.][]{Busso1999ARA&A..37..239B,Kappler2011RvMP...83..157K,Karakas2014PASA...31...30K}.
However, in the early Galaxy, AGB stars did not have time to significantly enrich the interstellar medium given their long lifetimes. 
Fast-rotating massive stars (FRMS) are expected to be a source of $s$-process at this epoch, but their contribution to the nucleosynthesis of elements around (and heavier than) Ba is predicted to be overall small  \citep[see e.g.][]{Frischknecht2016MNRAS.456.1803F}.  
Therefore, at metallicities below $\sim$$-2.5$, the $r$-process is more likely the primary production mechanism of Ba. 
Similarly to Ba ($s$-process: 81\%, A99), most La and Ce in the Solar neighbourhood were produced by the $s$-process (62\% and 77\%, A99), 
while Pr and Nd were more equally produced by the $s$- and $r$-processes ($s$-process: 49\% and 56\%, A99). 
It is therefore likely that the production of all those elements at low metallicities occurs also predominantly through the $r$-process.

Figures~\ref{Fig:correlations_bah} and C.2-C.6\footnote{\href{https://doi.org/10.5281/zenodo.14218032}{https://doi.org/10.5281/zenodo.14218032}}  in appendix C show the correlations between heavy elements pairs ([element1/element2] versus [element2/H]) for our sample of stars. 
In Fig.~\ref{Fig:correlations_bah}, [La/Ba], [Nd/Ba], [Sm/Ba], and [Eu/Ba] as a function of [Ba/H] are shown. 
We note that in each plot, for values of [Ba/H]$\approx$$-2.5$, there is a change in the trends: for [Ba/H]$\gtrsim$$-2.5$ the scatter in the abundance ratios increases, and a decreasing trend seems to appear as [Ba/H] increases. 
For [Ba/H]$\lesssim$$-2.5$, the observed trend is generally flat, and the stars tend to clump around the Solar System pure $r$-process value computed by A99. 
To estimate at which values of [Ba/H] and [Fe/H] the possible onset of the $s$-process occurs in the chemistry of our sample stars, we applied a mean shift clustering algorithm with a flat kernel using the Python module Scikit-learn\footnote{\url{https://scikit-learn.org/stable/index.html}} \citep{Meanshift,scikit-learn}.
Mean shift clustering is a non-parametric, density-based clustering algorithm that can be used to identify clusters in a data set.
Given a set of data points, the algorithm shifts each data point towards the maximum increase in the density of points within a certain radius. 
The operation is iterated until the points converge to a local maximum of the density function, which correspond to the cluster centroid. 
The main advantages of the mean shift clustering algorithm are that it does not require the number of clusters to be specified in advance and can handle arbitrary shapes and sizes of clusters.

\begin{figure}[h!]
   \centering
   \includegraphics[width=0.8\hsize]{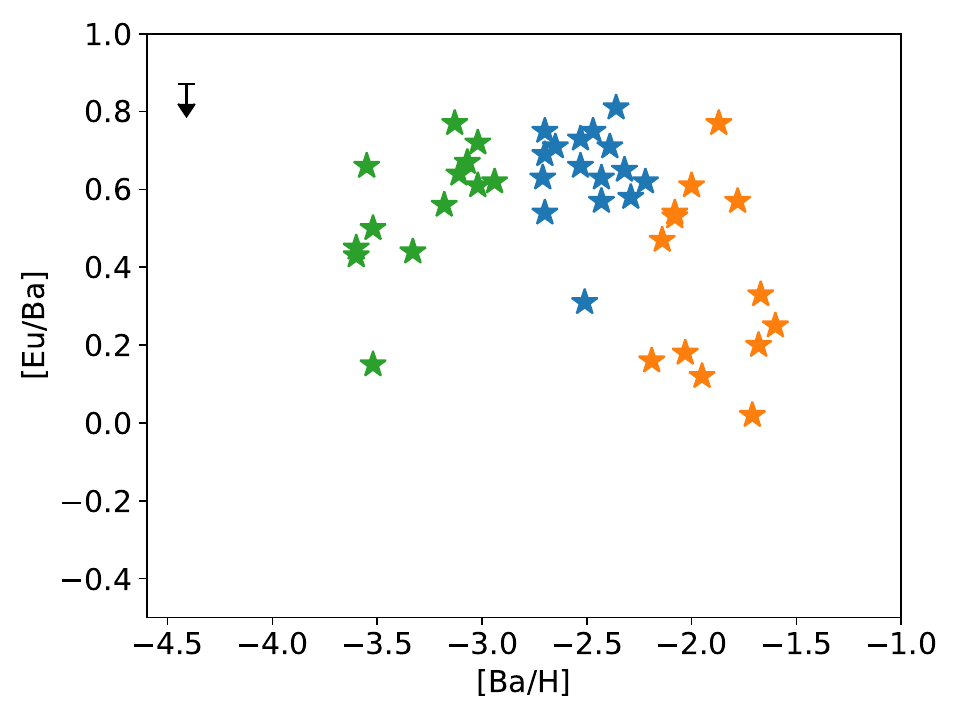}
      \caption{[Eu/Ba] as a function of [Ba/H] for our sample of stars after applying the mean shift clustering algorithm. The three clusters found by the algorithm are visible in different colours. The star CES1237+1922 is represented by a black arrow (upper limit), and it was not considered when applying the clustering algorithm.
              }
         \label{Fig:EuBa_BaH_meanshift}
   \end{figure}

We applied the mean shift clustering to [Eu/Ba] abundance ratios as a function of [Ba/H] and identified three clusters as shown in Fig.~\ref{Fig:EuBa_BaH_meanshift}.
We calculated the mean, median, and standard deviation of [Ba/H] and [Fe/H] for the stars belonging to the cluster that corresponds to the transition region between the region with [Eu/Ba]$\sim$constant and the region with a large dispersion in [Eu/Ba] (blue cluster in Fig.~\ref{Fig:EuBa_BaH_meanshift}). 
We then repeated the procedure for [La/Ba], [Nd/Ba], and [Sm/Ba] ratios as a function of [Ba/H] to check if the values were different adopting different abundance ratios. 
The results are listed in Table~\ref{tab:clustering}. 
The median of the median values we obtained for [Fe/H] is $-2.42$ ($\sigma$=0.02) and for [Ba/H] is $-2.42$ ($\sigma$=0.04). 
These results suggest that the change in the trend happens at [Ba/H]=$-2.4$, which corresponds to a metallicity of [Fe/H]=$-2.4$. 
For [Ba/H]$<$$-2.4$, the flat trend and the clustering around the solar pure $r$-process values support the scenario of the $r$-process as the primary production mechanism of Ba (and other heavy n-capture elements) at low metallicities. 
On the other hand, the larger scatter observed for [Ba/H]$>$$-2.4$ can be interpreted as the onset of the $s$-process contribution in Ba nucleosynthesis. 

Our results are in line with previous studies of neutron capture elements in the early Galaxy.
\citet{Burris2000ApJ...544..302B}, in the context of the Bond survey, found the presence of $s$-nuclei of Ba already in stars with a metallicity of about $-2.7$, but inferred finally that the bulk of the $s$-processing was delayed until [Fe/H]
$\simeq -2.2$.
From analysis of the [La/Eu] ratios, \citet{Simmerer2004ApJ...617.1091S} concluded that the $s$-process may be active as early as [Fe/H]=$-2.6$.
\citet{Hansen2012A&A...545A..31H} suggested that the contribution of the $s$-process might start at [Fe/H]=$-2.5$, given the change in the observed abundance trends of [Sr/Fe], [Y/Fe], [Zr/Fe], and [Ba/Fe] as a function of [Fe/H].

\begin{table*}[h!]
\centering
 \caption[]{\label{tab:clustering}Mean, median and standard deviations for [Fe/H] and [Ba/H]. }
\begin{tabular}{lcccccc}
 \hline \hline
  [X/Ba] & Mean$_{\rm[Fe/H]}$ & Median$_{\rm[Fe/H]}$ & $\sigma_{\rm[Fe/H]}$ & Mean$_{\rm[Ba/H]}$ & Median$_{\rm[Ba/H]}$ & $\sigma_{\rm[Ba/H]}$\\
 \hline
  La & $-$2.52 & $-$2.43 & 0.32 & $-$2.41 & $-$2.43 & 0.10\\
  Nd & $-$2.48 & $-$2.40 & 0.32 & $-$2.38 & $-$2.41 & 0.14\\
  Sm & $-$2.46 & $-$2.40 & 0.33 & $-$2.38 & $-$2.41 & 0.14\\
  Eu & $-$2.48 & $-$2.43 & 0.27 & $-$2.50 & $-$2.49 & 0.16\\
\hline
\end{tabular}
\tablefoot{These values were found by using mean shift clustering on [X/Ba] versus [Ba/H]}
\end{table*}

   \begin{figure}[h!]
   \centering
   \includegraphics[width=0.8\hsize]{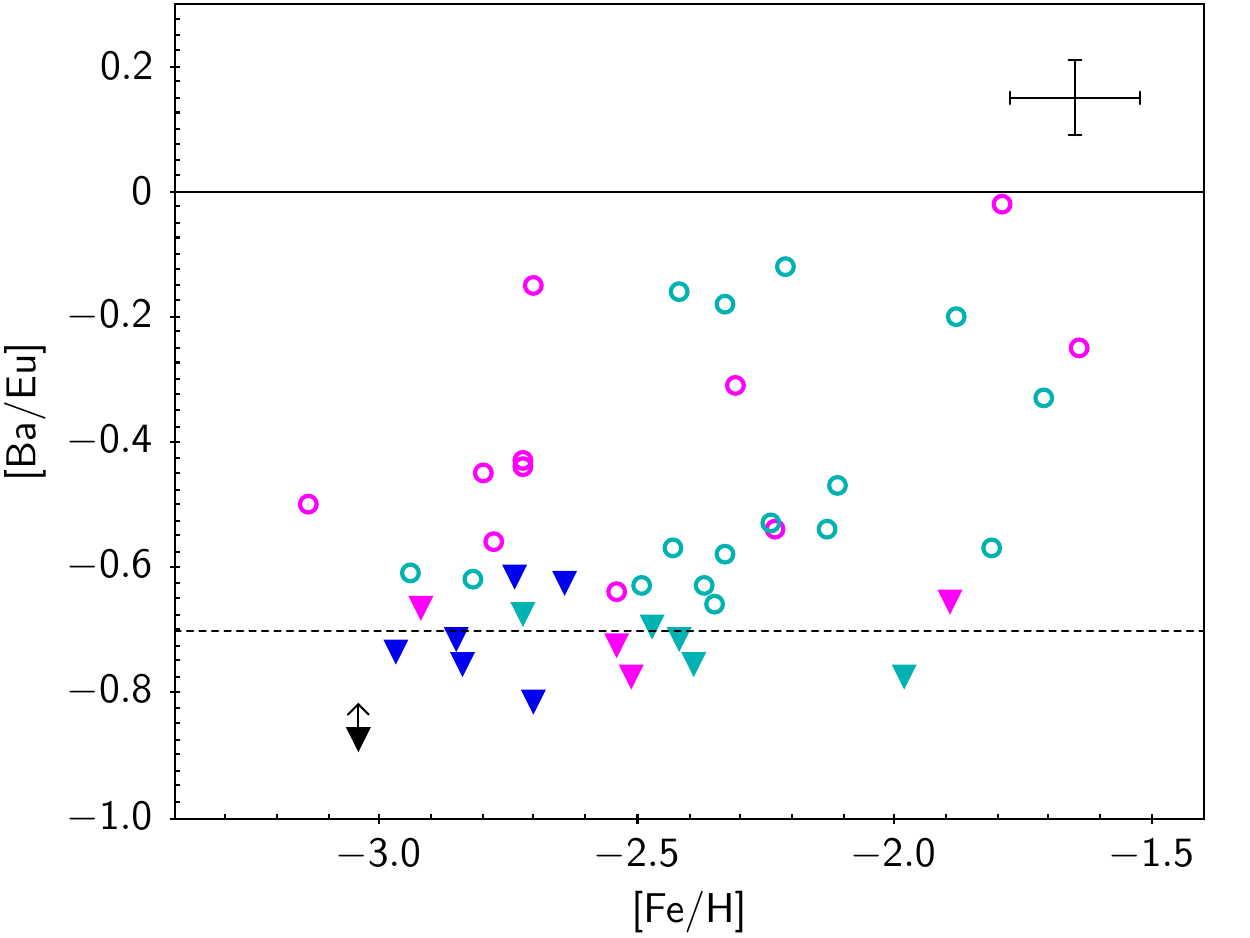}
      \caption{[Ba/Eu] abundance ratios as a function of [Fe/H]. Filled triangles are stars compatible within uncertainties with [Ba/Eu]$\leq$$-0.7$ (Solar System pure $r$-process according to A99). 
      Open circles are stars with [Ba/Eu]$>$$-0.7$.  
      Symbols are colour coded according to the \citet{Christlieb2004A&A...428.1027C} classification for $r$-process enhanced stars. Blue, cyan, and magenta symbols represent $r$-II, $r$-I, and 
      $r$-poor stars respectively. The black triangle indicates the star CES1237+1922.
              }
         \label{Fig:BaEu_FeH}
   \end{figure}

Figure~\ref{Fig:BaEu_FeH} shows [Ba/Eu] abundance ratios as a function of [Fe/H] for our sample of stars. 
We see that all the 43 stars with both Ba and Eu measurements in our sample have [Ba/Eu]$<$0. 
At these low [Fe/H], we expect the majority of stars to have [Ba/Eu]$<$0, because the $r$-process is the dominant process in the heavy elements' nucleosynthesis. 
Following the classification from \citet{Christlieb2004A&A...428.1027C}, 27 are $r$-process-rich stars ($r$-rich; [Eu/Fe]$\geq$0.3), of which six are $r$-II stars ([Eu/Fe]$>$1), and 21 are $r$-I stars (0.3$\leq$[Eu/Fe]$\leq$1.0). 
The remaining 16 stars have [Eu/Fe]$<$0.3, and we shall refer to them as $r$-process-poor or $r$-poor stars. 
In our sample, 16 out of 43 stars are compatible with [Ba/Eu]$\leq$$-0.7$ within uncertainties ($r$-pure), which is the Solar System pure $r$-process value from A99.
We note that all the $r$-II stars are compatible within errors with the Solar System pure $r$-process value. 
Among the $r$-I stars, only five out of 21 are compatible with the $r$-pure value. 
It is interesting to notice that also five 
$r$-poor stars ([Eu/Fe]$<$0.3) are compatible with the $r$-pure value. 
Figure~\ref{Fig:BaEu_FeH} also seems to indicate that stars with [Ba/Eu] ratios consistent with the $r$-pure value are found at metallicities at least as high as [Fe/H]=$-1.9$. 

\begin{figure*}[h!]
   \centering
   \includegraphics[width=0.40\hsize]{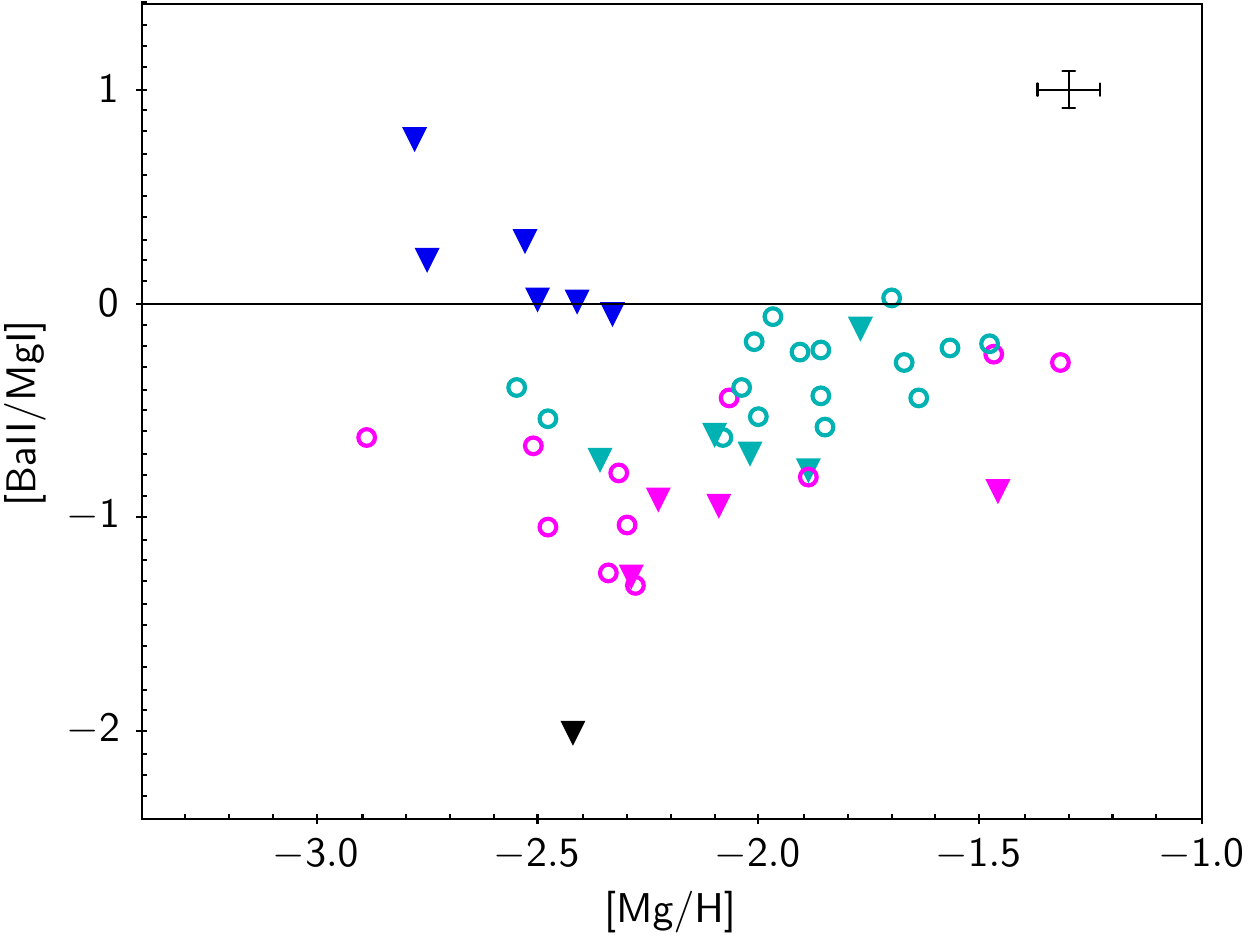}
   \includegraphics[width=0.40\hsize]{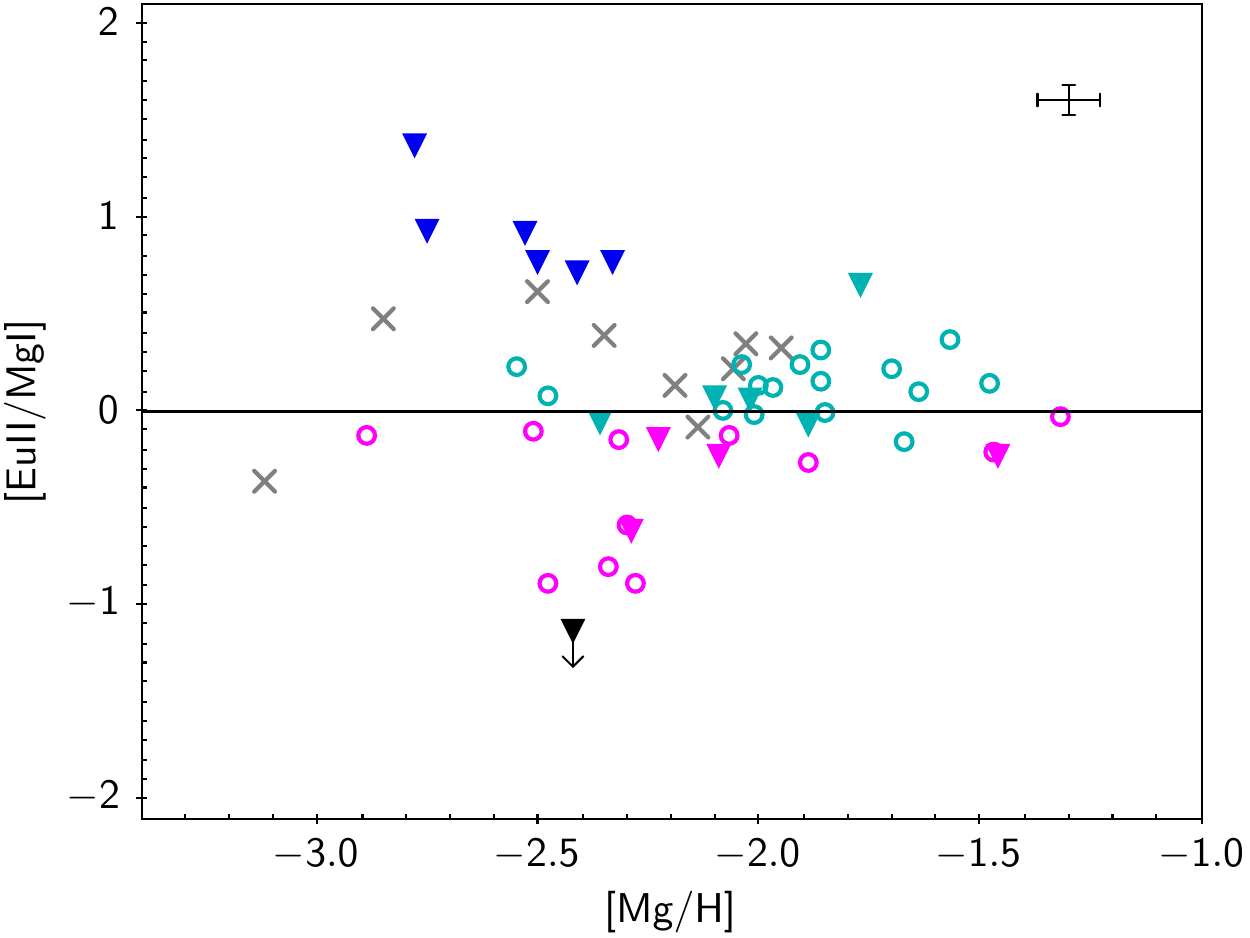}
      \caption{[Ba/Mg] and [Eu/Mg] abundance ratios as a function of [Mg/H]. Coloured symbols as in Fig.~\ref{Fig:BaEu_FeH}. Grey crosses represent stars without Ba abundance measurement. The black triangle indicates the star CES1237+1922.
              }
         \label{Fig:BaMg_EuMg_MgH}
   \end{figure*}

To understand if the stars compatible with $r$-pure [Ba/Eu] follow some trends or show a common behaviour, we checked for possible correlations between Ba and Eu with other elements. 
Mg is mainly produced in massive stars and released into the interstellar medium by core collapse supernovae (CC SNe or SNe II), contrary to Fe which can also be produced in SNe type Ia.   
For this reason, Mg has been suggested as an alternative `reference element' to Fe when investigating the Galactic chemical evolution \citep[see e.g.][]{Cayrel2004A&A...416.1117C, Mashonkina2017A&A...608A..89M}. 
Figure~\ref{Fig:BaMg_EuMg_MgH} shows [Ba/Mg] and [Eu/Mg] as a function of [Mg/H] for our sample of stars. 
Similarly to Fig.~\ref{Fig:BaFe_LaFe_CeFe_PrFe_NdFe_SmFe_EuFe_FeH}, the general trends of [Ba/Mg] and [Eu/Mg] seem to change at [Mg/H]$\sim$$-2.2$: for [Mg/H]$>$$-2.2$, [Ba/Mg] and [Eu/Mg] abundance ratios follow a flat trend, with an average of [Ba/Mg]=$-0.42$ ($\sigma$=0.26) and [Eu/Mg]=0.08 ($\sigma$=0.21) respectively, while for [Mg/H]$<$$-2.2$ the abundance ratios show a large scatter, with values spanning over 2 dex.
We note that in the [Ba/Mg] versus [Mg/H] plot, the $r$-II stars stand out from the overall decreasing trend of the other stars, showing [Ba/Mg]$\gtrsim$0 abundance ratios. 
The same is visible in the [Ba/Fe] versus [Fe/H] plot in Fig.~\ref{Fig:BaFe_LaFe_CeFe_PrFe_NdFe_SmFe_EuFe_FeH}. 
We also note that the $r$-pure stars do not show any particular trend compared to the other stars in the sample when comparing to Mg.

The wide dispersion observed at [Mg/H]$<$$-2.2$ for both Ba and Eu seems to suggest that these elements are not co-produced with Mg, implying that normal SNe II are unlikely to be the main or dominant formation sites of Ba and Eu.
This scenario is supported by nucleosynthesis models, which show that classic CC SNe would only be able to produce elements lighter than Ba \citep[see e.g.][and references therein]{Hansen2014ApJ...797..123H, ArconesThielemann2023A&ARv..31....1A}. 
However, some special (and rare) classes of SNe, such as MRD SNe, depending on the explosion dynamics, may produce heavy elements through the $r$-process, 
thus being able to explain the large dispersion observed at low metallicities \citep[see e.g.][]{Winteler2012ApJ...750L..22W, Nishimura2015ApJ...810..109N,Nishimura2017ApJ...836L..21N, Moesta2018ApJ...864..171M, Halevi2018MNRAS.477.2366H, Reichert2021MNRAS.501.5733R, Reichert2023MNRAS.518.1557R, Reichert2024MNRAS.529.3197R}. 
For this reason, MRD SNe together with NSMs, are considered to be some of the main sources of $r$-process at low metallicities \citep[see e.g.][]{Cowan2021RvMP...93a5002C}. 
All yields contribute to the next generation of stars via mixing processes in the ISM. To find the best tracers, mono-enrichment is desired \citep[see e.g.][]{Magg2020MNRAS.498.3703M,Hansen2020A&A...643A..49H} so further conclusions would require mono-enriched stars as well as more complete abundance patterns.

\subsection{Light versus heavy n-capture elements}

Light n-capture elements, such as Sr, are produced in different ways as a function of metallicity, but mainly by the $s$-process in the solar inventory (85\%, A99). 
Previous studies of metal-poor stars in the Milky Way pointed out that the large scatter observed in [Sr/Ba] at low [Ba/Fe] suggests the presence of an additional process to the main $r$-, that produce light n-capture elements (for example,  Sr, Y, Zr) but not the heavy ones, such as Ba  \citep{Travaglio2004ApJ...601..864T,Honda2004ApJ...607..474H,Francois2007A&A...476..935F, Qian2008ApJ...687..272Q,Andrievsky2011A&A...530A.105A,Hansen2012A&A...545A..31H, Roederer2013AJ....145...26R, Yong2013ApJ...762...26Y, Hansen2014ApJ...797..123H, Spite2014A&A...571A..40S, Spite2018A&A...611A..30S, Han2021RAA....21..111H}.
The nucleosynthesis processes and astrophysical sites associated with the production of light n-capture elements are still a matter of debate, as several processes could be involved \citep[see e.g. the review by][]{ArconesThielemann2023A&ARv..31....1A}. 
As discussed in Paper~I, neutrino-driven winds in CC SNe could be a possible formation site for light n-capture elements. In these environments, if the conditions are mildly neutron-rich, elements up to Z$\sim$50 can be produced by the weak $r$-process \citep[weak-$r$; see e.g.][]{Arcones2014JPhG...41d4005A}, whereas if the conditions are proton-rich, these nuclei can also be produced by the $\nu$p-process \citep[see e.g.][]{Wanajo2006ApJ...647.1323W, Froehlich2006PhRvL..96n2502F, Pruet2006ApJ...644.1028P}. 
Both processes are able to produce abundance patterns compatible with the ones observed for light n-capture elements in metal-poor stars \citep{Arcones2011ApJ...731....5A}, therefore the observationally derived abundances could be produced by the weak $r$-process, the $\nu$p-process, a main r-process or a combination of those \citep[see e.g.][and references therein]{Hansen2014ApJ...797..123H}. 
The FRMS could also be a possible formation site for light n-capture elements, as they are thought to be a 
source of $s$-process elements
through rotation-induced mixing, and they are expected to produce heavy elements up to Ba  \citep[`weak $s$-process'; see e.g.][]{Pignatari2008ApJ...687L..95P, Frischknecht2012A&A...538L...2F, Frischknecht2016MNRAS.456.1803F, LimongiChieffi2018ApJS..237...13L}.
Another possible process involved in the production of light n-capture elements at low metallicities is the lighter element primary process (LEPP, \citealt{Travaglio2004ApJ...601..864T,Montes2007ApJ...671.1685M}).

   \begin{figure*}[h!]
   \centering
   \includegraphics[width=0.40\hsize]{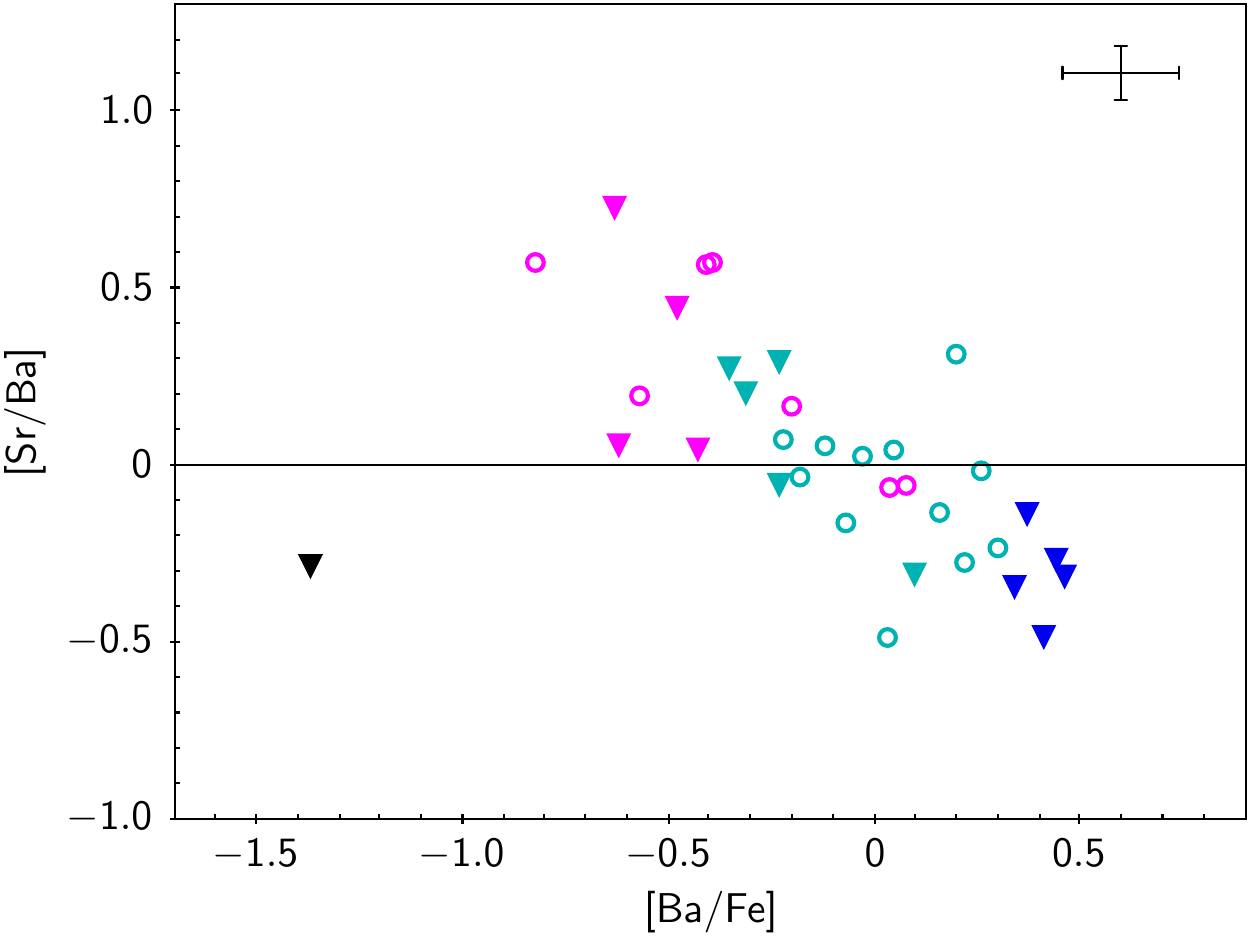}
   \includegraphics[width=0.40\hsize]{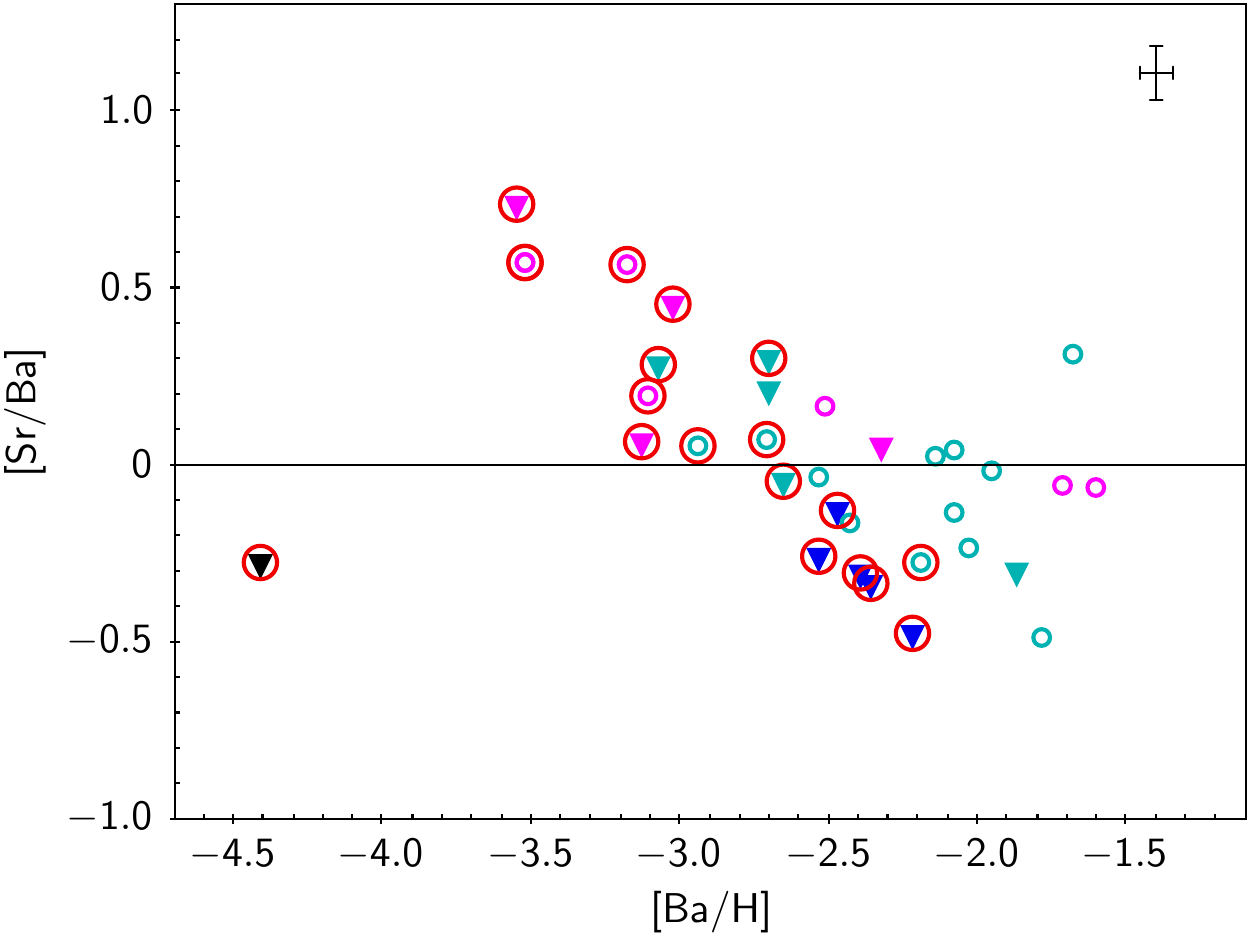}
   \includegraphics[width=0.40\hsize]{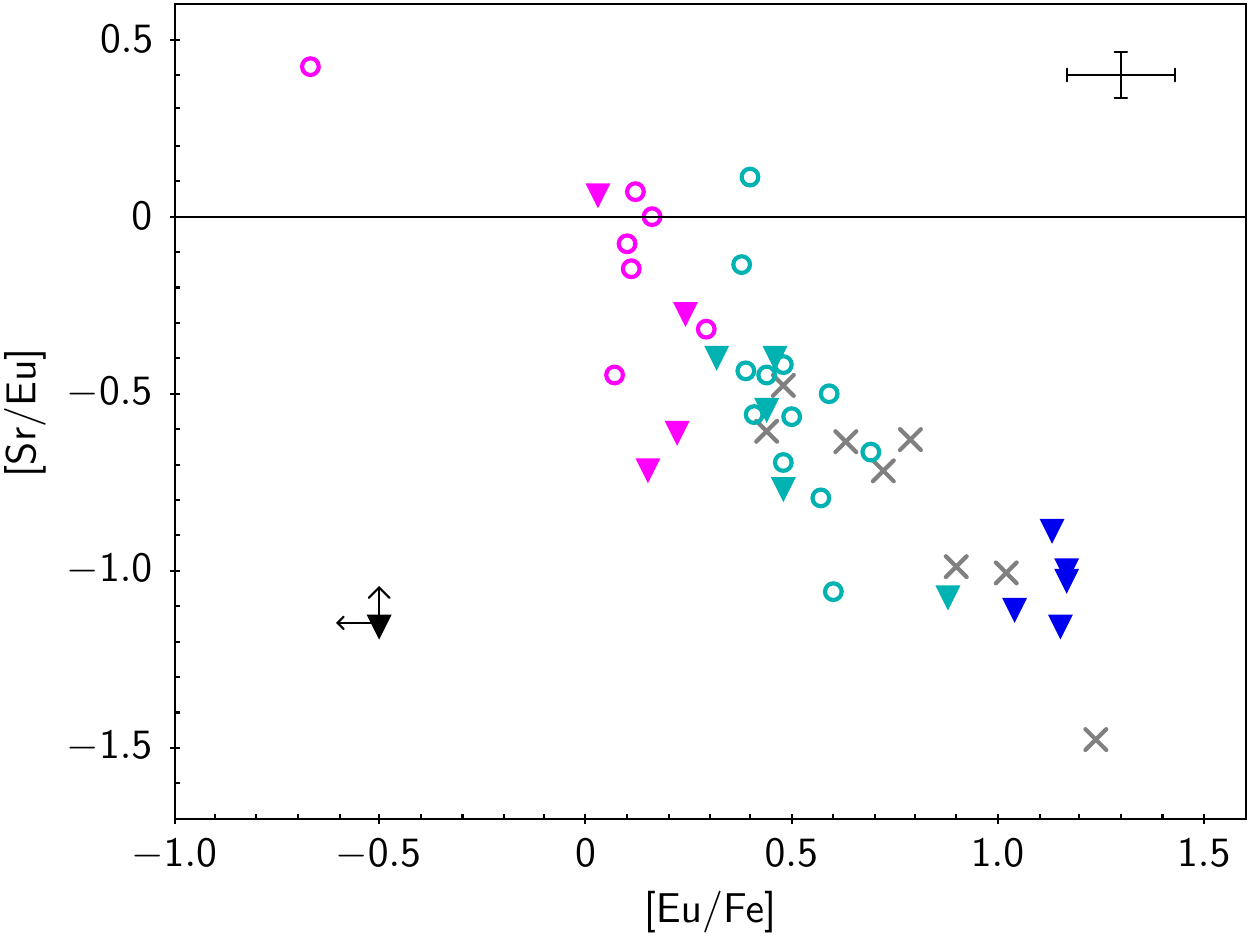}
   \includegraphics[width=0.40\hsize]{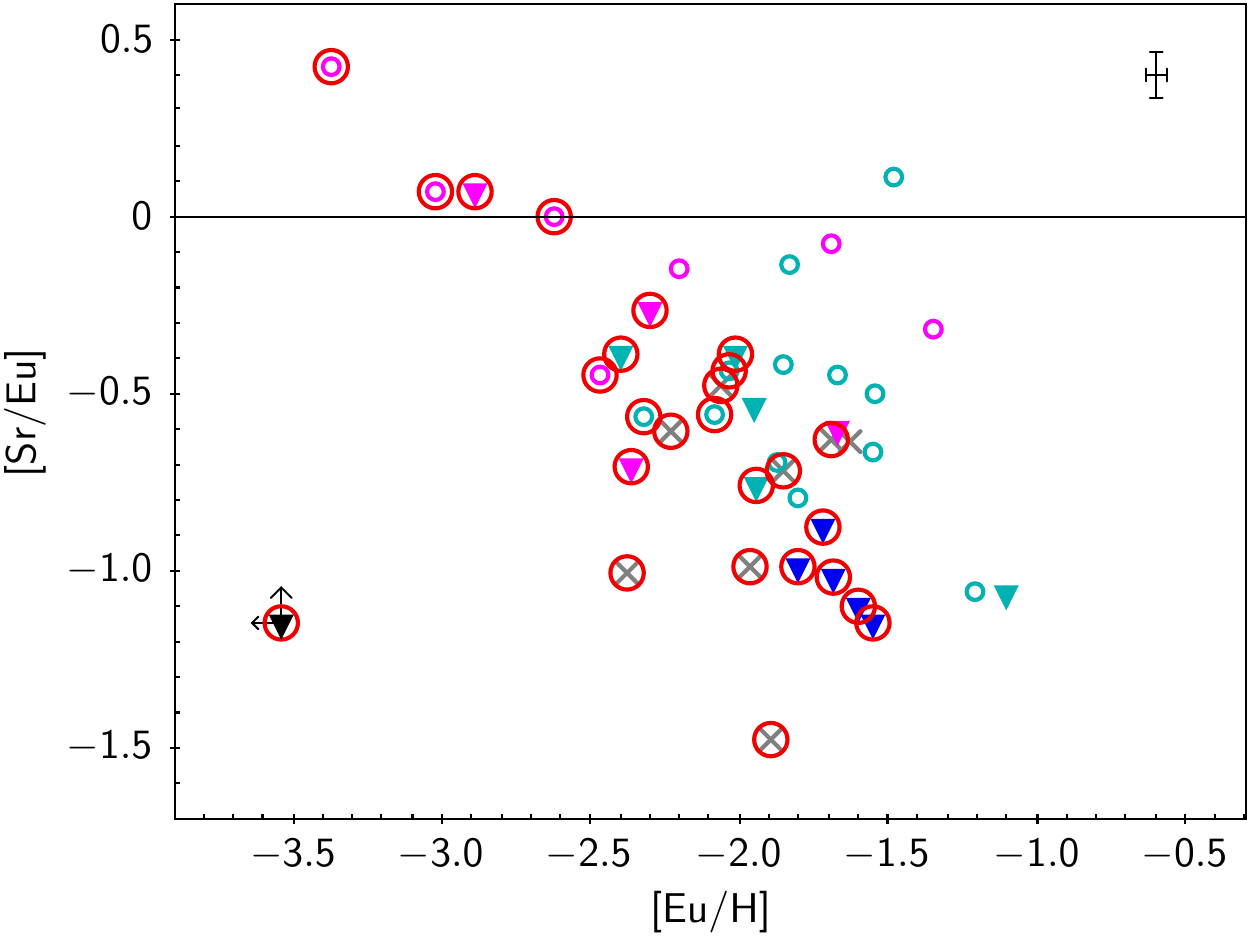}
      \caption{ [Sr/Ba] as a function of [Ba/Fe] (upper left panel) and [Ba/H] (upper right panel), and [Sr/Eu] as a function of [Eu/Fe] (lower left panel) and [Eu/H] (lower right panel) for our sample of stars. Coloured symbols as in Fig.~\ref{Fig:BaEu_FeH} and Fig.~\ref{Fig:BaMg_EuMg_MgH}. Red open circles highlight stars with [Fe/H]$<$$-2.4$. 
              }
         \label{Fig:SrBa_BaFe_BaH_SrEu_EuFe_EuH}
   \end{figure*}

Figure~\ref{Fig:SrBa_BaFe_BaH_SrEu_EuFe_EuH} shows [Sr/Ba] as a function of [Ba/Fe] (upper left panel) and [Sr/Eu] as a function of [Eu/Fe] (lower left panel) for our sample of stars. 
In our sample we do not see the large scatter in [Sr/Ba] observed in previous studies, for example in \citet{Spite2018A&A...611A..30S}, with the exception of the star CES1237+1922 (black triangle in 
Fig.~\ref{Fig:SrBa_BaFe_BaH_SrEu_EuFe_EuH}). 
However, looking at Figure 4 of \citet{Spite2018A&A...611A..30S}, 
we see that 
the scatter in [Sr/Ba] increases for [Ba/Fe]$\lesssim$$-0.5$. 
In our sample, only five stars with both Sr and Ba measurements have [Ba/Fe]$\lesssim$$-0.5$, and indeed the [Sr/Ba] for these stars vary from $-0.28$ (CES1237+1922) to $+0.73$ (CES2019-6130).
We observe that, for all stars except CES1237+1922, [Sr/Ba] (and [Sr/Eu]) increases when [Ba/Fe] (and [Eu/Fe]) decreases\footnote{Three stars in our sample show an abundance pattern consistent with the one of `limited-$r$' stars, which are characterised by [Eu/Fe]$<$0.3, [Sr/Ba]$>$0.5, and [Sr/Eu]$>$0 according to \citet{Frebel2018ARNPS..68..237F}.}. 
We also note that $r$-II stars have on average [Sr/Ba]$\sim$$-0.3$, which is close to the empirical $r$-process ratio [Sr/Ba]=$-0.4$ observed in strongly enhanced $r$-rich stars \citep[see e.g.][]{Barklem2005A&A...439..129B,Mashonkina2017A&A...608A..89M}.

When comparing [Sr/Ba] to [Ba/H] (upper right panel of Fig.~\ref{Fig:SrBa_BaFe_BaH_SrEu_EuFe_EuH}) and [Sr/Eu] to [Eu/H] (lower right panel of Fig.~\ref{Fig:SrBa_BaFe_BaH_SrEu_EuFe_EuH}) abundance ratios, the decreasing trend is still visible, with an increasing scatter towards the highest [Ba/H] and [Eu/H] abundances.
If we remove the stars with [Fe/H]$>$--2.4, the downward trend is very clean, and the scatter is drastically reduced
(stars identified by red open circles in Fig.~\ref{Fig:SrBa_BaFe_BaH_SrEu_EuFe_EuH}). 
This is likely due to the fact that stars with [Fe/H]$>$--2.4 can already show contamination from other processes, such as  the $s$-process, in Ba and Sr nucleosynthesis, therefore the scatter increases \citep[see e.g.][]{Hansen2011A&A...525L...5H, Hansen2012A&A...545A..31H, Hansen2014ApJ...797..123H}. 
The $r$-pure stars seem to show a scatter as well, and to behave similar to the other stars in the sample.
This behaviour seems to indicate that $r$-pure stars need not to be produced through different formation channels and/or scenarios. 

According to \citet{Spite2018A&A...611A..30S}, at a given [Ba/Fe], the star's [Sr/Ba] ratio depends on how strong the contribution is from the process that produces only light n-capture elements. 
This scenario implies that stars such as CES2019-6130 ([Ba/Fe]=$-0.63$, [Sr/Ba]=$+0.73$) are likely formed in a gas polluted by both mechanisms, but the contribution from the light n-capture component is dominating, while n-capture rich stars might be formed in a gas polluted by both mechanisms as well, but the contribution from the main $r$-process is so large that the contribution from light n-capture component would not alter the abundance pattern. 
Our results seem to support this scenario, although it is difficult to distinguish whether the two processes are independent, and thus generated in different astrophysical scenarios, or whether they reflect varying conditions in the micro physics of the same formation site \citep[see][and references therein]{ArconesThielemann2023A&ARv..31....1A}.

   \begin{figure*}[h!]
   \centering
   \includegraphics[width=0.40\hsize]{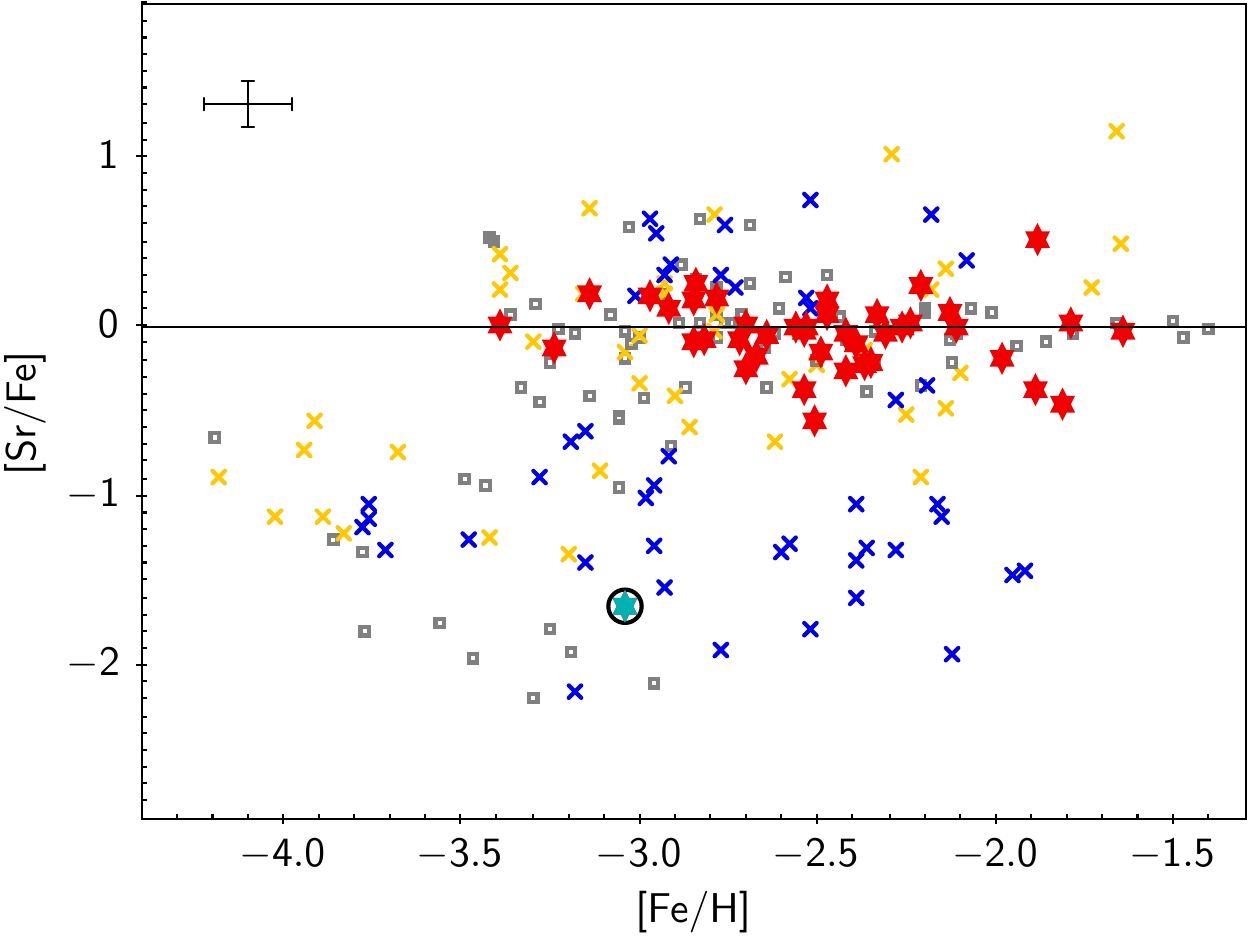}
   \includegraphics[width=0.40\hsize]{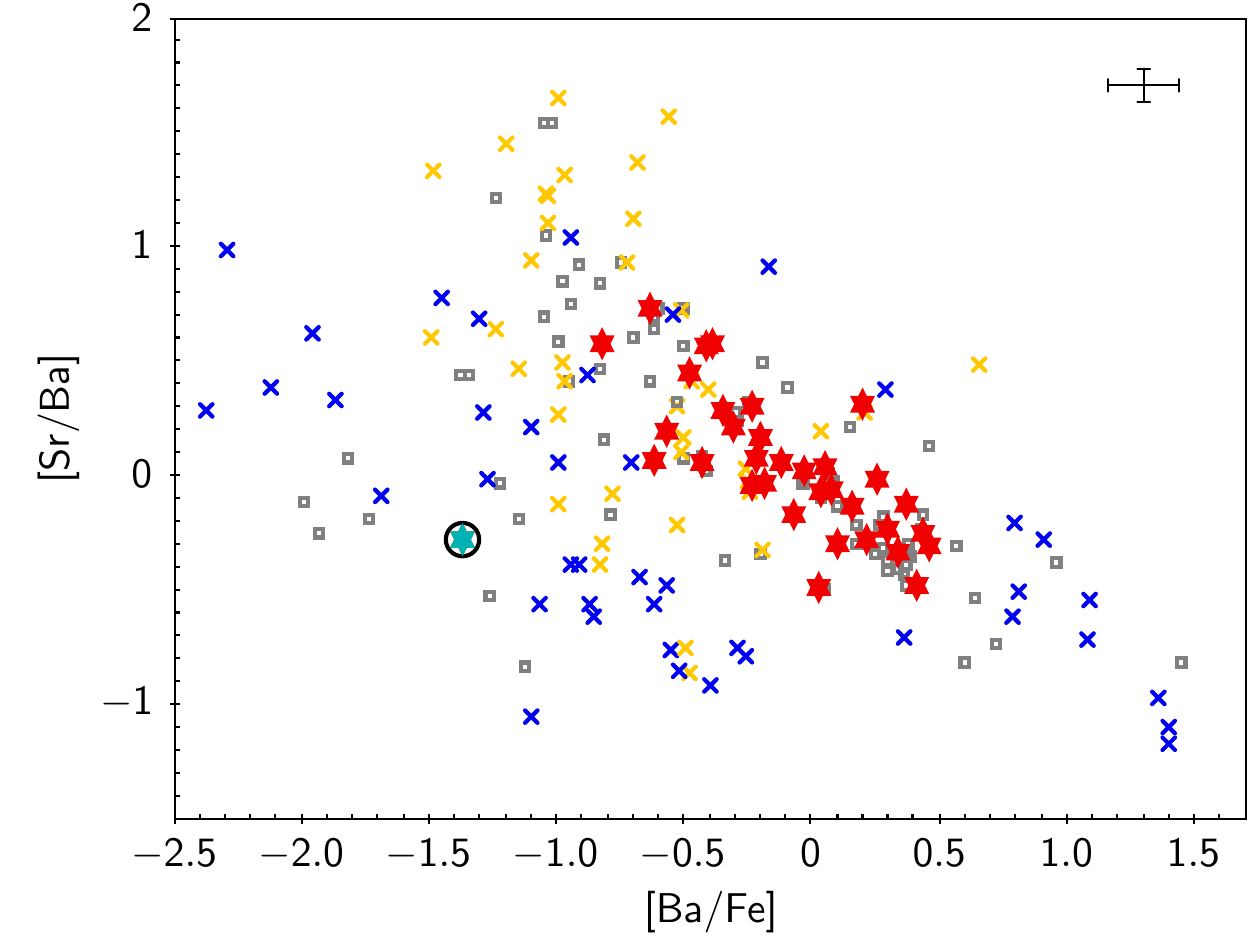}
      \caption{[Sr/Fe] as a function of [Fe/H] (upper panel) and [Sr/Ba] as a function of [Ba/Fe] abundance ratios (lower panel) for our sample of stars (red star symbols). The cyan star with a black open circle indicates the star CES1237+1922. Grey open squares are stars in the Milky Way halo and crosses are stars in dSph (yellow) and UFD galaxies (blue) in the literature. Literature abundance ratios are taken from: 
      \citealt{Francois2007A&A...476..935F} (Milky Way halo), 
      \citealt{Norris2010ApJ...711..350N} (Bootes I),
      \citealt{Frebel2010ApJ...708..560F} (Ursa Major II, Coma Berenices), 
      \citealt{Simon2010ApJ...716..446S} (Leo IV), 
      \citealt{Hansen2012A&A...545A..31H} (Milky Way halo),
      \citealt{Ishigaki2014A&A...562A.146I} (Bootes I),
      \citealt{Francois2016A&A...588A...7F} (Canes Venatici I, Hercules), 
      \citealt{Ji2016ApJ...830...93J} (Reticulum II), 
      \citealt{Roederer2016AJ....151...82R} (Reticulum II), 
      \citealt{Mashonkina2017A&A...608A..89M} (Milky Way halo, Sculptor, Fornax, Sextans, Ursa Minor, Bootes I, Ursa Major II, Leo IV), 
      \citealt{Reichert2020A&A...641A.127R} (Draco, Sculptor, Ursa Minor, Sextans, Sagittarius, Fornax, Ursa Major II, Bootes I, Segue 1, Triangulum II),
      \citealt{Sitnova2021MNRAS.504.1183S} (Segue 1, Triangulum II, Coma Berenices, Ursa Major II), and
      \citealt{Skuladottir2024A&A...681A..44S} (Sculptor)
      .
              }      
         \label{Fig:SrFe_FeH_SrBa_BaFe_dsph_ufd}
   \end{figure*}

\subsection{The peculiar star CES1237+1922}\label{sect:CES1237+1922}

In Paper~I, we found that CES1237+1922 (also known as BS~16085-0050) is deficient in Sr, Y, and Zr compared to the other stars in our sample. 
In Fig.~\ref{Fig:SrBa_BaFe_BaH_SrEu_EuFe_EuH}, we see that 
the star does not seem to follow the general trend observed for the rest of the stars in the sample. 
Figure~\ref{Fig:SrFe_FeH_SrBa_BaFe_dsph_ufd} shows [Sr/Fe] versus [Fe/H] and [Sr/Ba] versus [Ba/Fe] abundance ratios for our sample of stars and the ones observed in the Milky Way halo, dwarf Spheroidal (dSph), and Ultra-faint Dwarf (UFD) galaxies. 
The literature values for UFD galaxies were collected using the JINAbase database\footnote{\url{http://jinabase.pythonanywhere.com/}}. 
Stars in UFD galaxies show a large scatter in [Sr/Fe], as shown in the upper panel of Fig.~\ref{Fig:SrFe_FeH_SrBa_BaFe_dsph_ufd}, overlapping with both MW and dSph stars.
With the exception of stars in Canes Venatici I  
and Reticulum II, which are enhanced in n-capture elements, the other stars in UFDs show [Sr/Fe]$\lesssim$--0.6, while Milky Way halo stars with similar low [Sr/Fe] become more frequent for [Fe/H]$\lesssim$--3 \citep[see e.g.][]{Frebel2010ApJ...708..560F,Simon2010ApJ...716..446S,Koch2013A&A...554A...5K,Francois2016A&A...588A...7F, Roederer2016AJ....151...82R,Ji2016ApJ...830...93J, Mashonkina2017A&A...608A..89M, Sitnova2021MNRAS.504.1183S}. 
The enrichment history of such stars is still not clear yet, since for most of these extremely metal-poor stars only upper limits can usually be measured for most of the n-capture elements, with Sr and Ba being an exception due to their strong lines \citep[see e.g.][]{Roederer2013AJ....145...26R, Hansen2013A&A...551A..57H,Spite2018A&A...611A..30S}.

In the lower panel of  Fig.~\ref{Fig:SrFe_FeH_SrBa_BaFe_dsph_ufd}, we note that 
other halo stars similar to CES1237+1922 exist in the literature, and are characterised by [Fe/H]$\lesssim$--3, [Sr/Fe]$<$--1.5, and [Ba/Fe]$<$--1. 
Since these stars lie in the region of the [Sr/Ba] versus [Ba/Fe] diagram mainly occupied by UFD stars, it has been suggested that halo stars such as CES1237+1922 could be formed in UFDs that were later accreted by the Milky Way \citep[see][and references therein]{Andales2024MNRAS.530.4712A}. 
However, it is also possible that these stars formed in the Milky Way through the same mechanism that is producing low [Sr/Ba] stars in UFDs. 
We defer further discussion on our low [Sr/Ba] star to Lombardo et al. (in prep).

\subsection{Comparison with Galactic chemical evolution models}

\begin{figure*}[h!]
   \centering
   \includegraphics[width=0.8\hsize]{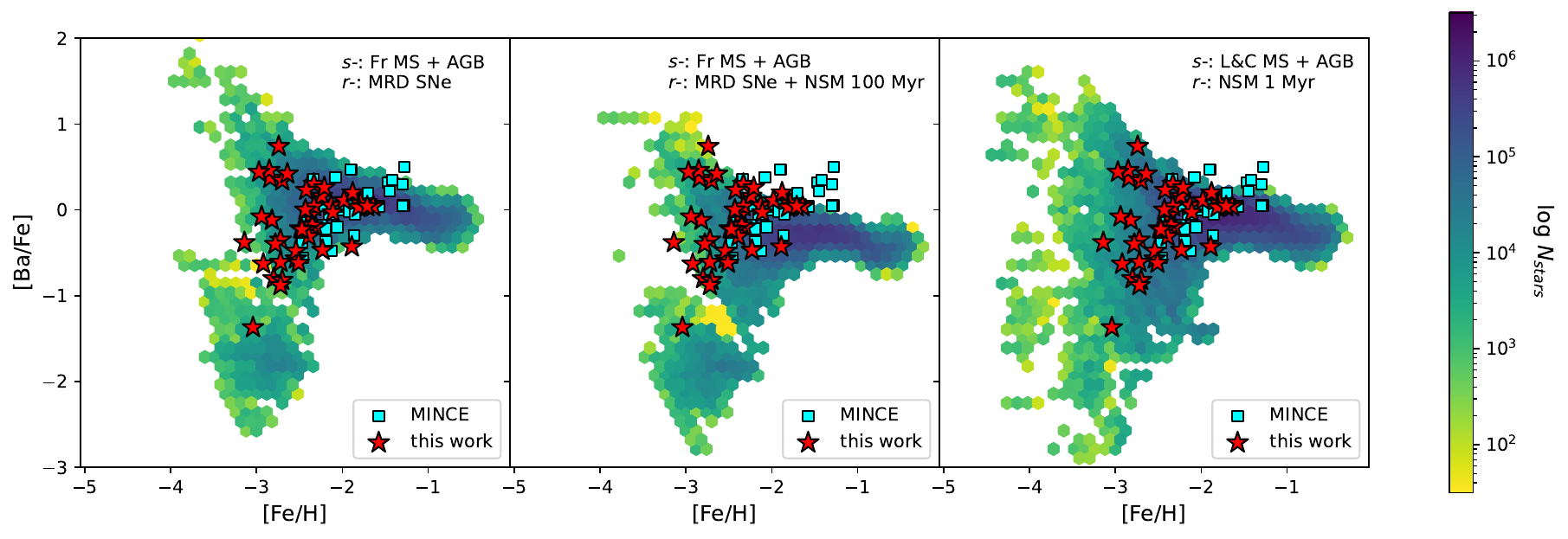}
      \caption{[Ba/Fe] abundance ratios as a function of [Fe/H] for our sample of stars (red stars) and MINCE data \citep[][cyan squares]{2024A&A...686A.295F} compared to GCE models. Left panel: GCE model from \cite{Cescutti2014A&A...565A..51C}. Centre panel: GCE model from \cite{Cescutti2015A&A...577A.139C}. Right panel: GCE model from \cite{Rizzuti2021MNRAS.502.2495R} (see text for more details).
              }
         \label{Fig:BaFe_FeH_GCE}
   \end{figure*}

\begin{figure*}[h!]
   \centering
      \includegraphics[width=0.392\hsize]{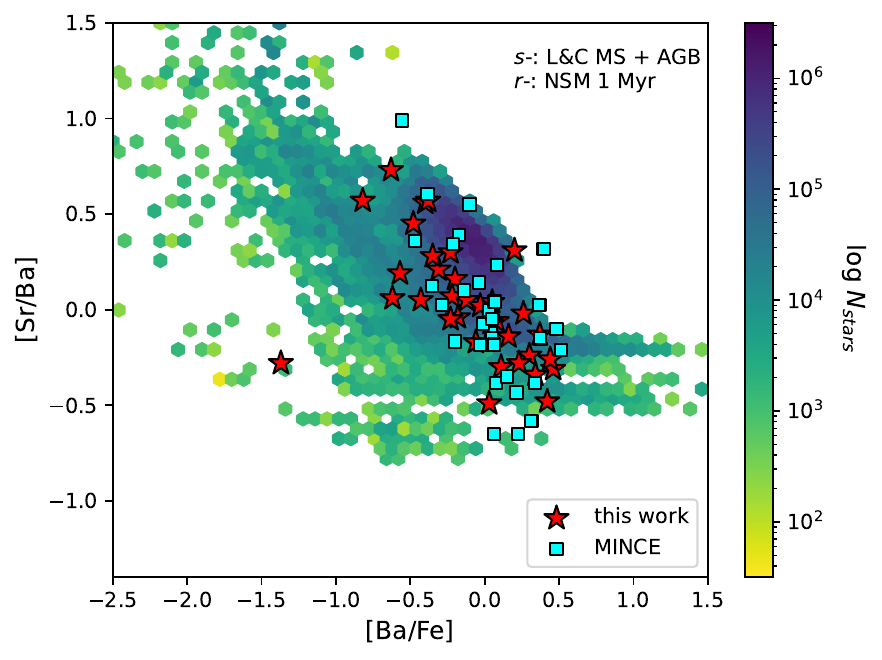}
      \includegraphics[width=0.392\hsize]{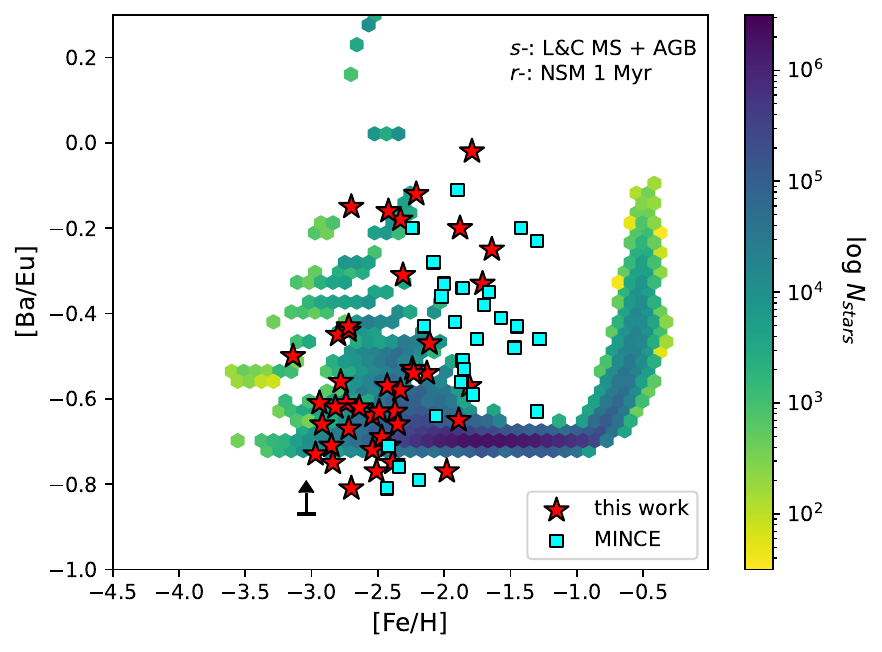}

      \caption{[Sr/Ba] (left) and [Ba/Eu] (right) abundance ratios as a function of [Ba/Fe] and [Fe/H], respectively, for our sample of stars  (red stars) and MINCE data \citep[][cyan squares]{2024A&A...686A.295F} compared to the GCE model of \cite{Rizzuti2021MNRAS.502.2495R}.}
         \label{Fig:EuBa_FeH_GCE_2021}
   \end{figure*}

In Figs.~\ref{Fig:BaFe_FeH_GCE} and \ref{Fig:EuBa_FeH_GCE_2021}, we compare our measured abundance ratios to the ones predicted by the GCE models from \citet{Cescutti2014A&A...565A..51C}, \citet{Cescutti2015A&A...577A.139C} and \citet{Rizzuti2021MNRAS.502.2495R}.
These models consider the stochastic formation of stars and $r$-process events, a concept described in  \citet{Cescutti2008A&A...481..691C}. 
The results show the dispersion created by different nucleosynthesis sites and at the same time reproduce the main trend of chemical evolution of the stellar system considered. 
To recover the possible combinations of enrichments, we assume several isolated volumes ($>$100), each of them containing the typical mass of gas swept by a SNe II explosion, the minimum mass that we can consider isolated.
We recall that the chemical evolution models are not ab initio modelling, therefore parameters such as infall, outflow of gas, and star formation efficiency should be tuned to reproduce the observed quantities of the stellar system we intend to reproduce. 

In \citet{Cescutti2014A&A...565A..51C}, the stellar system analysed was at the more metal-poor tail of the Galactic halo. 
Concerning nucleosynthesis, the model assumed three main sources for the neutron capture elements. 
The main $s$-process is produced by low-mass AGB stars with the yields of the F.R.U.I.T.Y. database\footnote{\url{http://fruity.oa-teramo.inaf.it/}} \citep{Fruity2011ApJS..197...17C}. 
The $r$-process site is MRD SNe and the model considers 10\% of all the exploding SNe II to enrich the interstellar medium with $r$-process material. 
The mean yields 
for Ba are  $8.0\times 10^{-6}$ \Msol{} (obtained from fine-tuning the model of \citealp{Cescutti2013A&A...553A..51C}), 
and they also consider a possible variation in the ejecta of the single event \citep[for details see][]{Cescutti2014A&A...565A..51C}.
The other chemical elements (for example Eu) are simply scaled using the Solar System $r$-process contribution as determined by \citet{Simmerer2004ApJ...617.1091S}.
Finally, they also assume the production of $s$-process from massive stars, thanks to FRMS 
considered in \citet{Frischknecht2016MNRAS.456.1803F}, with fixed rotation velocity. 
These yields can produce $s$-process elements up to the second peak (barium and lanthanum). 

On the other hand, \cite{Cescutti2015A&A...577A.139C} assumed the same sources for $s$-process, but investigated a different scenario for $r$-process, namely a contribution from MRD SNe, NSMs or both. 
They found that the synthesis of Eu in the Galaxy can be explained by either only NSMs with a short time delay of 1 Myr, or both MRD SNe and NSMs assuming a fixed delay of 100 Myr, with similar results. 
This is in agreement with the results from \citet{Matteucci2014MNRAS.438.2177M}, who found that NSMs can be the sole responsible of $r$-process enrichment only if they have a very short time-scale.

\cite{Rizzuti2021MNRAS.502.2495R} focused instead on $s$-process sources, having fixed the $r$-process to NSMs with 1 Myr time delay. 
They have employed the FRMS from \citet{LimongiChieffi2018ApJS..237...13L} for low-metallicity $s$-process, and calibrated the rotation velocity distribution in order to reproduce at best the dispersion in Sr and Ba. 
They showed that these assumptions can also explain the chemical evolution of Y, Zr, and La, despite the fact that these have not been used for calibration. 

The models run with different $r$-process sources show that, apart from an offset, the observed trends for [Ba/Fe] and [Eu/Fe] can be reproduced by all models, validating both scenarios with MRD SNe and/or NSMs (see Fig.~\ref{Fig:BaFe_FeH_GCE} and C.7\footnote{\href{https://doi.org/10.5281/zenodo.14218032}{https://doi.org/10.5281/zenodo.14218032}} in appendix C).
At the lowest metallicity,  elements such as Sr may receive a contribution from the FRMS that contribute to the early star-to-star scatter of these  elements (typically up to the second peak). FRMS also contribute to the large variation in [Sr/Ba] from $-0.5$ to $+1.0$ at very low [Ba/Fe] (see Fig.~\ref{Fig:EuBa_FeH_GCE_2021}, left panel).
Finally, FRMS may also explain part of  
the observed spread in [Ba/Eu] at low metallicity, shown in the right panel of Fig.~\ref{Fig:EuBa_FeH_GCE_2021}. 
In the same figure, the contribution from AGB stars is also visible, 
where they produce the sharp increase in [Ba/Eu] ratio at [Fe/H]$>$$-1.0$. 
In the [Ba/Fe] versus [Fe/H] plot (Fig.~\ref{Fig:BaFe_FeH_GCE}), their enrichment is neutralised by the iron produced by SNe Ia with similar timescales. 
The predictions of the GCE model for [Fe/H]$>$$-1.5$ are less reliable, because the model has been created to explain the most metal-poor part of the halo, and has not been calibrated yet for this metallicity range. 
It is possible that the contribution of $s$-process from FRMS also extends above $-1.5$, while in \cite{Rizzuti2021MNRAS.502.2495R} stars stop rotating at higher metallicity and therefore create the visible plateau, or that the AGB contribution begins at lower metallicities. 
Overall, the GCE models show that the observational data provide an excellent constraint to the neutron capture nucleosynthesis sites, showing that the strong $r$-process contribution dominates the lowest metallicity, but also the need for additional production sites, such as FRMS investigated here, to reproduce the dispersion observed in heavy element ratios.

\section{Conclusions}\label{Sect:conclusions}
In this study we present the chemical abundances of the heavy n-capture elements from Ba to Eu for the CERES sample, adopting the stellar parameters and the abundances of light elements derived in Paper I and II. 
The main conclusions of this study are the following: 
\begin{itemize}
    \item We derived abundances or upper limits of Ba, La, Ce, Pr, Nd, Sm, and Eu for a sample of 52 Milky Way halo stars. The general trends observed for heavy n-capture element abundance ratios ([X/Fe]) as a function of [Fe/H] are in good agreement with the ones found in previous studies. 
    \item We applied 1D NLTE corrections to Ba and Eu abundances. We found that the corrections for Ba tend to be larger for stars with [Fe/H]$>$--2.5 and [Ba/Fe]$>$0, and to decrease the Ba abundances. We also applied 3D NLTE corrections for seven stars in our sample, and found that they are on average smaller than 1D NLTE corrections and positive in sign, thus increasing the Ba abundances. We found that 1D NLTE corrections for Eu are all positive in sign and increase the Eu abundances. This is in agreement with previous studies. 
    \item We 
    estimated at which values of [Ba/H] and [Fe/H] the onset of the $s$-process occurs in our sample of stars using the mean shift clustering algorithm. We found that the change in the trend happens at [Ba/H]=$-2.4$, which corresponds to a metallicity of [Fe/H]=$-2.4$. This suggests that, for [Ba/H]$<$$-2.4$, the $r$-process is likely the primary production mechanism of Ba, and the large scatter observed at [Ba/H]$>$$-2.4$ is probably due to the onset of the $s$-process in Ba nucleosynthesis suffering from time delay in the Milky Way halo. 
    \item We selected stars with [Ba/Eu] compatible with the Solar System pure $r$-process values ($r$-pure), to check for possible correlations with other elements. The $r$-pure stars did not show any particular trend compared to the others in the sample. This seems to suggest that $r$-pure stars might be produced through similar formation channels and/or scenarios similar to stars with other $r$-process enrichments. 
    \item The star CES1237+1922 does not follow the general trend observed for other stars in the sample, and it is characterised by very low n-capture elements abundances. Other stars in the literature show a similar chemistry, and they lie in the region of the [Sr/Ba] versus [Ba/Fe] diagram mainly occupied by UFD stars. The origin of such stars is still uncertain, as they could form in situ or UFD galaxies and later be accreted by the Milky Way. 
    \item The comparison of the abundances obtained in this study to up-to-date Galactic Chemical Evolution models show the crucial role of $r$-process sources at low metallicities, whether they are NSMs or MRD SNe, to explain the measured abundances of heavy elements. The large scatter in the abundance ratios between elements produced by $s$- and $r$-processes 
    seems to suggest that other sources may contribute at these low metallicities, one of which could be FRMS.  

\end{itemize}

\section{Data availability}
Table~\ref{tab:abundances} is only available in electronic form at the CDS via anonymous ftp to cdsarc.u-strasbg.fr (130.79.128.5) or via http://cdsweb.u-strasbg.fr/cgi-bin/qcat?J/A+A/.
Appendix A and C are only available in electronic form at \url{https://doi.org/10.5281/zenodo.14218032}.

\begin{acknowledgements}
      We wish to thank T.M.Sitnova, A.Arcones and M.Reichert, M.Hanke and M.Eichler for the fruitful discussions that helped develop the project. LL,CJH,AAP,RFM acknowledge the support by the State of Hesse within the Research Cluster ELEMENTS (Project ID 500/10.006).  FR and GC acknowledge the grant PRIN project No. 2022X4TM3H ‘Cosmic POT’ from Ministero dell’Università e della Ricerca (MUR). ÁS acknowledges funding from the European Research Council (ERC) under the European Union’s Horizon 2020 research and innovation programme (grant agreement No. 101117455). This work has made use of data from the European Space Agency (ESA) mission {\it Gaia} (\url{https://www.cosmos.esa.int/gaia}), processed by the {\it Gaia} Data Processing and Analysis Consortium (DPAC, \url{https://www.cosmos.esa.int/web/gaia/dpac/consortium}). Funding for the DPAC has been provided by national institutions, in particular the institutions participating in the {\it Gaia} Multilateral Agreement. This work was also partially supported by the European Union (ChETEC-INFRA, project no. 101008324).

\end{acknowledgements}

%
%
\bibliographystyle{aa} 
\bibliography{ceres_heavy} 

\begin{appendix} 

\section{Additional tables}

\begin{table}[ht]
\centering
 \caption[]{\label{tab:stelpar}Stellar parameters for our sample stars as derived in Paper I.}
\begin{tabular}{lcccc}
 \hline \hline
 Star            & \teff &	 \logg &       \vt	& [Fe$/$H]    \\
                 & K	 &	 dex   &       \kms	& dex	      \\
 \hline
  CES0031--1647 & 4960 & 1.83 & 1.91 & $-$2.49\\
  CES0045--0932 & 5023 & 2.29 & 1.76 & $-$2.95\\
  CES0048--1041 & 4856 & 1.68 & 1.93 & $-$2.48\\
  CES0055--3345 & 5056 & 2.45 & 1.66 & $-$2.36\\
  CES0059--4524 & 5129 & 2.72 & 1.56 & $-$2.39\\
  CES0102--6143 & 5083 & 2.37 & 1.75 & $-$2.86\\
  CES0107--6125 & 5286 & 2.97 & 1.54 & $-$2.59\\
  CES0109--0443 & 5206 & 2.74 & 1.69 & $-$3.23\\
  CES0215--2554 & 5077 & 2.00 & 1.91 & $-$2.73\\
  CES0221--2130 & 4908 & 1.84 & 1.84 & $-$1.99\\
  CES0242--0754 & 4713 & 1.36 & 2.03 & $-$2.90\\
  CES0301+0616  & 5224 & 3.01 & 1.51 & $-$2.93\\
  CES0338--2402 & 5244 & 2.78 & 1.62 & $-$2.81\\
  CES0413+0636  & 4512 & 1.10 & 2.01 & $-$2.24\\
  CES0419--3651 & 5092 & 2.29 & 1.78 & $-$2.81\\
  CES0422--3715 & 5104 & 2.46 & 1.68 & $-$2.45\\
  CES0424--1501 & 4646 & 1.74 & 1.74 & $-$1.79\\
  CES0430--1334 & 5636 & 3.07 & 1.63 & $-$2.09\\
  CES0444--1228 & 4575 & 1.40 & 1.92 & $-$2.54\\
  CES0518--3817 & 5291 & 3.06 & 1.49 & $-$2.49\\
  CES0527--2052 & 4772 & 1.81 & 1.84 & $-$2.75\\
  CES0547--1739 & 4345 & 0.90 & 2.01 & $-$2.05\\
  CES0747--0405 & 4111 & 0.54 & 2.08 & $-$2.25\\
  CES0900--6222 & 4329 & 0.94 & 1.98 & $-$2.11\\
  CES0908--6607 & 4489 & 0.90 & 2.12 & $-$2.62\\
  CES0919--6958 & 4430 & 0.70 & 2.17 & $-$2.46\\
  CES1116--7250 & 4106 & 0.48 & 2.14 & $-$2.74\\
  CES1221--0328 & 5145 & 2.76 & 1.60 & $-$2.96\\
  CES1222+1136  & 4832 & 1.72 & 1.93 & $-$2.91\\
  CES1226+0518  & 5341 & 2.84 & 1.60 & $-$2.38\\
  CES1228+1220  & 5089 & 2.04 & 1.87 & $-$2.32\\
  CES1237+1922  & 4960 & 1.86 & 1.95 & $-$3.19\\
  CES1245--2425 & 5023 & 2.35 & 1.72 & $-$2.85\\
  CES1322--1355 & 4960 & 1.81 & 1.96 & $-$2.93\\
  CES1402+0941  & 4682 & 1.35 & 2.01 & $-$2.79\\
  CES1405--1451 & 4642 & 1.58 & 1.81 & $-$1.87\\
  CES1413--7609 & 4782 & 1.72 & 1.87 & $-$2.52\\
  CES1427--2214 & 4913 & 1.99 & 1.85 & $-$3.05\\
  CES1436--2906 & 5280 & 3.15 & 1.42 & $-$2.15\\
  CES1543+0201  & 5157 & 2.77 & 1.57 & $-$2.65\\
  CES1552+0517  & 5013 & 2.30 & 1.72 & $-$2.60\\
  CES1732+2344  & 5370 & 2.82 & 1.65 & $-$2.57\\
  CES1804+0346  & 4390 & 0.80 & 2.12 & $-$2.48\\
  CES1942--6103 & 4748 & 1.53 & 2.01 & $-$3.34\\
  CES2019--6130 & 4590 & 1.13 & 2.09 & $-$2.97\\
  CES2103--6505 & 4916 & 2.05 & 1.85 & $-$3.58\\
  CES2231--3238 & 5222 & 2.67 & 1.67 & $-$2.77\\
  CES2232--4138 & 5194 & 2.76 & 1.59 & $-$2.58\\
  CES2250--4057 & 5634 & 2.51 & 1.88 & $-$2.14\\
  CES2254--4209 & 4805 & 1.98 & 1.79 & $-$2.88\\
  CES2330--5626 & 5028 & 2.31 & 1.75 & $-$3.10\\
  CES2334--2642 & 4640 & 1.42 & 2.02 & $-$3.48\\
\hline
\end{tabular}
\tablefoot{[Fe/H] is computed with the Solar abundance from \citet{CaffauSun}.}
\end{table}

\begin{table}[h]
\caption{List of atomic data without hyperfine and isotopic structure.}\label{all_lines}
\centering
\begin{tabular}{lccr}
\hline \hline
Element & Wavelength (\AA) & $\chi_{exc}$ & $\log gf$ \\
 & \AA & & \\
 \hline
\ion{Ba}{II} & 5853.668  & 0.604 &  --1.01 \\ 
\ion{Ba}{II} & 6141.713  & 0.703 &  --0.08 \\
\ion{Ba}{II} & 6496.897  & 0.604 &  --0.38 \\ 
\hline
\ion{La}{II} & 3949.100  & 0.403 &   0.49 \\ 
\ion{La}{II} & 4086.710  & 0.000 &  --0.07 \\ 
\ion{La}{II} & 4123.220  & 0.321 &   0.13 \\ 
\ion{La}{II} & 4920.980  & 0.126 &  --0.58 \\ 
\hline
\ion{Ce}{II} & 3577.456  & 0.470 &   0.14 \\
\ion{Ce}{II} & 3999.237  & 0.295 &   0.06 \\
\ion{Ce}{II} & 4073.474  & 0.477 &   0.21 \\
\ion{Ce}{II} & 4083.222  & 0.700 &   0.27 \\
\ion{Ce}{II} & 4118.143  & 0.696 &   0.13 \\
\ion{Ce}{II} & 4120.827  & 0.320 &  --0.37 \\
\ion{Ce}{II} & 4137.645  & 0.516 &   0.40 \\ 
\ion{Ce}{II} & 4165.599  & 0.909 &   0.52 \\
\ion{Ce}{II} & 5274.229  & 1.044 &   0.13 \\
\hline
\ion{Pr}{II} & 4408.810  & 0.000 &   0.05 \\
\ion{Pr}{II} & 5259.731  & 0.633 &   0.12 \\ 
\ion{Pr}{II} & 5322.770  & 0.482 &  --0.12 \\
\hline
\ion{Nd}{II} & 3784.240  & 0.380 &   0.15 \\
\ion{Nd}{II} & 3826.410  & 0.064 &  --0.41 \\
\ion{Nd}{II} & 4021.330  & 0.320 &  --0.10 \\
\ion{Nd}{II} & 4446.380  & 0.204 &  --0.35 \\
\ion{Nd}{II} & 4959.120  & 0.064 &  --0.80 \\
\ion{Nd}{II} & 5255.510  & 0.204 &  --0.67 \\
\ion{Nd}{II} & 5293.160  & 0.822 &   0.10 \\
\ion{Nd}{II} & 5319.810  & 0.550 &  --0.14 \\
\hline
\ion{Sm}{II} & 4434.320  & 0.378 &  --0.07 \\
\ion{Sm}{II} & 4704.400  & 0.000 &  --0.86 \\
\hline
\ion{Eu}{II} & 3819.670  & 0.000 &   0.51 \\ 
\ion{Eu}{II} & 4129.720  & 0.000 &   0.22 \\ 
\ion{Eu}{II} & 6645.060  & 1.379 &   0.12 \\ 
\hline
\end{tabular}
\end{table}

\begin{table}[h!]
\caption{Hyperfine and isotopic structure for \ion{Ba}{II} lines (Z=56).}\label{ba_lines}
\centering
\begin{tabular}{cccc}
\hline \hline
Wavelength & Isotope & $\chi_{exc}$ & $\log gf$ \\
\AA & & & \\
\hline
5853.686 & 137  &  0.604 &  --2.066 \\ 
5853.687 & 135  &  0.604 &  --2.066 \\ 
5853.687 & 137  &  0.604 &  --2.009 \\ 
5853.688 & 135  &  0.604 &  --2.009 \\ 
5853.689 & 135  &  0.604 &  --2.215 \\ 
5853.689 & 137  &  0.604 &  --2.215 \\ 
5853.690 & 134  &  0.604 &  --1.010 \\ 
5853.690 & 135  &  0.604 &  --2.620 \\ 
5853.690 & 135  &  0.604 &  --1.914 \\ 
5853.690 & 135  &  0.604 &  --1.466 \\ 
5853.690 & 136  &  0.604 &  --1.010 \\ 
5853.690 & 137  &  0.604 &  --2.620 \\ 
5853.690 & 137  &  0.604 &  --1.914 \\ 
5853.690 & 137  &  0.604 &  --1.466 \\ 
5853.690 & 138  &  0.604 &  --1.010 \\ 
5853.691 & 135  &  0.604 &  --2.215 \\ 
5853.692 & 137  &  0.604 &  --2.215 \\ 
5853.693 & 135  &  0.604 &  --2.009 \\ 
5853.693 & 137  &  0.604 &  --2.009 \\ 
5853.694 & 135  &  0.604 &  --2.066 \\ 
5853.694 & 137  &  0.604 &  --2.066 \\ 
\hline
6141.725 & 135  &  0.704 &  --2.456 \\ 
6141.725 & 137  &  0.704 &  --2.456 \\ 
6141.727 & 135  &  0.704 &  --1.311 \\ 
6141.727 & 137  &  0.704 &  --1.311 \\ 
6141.728 & 135  &  0.704 &  --2.284 \\ 
6141.728 & 137  &  0.704 &  --2.284 \\ 
6141.729 & 135  &  0.704 &  --1.214 \\ 
6141.729 & 135  &  0.704 &  --0.503 \\ 
6141.729 & 137  &  0.704 &  --1.214 \\ 
6141.729 & 137  &  0.704 &  --0.503 \\ 
6141.730 & 134  &  0.704 &  --0.077 \\ 
6141.730 & 136  &  0.704 &  --0.077 \\ 
6141.730 & 138  &  0.704 &  --0.077 \\ 
6141.731 & 135  &  0.704 &  --1.327 \\ 
6141.731 & 135  &  0.704 &  --0.709 \\ 
6141.731 & 137  &  0.704 &  --1.327 \\ 
6141.731 & 137  &  0.704 &  --0.709 \\ 
6141.732 & 135  &  0.704 &  --1.281 \\ 
6141.732 & 135  &  0.704 &  --0.959 \\ 
6141.732 & 137  &  0.704 &  --0.959 \\ 
6141.733 & 137  &  0.704 &  --1.281 \\ 
\hline
6496.898 & 137  &  0.604 &  --1.886 \\ 
6496.899 & 135  &  0.604 &  --1.886 \\ 
6496.901 & 137  &  0.604 &  --1.186 \\ 
6496.902 & 135  &  0.604 &  --1.186 \\ 
6496.906 & 135  &  0.604 &  --0.739 \\ 
6496.906 & 137  &  0.604 &  --0.739 \\ 
6496.910 & 134  &  0.604 &  --0.380 \\ 
6496.910 & 136  &  0.604 &  --0.380 \\ 
6496.910 & 138  &  0.604 &  --0.380 \\ 
6496.916 & 135  &  0.604 &  --1.583 \\ 
6496.916 & 137  &  0.604 &  --1.583 \\ 
6496.917 & 135  &  0.604 &  --1.186 \\ 
6496.918 & 137  &  0.604 &  --1.186 \\ 
6496.920 & 135  &  0.604 &  --1.186 \\ 
6496.922 & 137  &  0.604 &  --1.186 \\ 
\hline
\end{tabular}
\end{table}

\begin{table}[h!]
\caption{Hyperfine and isotopic structure for \ion{La}{II} lines (Z=57).}\label{la_lines}
\centering
\begin{tabular}{cccc}
\hline \hline
Wavelength & Isotope & $\chi_{exc}$ & $\log gf$ \\
\AA & & & \\
\hline
 3949.0377 & 139  &  0.403 &  --1.337 \\
 3949.0387 & 139  &  0.403 &  --1.191 \\
 3949.0444 & 139  &  0.403 &  --0.995 \\
 3949.0460 & 139  &  0.403 &  --1.008 \\
 3949.0470 & 139  &  0.403 &  --1.669 \\
 3949.0559 & 139  &  0.403 &  --0.761 \\
 3949.0582 & 139  &  0.403 &  --0.886 \\
 3949.0598 & 139  &  0.403 &  --1.559 \\
 3949.0723 & 139  &  0.403 &  --0.576 \\
 3949.0752 & 139  &  0.403 &  --0.825 \\
 3949.0775 & 139  &  0.403 &  --1.581 \\
 3949.0936 & 139  &  0.403 &  --0.420 \\
 3949.0972 & 139  &  0.403 &  --0.821 \\
 3949.1001 & 139  &  0.403 &  --1.690 \\
 3949.1199 & 139  &  0.403 &  --0.284 \\
 3949.1241 & 139  &  0.403 &  --0.887 \\
 3949.1277 & 139  &  0.403 &  --1.901 \\
 3949.1512 & 139  &  0.403 &  --0.163 \\
 3949.1561 & 139  &  0.403 &  --1.092 \\
 3949.1603 & 139  &  0.403 &  --2.306 \\
\hline
 4086.6947 & 139  &  0.000 &  --1.266 \\
 4086.6986 & 139  &  0.000 &  --1.108 \\
 4086.7022 & 139  &  0.000 &  --1.119 \\
 4086.7054 & 139  &  0.000 &  --1.292 \\
 4086.7070 & 139  &  0.000 &  --0.696 \\
 4086.7086 & 139  &  0.000 &  --1.094 \\
 4086.7099 & 139  &  0.000 &  --1.790 \\
 4086.7109 & 139  &  0.000 &  --3.219 \\
 4086.7116 & 139  &  0.000 &  --1.468 \\
 4086.7171 & 139  &  0.000 &  --1.292 \\
 4086.7186 & 139  &  0.000 &  --1.119 \\
 4086.7198 & 139  &  0.000 &  --1.108 \\
 4086.7208 & 139  &  0.000 &  --1.266 \\
\hline
 4123.2021 & 139  &  0.321 &  --0.472 \\
 4123.2120 & 139  &  0.321 &  --1.212 \\
 4123.2121 & 139  &  0.321 &  --0.643 \\
 4123.2201 & 139  &  0.321 &  --2.212 \\
 4123.2202 & 139  &  0.321 &  --1.030 \\
 4123.2204 & 139  &  0.321 &  --0.850 \\
 4123.2267 & 139  &  0.321 &  --1.794 \\
 4123.2270 & 139  &  0.321 &  --0.996 \\
 4123.2273 & 139  &  0.321 &  --1.121 \\
 4123.2319 & 139  &  0.321 &  --1.560 \\
 4123.2322 & 139  &  0.321 &  --1.055 \\
 4123.2325 & 139  &  0.321 &  --1.539 \\
 4123.2357 & 139  &  0.321 &  --1.414 \\
 4123.2360 & 139  &  0.321 &  --1.238 \\
 4123.2381 & 139  &  0.321 &  --1.317 \\
\hline
\end{tabular}
\end{table}

\begin{table}[h!]
\caption{Hyperfine and isotopic structure for \ion{Pr}{II} lines (Z=59).}\label{pr_lines}
\centering
\begin{tabular}{cccc}
\hline \hline
Wavelength & Isotope & $\chi_{exc}$ & $\log gf$ \\
\AA & & & \\
\hline
 4408.7323 & 141  &  0.000 &  --0.562 \\
 4408.7696 & 141  &  0.000 &  --1.734 \\
 4408.7797 & 141  &  0.000 &  --0.655 \\
 4408.8019 & 141  &  0.000 &  --3.233 \\
 4408.8119 & 141  &  0.000 &  --1.544 \\
 4408.8204 & 141  &  0.000 &  --0.754 \\
 4408.8392 & 141  &  0.000 &  --2.905 \\
 4408.8478 & 141  &  0.000 &  --1.502 \\
 4408.8547 & 141  &  0.000 &  --0.859 \\
 4408.8701 & 141  &  0.000 &  --2.818 \\
 4408.8771 & 141  &  0.000 &  --1.556 \\
 4408.8825 & 141  &  0.000 &  --0.968 \\
 4408.8945 & 141  &  0.000 &  --2.964 \\
 4408.8999 & 141  &  0.000 &  --1.746 \\
 4408.9037 & 141  &  0.000 &  --1.077 \\
\hline
 5259.6145 & 141  &  0.633 &  --3.727 \\
 5259.6329 & 141  &  0.633 &  --3.418 \\
 5259.6498 & 141  &  0.633 &  --3.356 \\
 5259.6653 & 141  &  0.633 &  --3.539 \\
 5259.6667 & 141  &  0.633 &  --1.961 \\
 5259.6789 & 141  &  0.633 &  --1.763 \\
 5259.6897 & 141  &  0.633 &  --1.716 \\
 5259.6991 & 141  &  0.633 &  --1.767 \\
 5259.7070 & 141  &  0.633 &  --1.965 \\
 5259.7251 & 141  &  0.633 &  --0.538 \\
 5259.7312 & 141  &  0.633 &  --0.603 \\
 5259.7358 & 141  &  0.633 &  --0.669 \\
 5259.7390 & 141  &  0.633 &  --0.737 \\
 5259.7408 & 141  &  0.633 &  --0.806 \\
 5259.7411 & 141  &  0.633 &  --0.874 \\
\hline
 5322.6702 & 141  &  0.482 &  --3.392 \\
 5322.6704 & 141  &  0.482 &  --3.320 \\
 5322.6714 & 141  &  0.482 &  --3.710 \\
 5322.6718 & 141  &  0.482 &  --3.488 \\
 5322.7044 & 141  &  0.482 &  --2.073 \\
 5322.7102 & 141  &  0.482 &  --1.878 \\
 5322.7173 & 141  &  0.482 &  --1.826 \\
 5322.7257 & 141  &  0.482 &  --1.871 \\
 5322.7297 & 141  &  0.482 &  --1.164 \\
 5322.7354 & 141  &  0.482 &  --2.066 \\
 5322.7427 & 141  &  0.482 &  --1.082 \\
 5322.7571 & 141  &  0.482 &  --0.998 \\
 5322.7727 & 141  &  0.482 &  --0.915 \\
 5322.7897 & 141  &  0.482 &  --0.836 \\
 5322.8079 & 141  &  0.482 &  --0.760 \\
\hline
\end{tabular}
\end{table}

\begin{table}[h!]
\caption{Hyperfine and isotopic structure for \ion{Nd}{II} lines (Z=60).}\label{nd_lines}
\centering
\begin{tabular}{cccc}
\hline \hline
Wavelength & Isotope & $\chi_{exc}$ & $\log gf$ \\
\AA & & & \\
\hline
 4446.3635 & 143  &  0.204 &  --3.228 \\
 4446.3647 & 143  &  0.204 &  --3.021 \\ 
 4446.3648 & 143  &  0.204 &  --2.228 \\ 
 4446.3657 & 143  &  0.204 &  --1.776 \\ 
 4446.3664 & 143  &  0.204 &  --2.017 \\ 
 4446.3665 & 143  &  0.204 &  --2.993 \\ 
 4446.3677 & 143  &  0.204 &  --1.658 \\ 
 4446.3686 & 143  &  0.204 &  --1.921 \\ 
 4446.3690 & 143  &  0.204 &  --3.073 \\ 
 4446.3703 & 143  &  0.204 &  --1.540 \\ 
 4446.3714 & 143  &  0.204 &  --1.885 \\ 
 4446.3722 & 143  &  0.204 &  --3.265 \\ 
 4446.3735 & 143  &  0.204 &  --1.428 \\ 
 4446.3749 & 143  &  0.204 &  --1.901 \\ 
 4446.3750 & 145  &  0.204 &  --3.228 \\ 
 4446.3758 & 145  &  0.204 &  --2.228 \\ 
 4446.3758 & 145  &  0.204 &  --3.021 \\ 
 4446.3763 & 143  &  0.204 &  --3.654 \\ 
 4446.3764 & 145  &  0.204 &  --1.776 \\ 
 4446.3769 & 145  &  0.204 &  --2.017 \\ 
 4446.3769 & 145  &  0.204 &  --2.993 \\ 
 4446.3773 & 143  &  0.204 &  --1.324 \\ 
 4446.3774 & 142  &  0.204 &  --0.350 \\ 
 4446.3777 & 145  &  0.204 &  --1.658 \\ 
 4446.3782 & 145  &  0.204 &  --1.921 \\ 
 4446.3785 & 145  &  0.204 &  --3.073 \\ 
 4446.3791 & 143  &  0.204 &  --1.983 \\ 
 4446.3793 & 145  &  0.204 &  --1.540 \\ 
 4446.3800 & 145  &  0.204 &  --1.885 \\ 
 4446.3805 & 145  &  0.204 &  --3.265 \\ 
 4446.3813 & 145  &  0.204 &  --1.428 \\ 
 4446.3817 & 143  &  0.204 &  --1.228 \\ 
 4446.3821 & 145  &  0.204 &  --1.901 \\ 
 4446.3829 & 144  &  0.204 &  --0.350 \\ 
 4446.3830 & 145  &  0.204 &  --3.654 \\ 
 4446.3836 & 145  &  0.204 &  --1.324 \\ 
 4446.3840 & 143  &  0.204 &  --2.201 \\ 
 4446.3847 & 145  &  0.204 &  --1.983 \\ 
 4446.3864 & 145  &  0.204 &  --1.228 \\ 
 4446.3869 & 143  &  0.204 &  --1.138 \\ 
 4446.3878 & 145  &  0.204 &  --2.201 \\ 
 4446.3882 & 146  &  0.204 &  --0.350 \\ 
 4446.3896 & 145  &  0.204 &  --1.138 \\ 
 4446.3928 & 143  &  0.204 &  --1.054 \\ 
 4446.3932 & 145  &  0.204 &  --1.054 \\ 
 4446.3937 & 148  &  0.204 &  --0.350 \\ 
 4446.4000 & 150  &  0.204 &  --0.350 \\ 
\hline
\end{tabular}
\end{table}

\onecolumn
{
\centering
\begin{longtable}{cccc}
\caption{\label{eu_lines} Hyperfine and isotopic structure for \ion{Eu}{II} lines (Z=63).}\\
\hline\hline
Wavelength (\AA) & Isotope & $\chi_{exc}$ & $\log gf$ \\
\hline
\endfirsthead
\caption{continued.}\\
\hline\hline
Wavelength (\AA) & Isotope & $\chi_{exc}$ & $\log gf$ \\
\hline
\endhead
\hline
\endfoot
 3819.5576 & 151  &  0.000 &  --0.620 \\
 3819.5746 & 151  &  0.000 &  --0.511 \\
 3819.5763 & 151  &  0.000 &  --1.289 \\
 3819.5983 & 151  &  0.000 &  --0.402 \\
 3819.6008 & 151  &  0.000 &  --1.099 \\
 3819.6026 & 151  &  0.000 &  --2.507 \\
 3819.6243 & 153  &  0.000 &  --0.620 \\
 3819.6285 & 151  &  0.000 &  --0.297 \\
 3819.6320 & 151  &  0.000 &  --1.045 \\
 3819.6326 & 153  &  0.000 &  --1.290 \\
 3819.6333 & 153  &  0.000 &  --0.511 \\
 3819.6345 & 151  &  0.000 &  --2.363 \\
 3819.6443 & 153  &  0.000 &  --2.509 \\
 3819.6450 & 153  &  0.000 &  --1.099 \\
 3819.6452 & 153  &  0.000 &  --0.402 \\
 3819.6593 & 153  &  0.000 &  --0.297 \\
 3819.6600 & 153  &  0.000 &  --2.363 \\
 3819.6602 & 153  &  0.000 &  --1.045 \\
 3819.6649 & 151  &  0.000 &  --0.198 \\
 3819.6697 & 151  &  0.000 &  --1.087 \\
 3819.6732 & 151  &  0.000 &  --2.445 \\
 3819.6752 & 153  &  0.000 &  --0.198 \\
 3819.6777 & 153  &  0.000 &  --1.087 \\
 3819.6785 & 153  &  0.000 &  --2.444 \\
 3819.6919 & 153  &  0.000 &  --0.106 \\
 3819.6968 & 153  &  0.000 &  --1.277 \\
 3819.6993 & 153  &  0.000 &  --2.776 \\
 3819.7074 & 151  &  0.000 &  --0.105 \\
 3819.7137 & 151  &  0.000 &  --1.277 \\
 3819.7185 & 151  &  0.000 &  --2.771 \\
\hline
 4129.5966 & 151  &  0.000 &  --1.512 \\
 4129.6001 & 151  &  0.000 &  --1.035 \\
 4129.6137 & 151  &  0.000 &  --1.316 \\
 4129.6185 & 151  &  0.000 &  --0.977 \\
 4129.6220 & 151  &  0.000 &  --1.512 \\
 4129.6387 & 151  &  0.000 &  --1.257 \\
 4129.6444 & 151  &  0.000 &  --0.847 \\
 4129.6492 & 151  &  0.000 &  --1.316 \\
 4129.6716 & 151  &  0.000 &  --1.294 \\
 4129.6774 & 153  &  0.000 &  --1.513 \\
 4129.6781 & 151  &  0.000 &  --0.696 \\
 4129.6801 & 153  &  0.000 &  --1.035 \\
 4129.6838 & 151  &  0.000 &  --1.257 \\
 4129.6838 & 153  &  0.000 &  --1.316 \\
 4129.6871 & 153  &  0.000 &  --0.977 \\
 4129.6898 & 153  &  0.000 &  --1.513 \\
 4129.6941 & 153  &  0.000 &  --1.257 \\
 4129.6974 & 153  &  0.000 &  --0.847 \\
 4129.7007 & 153  &  0.000 &  --1.316 \\
 4129.7091 & 153  &  0.000 &  --1.294 \\
 4129.7117 & 153  &  0.000 &  --0.697 \\
 4129.7130 & 151  &  0.000 &  --1.480 \\
 4129.7150 & 153  &  0.000 &  --1.257 \\
 4129.7198 & 151  &  0.000 &  --0.545 \\
 4129.7263 & 151  &  0.000 &  --1.294 \\
 4129.7295 & 153  &  0.000 &  --1.480 \\
 4129.7305 & 153  &  0.000 &  --0.545 \\
 4129.7331 & 153  &  0.000 &  --1.294 \\
 4129.7548 & 153  &  0.000 &  --0.401 \\
 4129.7558 & 153  &  0.000 &  --1.480 \\
 4129.7700 & 151  &  0.000 &  --0.401 \\
 4129.7769 & 151  &  0.000 &  --1.480 \\
\hline
 6645.0717 & 151  &  1.379 &  --0.517 \\
 6645.0727 & 153  &  1.379 &  --1.823 \\
 6645.0744 & 153  &  1.379 &  --0.517 \\
 6645.0749 & 153  &  1.379 &  --3.452 \\
 6645.0785 & 151  &  1.379 &  --1.823 \\
 6645.0859 & 151  &  1.379 &  --3.480 \\
 6645.0876 & 153  &  1.379 &  --0.593 \\
 6645.0898 & 153  &  1.379 &  --1.628 \\
 6645.0945 & 153  &  1.379 &  --3.151 \\
 6645.0974 & 153  &  1.379 &  --0.672 \\
 6645.0975 & 151  &  1.379 &  --0.593 \\
 6645.1021 & 153  &  1.379 &  --1.583 \\
 6645.1047 & 153  &  1.379 &  --0.755 \\
 6645.1050 & 151  &  1.379 &  --1.628 \\
 6645.1081 & 153  &  1.379 &  --3.079 \\
 6645.1101 & 153  &  1.379 &  --0.839 \\
 6645.1107 & 153  &  1.379 &  --1.635 \\
 6645.1125 & 151  &  1.379 &  --3.144 \\
 6645.1144 & 153  &  1.379 &  --0.921 \\
 6645.1164 & 153  &  1.379 &  --1.830 \\
 6645.1170 & 153  &  1.379 &  --3.236 \\
 6645.1194 & 151  &  1.379 &  --0.672 \\
 6645.1270 & 151  &  1.379 &  --1.583 \\
 6645.1341 & 151  &  1.379 &  --3.082 \\
 6645.1376 & 151  &  1.379 &  --0.754 \\
 6645.1448 & 151  &  1.379 &  --1.635 \\
 6645.1513 & 151  &  1.379 &  --3.237 \\
 6645.1525 & 151  &  1.379 &  --0.839 \\
 6645.1590 & 151  &  1.379 &  --1.829 \\
 6645.1643 & 151  &  1.379 &  --0.921 \\
\end{longtable}
}

\balance

\twocolumn

{

\justifying

\section{Comparison with previous analysis}\label{App:comparison}

In Paper\,I we have already compared our \teff , \logg, [\ion{Fe}{i}/H] and [\ion{Fe}{ii}/H] with literature data (see Fig.~1 in Paper\,I). 
We concluded that, by and large our parameters agree with those of other analyses with the exception of \citet{Roederer2014AJ....147..136R} whose effective temperatures are systematically lower (by $\sim 300\,K$) as well as surface gravities (by $\sim 0.75$\,dex).
We here concentrate on the abundances of the seven elements analysed in this paper.

\begin{figure}[h!]
    \centering
    \includegraphics[width=\hsize]{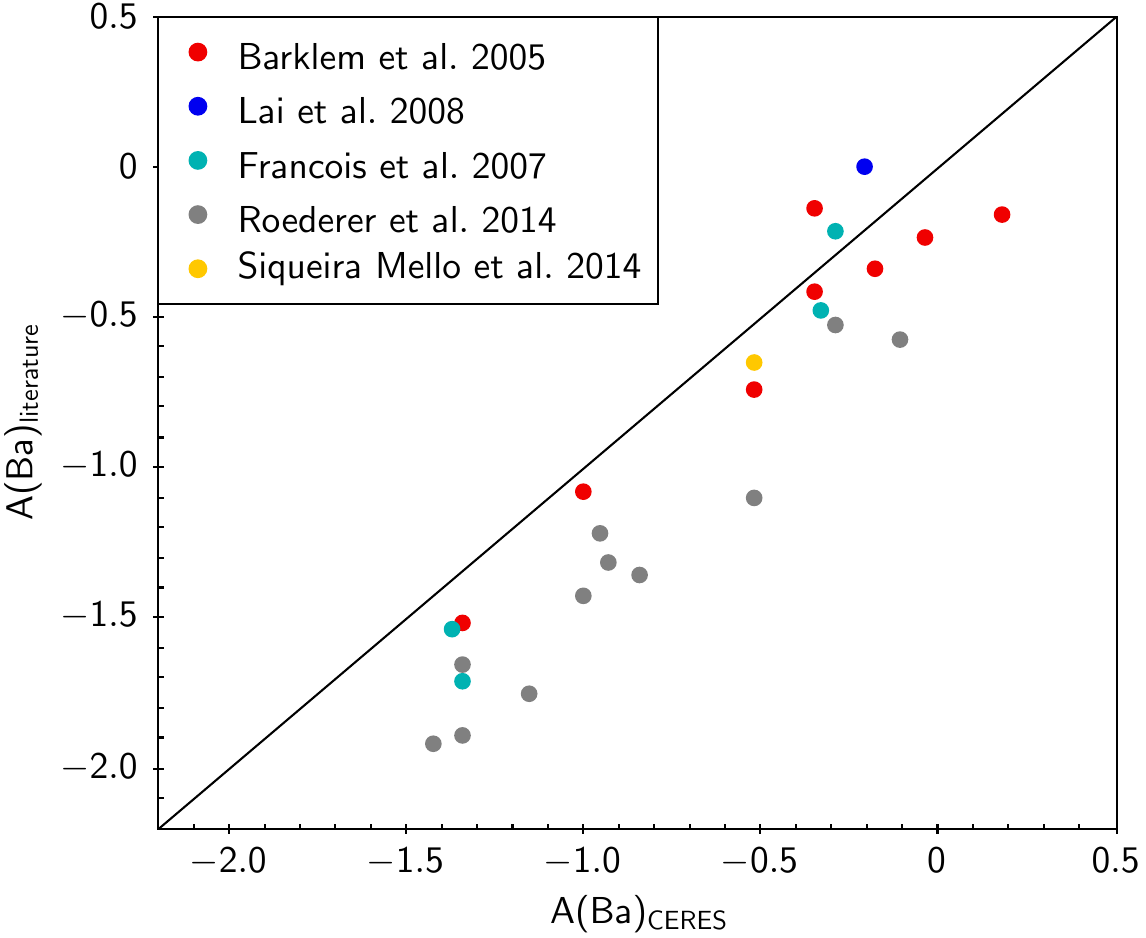}
    \includegraphics[width=\hsize]{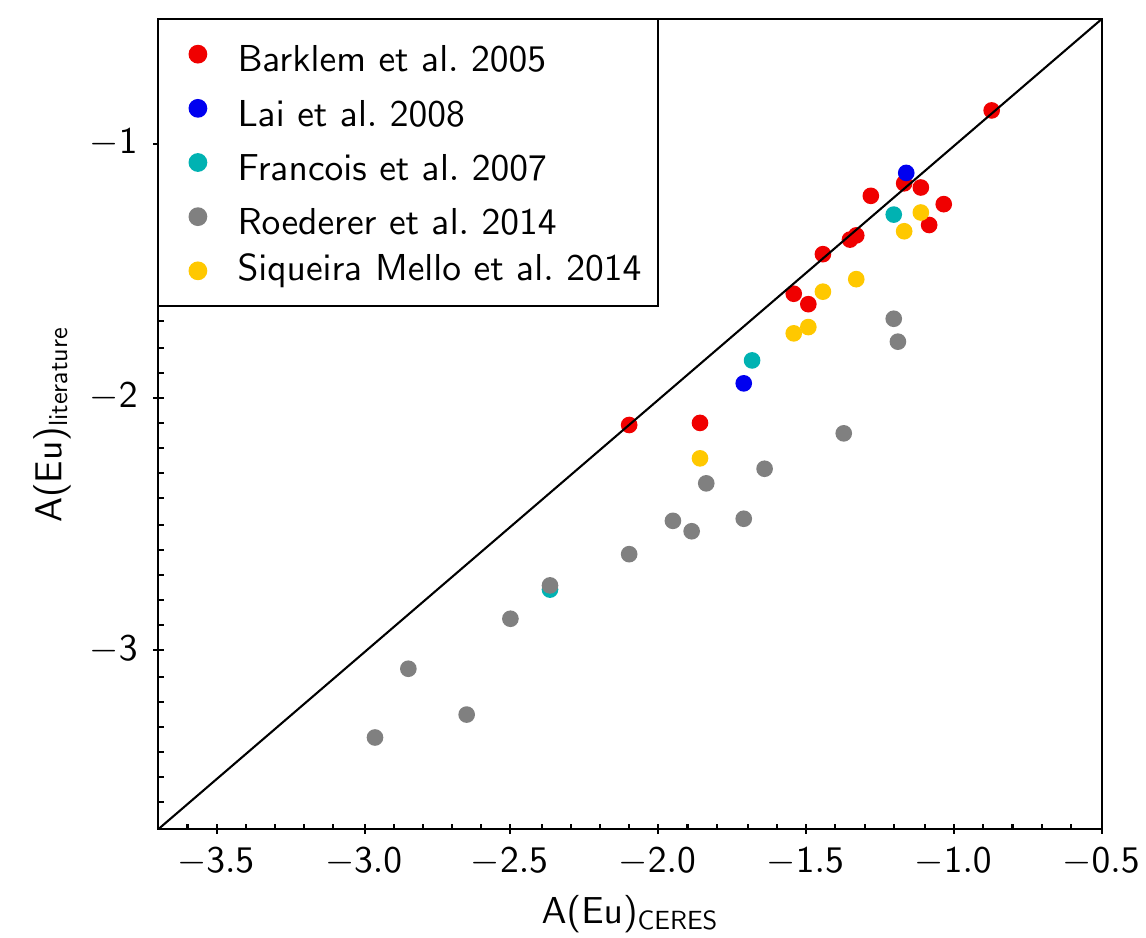}
    \caption{Comparison of our Ba and Eu abundances for the stars in common with five
    high resolution surveys. The black lines represent the identity function.}
    \label{fig:ba_eu_lit}
\end{figure}

\subsection{High resolution surveys}
It is interesting to compare the yield of the CERES surveys with other surveys of comparable high resolution.
The first obvious comparison is with the First Stars survey in which \citet{Francois2007A&A...476..935F} provided abundances of neutron capture elements for 32 giant extremely metal-poor stars.
In our sample we have included five stars already analysed by   \citet{Francois2007A&A...476..935F}.
Another survey to compare to is HERES \citep{Barklem2005A&A...439..129B} that provided abundances for 253 metal-poor halo stars, almost equally split between giants and dwarfs, or subgiants.
The survey was a snapshot survey, conducted with UVES at the 8.2\,m VLT, but with wide slit ($2''$), low resolving power $R\sim 20\, 000$ and low S/N ratio.
We have in common eight stars with this survey. 
A third survey to compare with is that of \citet{2008ApJ...681.1524L} who determined the abundances for a sample of 28 metal-poor stars based on high-resolution spectra obtained with HIRES at the Keck 10\,m telescope. 
Again the sample is almost equally split between giants and dwarfs, or subgiants.
We have two stars in common with this survey.
\citet[][0Z project]{2013ApJ...778...56C} determined detailed abundances for 146 metal-poor stars from spectra obtained with HIRES at the Keck Telescope and with Mike at the 6.5\,m Magellan Telescope. 
About half of the sample is constituted by giant stars.
We have no stars in common with this sample.
\citet{Roederer2014AJ....147..136R} determined abundances for 313 stars based on spectra obtained with Mike at the Magellan telescope, with the Robert G. Tull Coud{\' e} Spectrograph on the 2.7\,m Harlan J. Smith Telescope at McDonald Observatory and with the High Resolution Spectrograph on the 9.2 m Hobby–Eberly Telescope at McDonald Observatory during a ten year long observing campaign. 
The sample is constituted for 57\% of giant and horizontal branch (HB) stars, including some blue HB stars.
We have 14 stars in common with this survey.
\citet{2022ApJ...931..147L} determined abundances for 385 metal-poor stars, almost equally split among giants and dwarfs, or subgiants.
The candidates were selected from the LAMOST \citep{deng_lamost_2012,liu_k_2014} survey and followed up with the HDS spectrograph on the 8.2\,m Subaru telescope.
We have only one star in common with this survey. 
Of the seven elements here examined they provide only Ba, La, and Eu.
\citet{Hansen2012A&A...545A..31H} derived abundances of seven heavy elements (Sr, Y, Zr, Pd, Ag, Ba, and Eu) for a sample of 71 giant and dwarf stars observed at high resolution ($R>40\, 000$) and high S/N ratio (>100 per pixel at 320 nm) with UVES at VLT. We have nine stars in common with this study. 
Another useful comparison is with the sample of  \citet{SiqueiraMello2014A&A...565A..93S}, who derived the abundances of the elements from Li to Th in a sample of seven stars observed at high-resolution ($R\sim 40\, 000$) and high S/N ratio with UVES at VLT.
All seven stars in the sample of \citet{SiqueiraMello2014A&A...565A..93S} were also analysed in this study.

In Table \ref{surveys} we compare the number of measures (not including upper limits) of the different surveys.
It is clear that CERES is providing more detailed abundances that most of the other surveys, for most elements. 
The Survey with most measurements is that of \citet{Roederer2014AJ....147..136R}, that is also the one that can count on the largest investment of telescope time, and provides more measurements than we do for all elements.
Ba is provided by HERES,  by the 0Z project and by \citet{2022ApJ...931..147L} for a sample that
is an order of magnitude larger than ours. 
For Eu, however,  HERES and \citet{2022ApJ...931..147L} provide a larger sample, but in this case of the same order of magnitude as CERES, while the 0Z project provides
less measurements.

In Fig.~\ref{fig:ba_eu_lit} we show the comparison between our abundances of Ba and Eu, respectively, and literature values for the stars in common.
While the agreement with \citet{Barklem2005A&A...439..129B}, \citet{2008ApJ...681.1524L}, \citet{Francois2007A&A...476..935F}, and \citet{SiqueiraMello2014A&A...565A..93S} is reasonably good, the values of \citet{Roederer2014AJ....147..136R} are systematically lower than ours. 
This is also true for Fe, as pointed out in Paper\,I and is a result of the lower effective temperatures and gravities adopted by \citet{Roederer2014AJ....147..136R}.

One further thing to be considered is that CERES is designed to provide the measurement of as many neutron capture elements as possible, so that the real throughput of the survey in terms of achieved chemical inventory should be evaluated after all the papers in the series are published.
One success of CERES, is the high S/N ratio of the observations obtained in program 0104.D-0059, and the fact that for archival spectra we benefit from high S/N ratio spectra and in some cases we could combine several spectra, as detailed in Paper I (see table A1 of that paper).
The high quality of the data allows for more heavy element abundance derivations of poorly studied elements exceeding Ba and Eu (\citealt{paperIV}; Lombardo et al, in prep.) and to compare them with rare earth elements.
Another reason of success in measuring abundances of neutron capture elements
is that we concentrated on giant stars, similar to what done by \citet{Francois2007A&A...476..935F},
unlike the other quoted surveys.

\subsection{Individual high resolution studies}
In this section we compare the results obtain in this study with the ones from individual literature studies that analysed in detail some of the stars in our sample. 
In Fig.~\ref{Fig:single_stars} we show the comparison between our derived abundances of Ba, La, Ce, Pr, Nd, Sm, and Eu for the stars CES1221-0328 (HE1219-0312), CES1402+0941 (HD122563), CES2231-3238 (CS\,29491-069), CES2254-4209 (HE2252-4225), and CES2330-5626 (HE2327-5642) and the ones obtained by \citet{Honda2006ApJ...643.1180H}, \citet{Hayek2009A&A...504..511H}, \citet{Mashonkina2010A&A...516A..46M}, and \citet{Mashonkina2014A&A...569A..43M}.
We note that for the stars CS\,29491-069, HE2327-5642, and HE2252-4225 the values are overall in good agreement, while for the stars HD122563 and HE1219-0312 we tend to have higher abundances compared to the ones in the literature. 
This is mostly due to the higher \logg\ derived by us for these two
stars.
In Table~\ref{tab:single_stars} the stellar parameters derived in this study are compared to the literature. 
We see that for the stars for which we observe the largest difference in the abundances, HD122563 and HE1219-0312, there is also a difference in the stellar parameters, particularly in \logg\ and microturbulence.

}

\begin{table}
\caption{Number of abundance measures for the different surveys}
\label{surveys}
\centering
\resizebox{0.5\textwidth}{!}{%
\begin{tabular}{lrrrrrrrr}
\hline\hline
Survey & Ba & La & Ce & Pr & Nd & Sm & Eu \\
\hline
This paper & 43 & 45 & 46 & 29 & 43 & 38 & 51 \\
Barklem et al. 2005 & 220 & 33 & 13 & 0 & 35 & 9 & 68 \\
Fran\c{c}ois et al. 2007 & 31 & 17 & 10 & 14 & 19 & 7 & 17 \\
Lai et al. 2008 & 24 & 3 & 0 & 0 & 0 & 0 & 5 \\
Hansen et al. 2012 & 60 & 0 & 0 & 0 & 0 & 0 & 51 \\
Cohen et al. 2013 & 111 & 30 & 21 & 6 & 27 & 4 & 27 \\
Roederer et al. 2014 & 297 & 112 & 65 & 43 & 96 & 50 & 141 \\
Siqueira Mello et al. 2014 & 7 & 7 & 6 & 4 & 7 & 6 & 7 \\
Li et al. 2022 & 318 & 72 & 0 & 0 & 0 & 0 & 79 \\
\hline
\end{tabular}
}
\end{table}

   \begin{figure*}[h!]
   \centering
   \includegraphics[width=0.33\hsize]{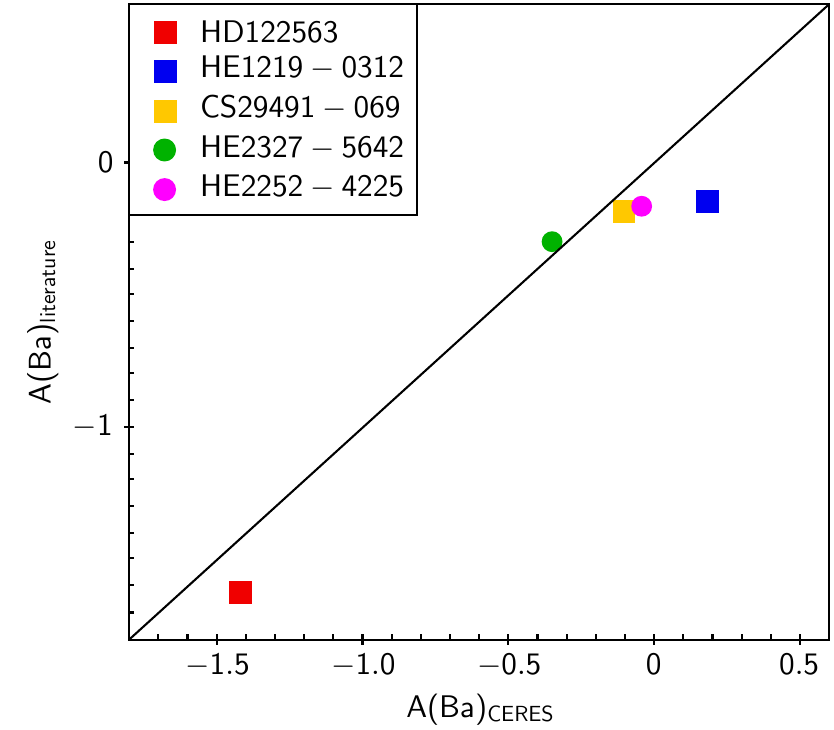}
   \includegraphics[width=0.33\hsize]{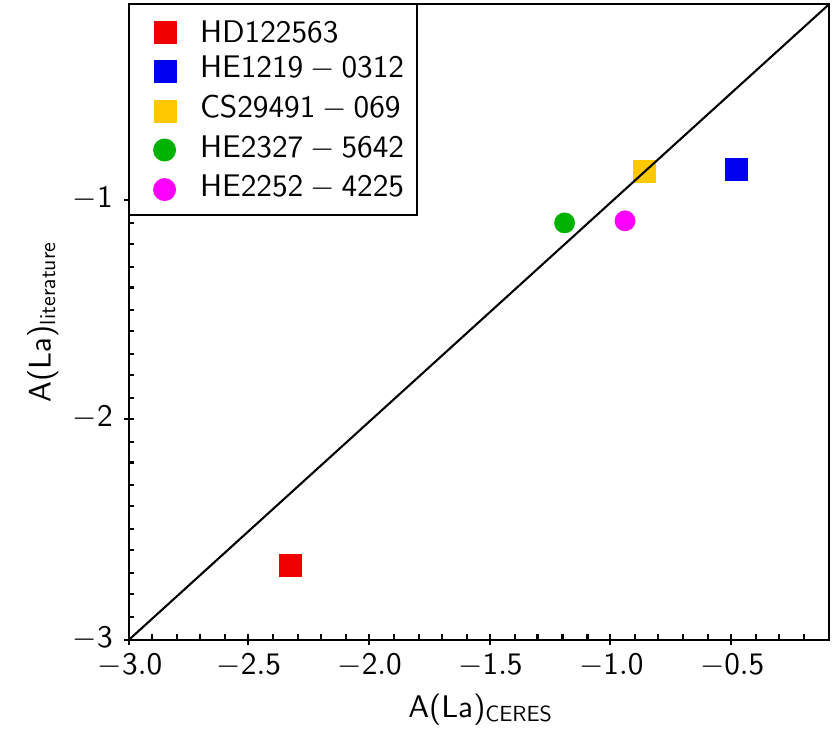}
   \includegraphics[width=0.33\hsize]{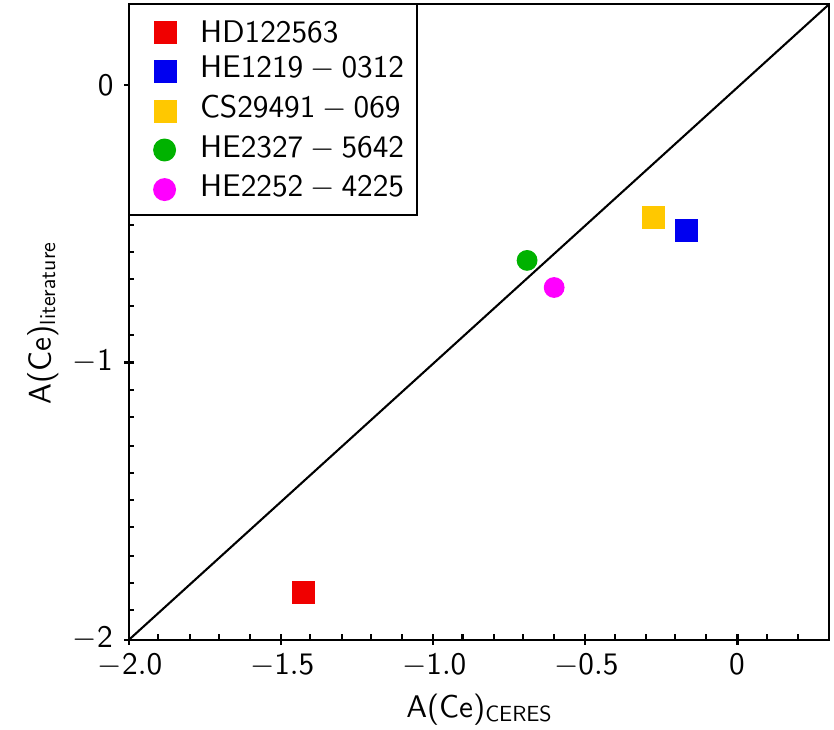}
   \includegraphics[width=0.33\hsize]{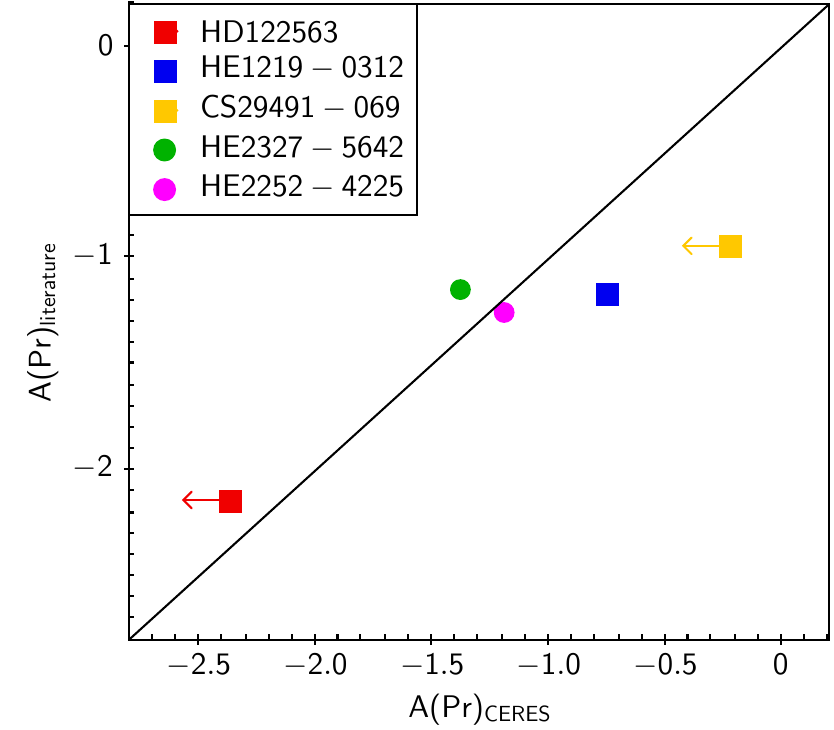}
   \includegraphics[width=0.33\hsize]{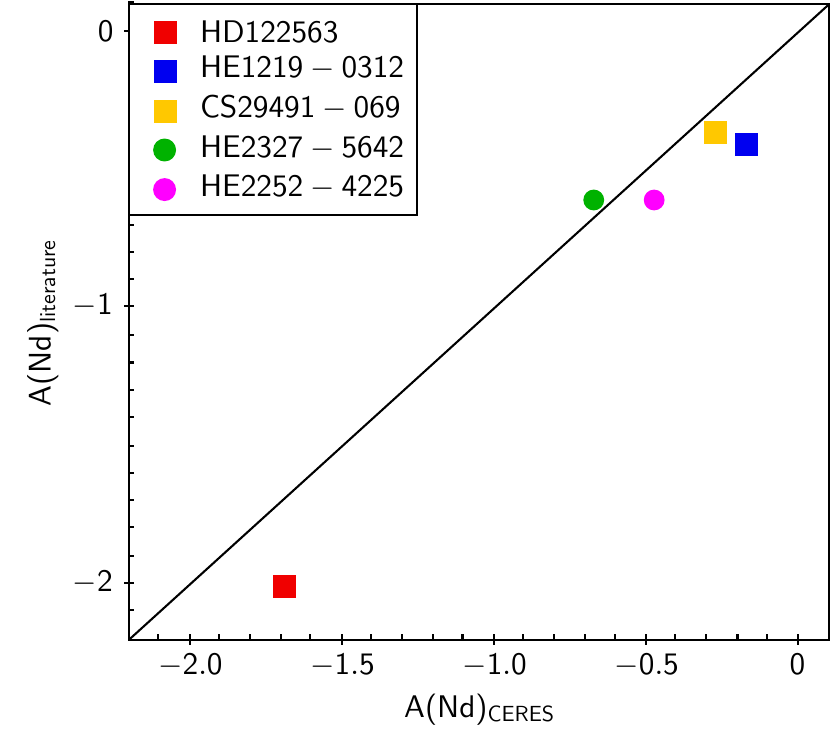}
   \includegraphics[width=0.33\hsize]{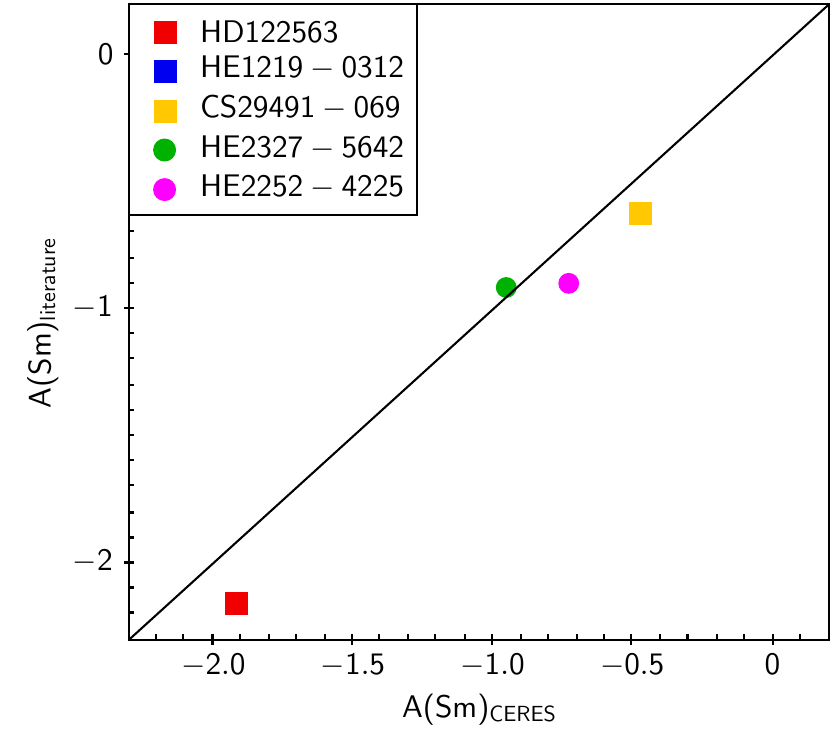}
   \includegraphics[width=0.33\hsize]{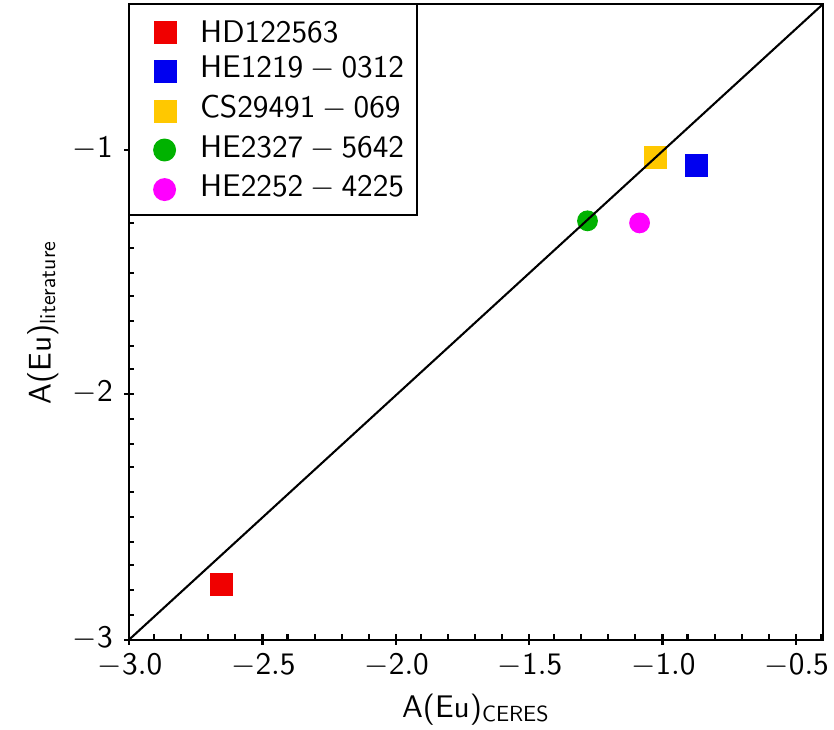}
      \caption{Comparison of our Ba, La, Ce, Pr, Nd, Sm, and Eu abundances for the stars in common with \citealt{Honda2006ApJ...643.1180H} (HD122563), \citealt{Hayek2009A&A...504..511H} (HE1219-0312, CS\,29491-069), \citealt{Mashonkina2010A&A...516A..46M} (HE2327-5642), and  \citealt{Mashonkina2014A&A...569A..43M} (HE2252-4225). The black lines represent the identity function.
             }
         \label{Fig:single_stars}
   \end{figure*}

\begin{table*}
\caption{Comparison between stellar parameters in literature and in this paper for selected stars.}
\label{tab:single_stars}
\centering
\resizebox{\textwidth}{!}{%
\begin{tabular}{llrrrrrrrrr}
\hline\hline
Star & Simbad name & \teff(lit.) & \teff(CERES) & logg(lit.) & logg(CERES) & vt(lit.) & vt(CERES) & [Fe/H](lit.) & [Fe/H](CERES) & ref.  \\
\hline
  CES1221-0328 & HE1219-0312 & 5060 & 5145 & 2.30 & 2.76 & 1.60& 1.60 & $-$2.96 & $-$2.96 & 1\\
  CES1402+0941 & HD122563 & 4570 & 4682 & 1.10 & 1.35 & 2.20 & 2.01 & $-$2.77 & $-$2.79 & 2\\
  CES2231-3238 & BPS CS29491-069 & 5300 & 5222 & 2.80 & 2.67 & 1.60 & 1.67 & $-$2.51 & $-$2.77 & 1\\
  CES2254-4209 & HE2252-4225 & 4710 & 4805 & 1.65 & 1.98 & 1.70 & 1.79 & $-$2.63 & $-$2.88 & 3\\
  CES2330-5626 & HE2327-5642 & 5050 & 5028 & 2.34 & 2.31 & 1.80 & 1.75 & $-$2.78 & $-$3.10 & 4\\
\hline
\end{tabular}
}
\tablebib{(1)~\citet{Hayek2009A&A...504..511H};
(2) \citet{Honda2006ApJ...643.1180H}; (3) \citet{Mashonkina2014A&A...569A..43M}; (4) \citet{Mashonkina2010A&A...516A..46M}.
}

\end{table*}

\clearpage

\section{Additional plots}

   \begin{figure*}[h!]
   \centering
   \includegraphics[width=0.49\hsize]{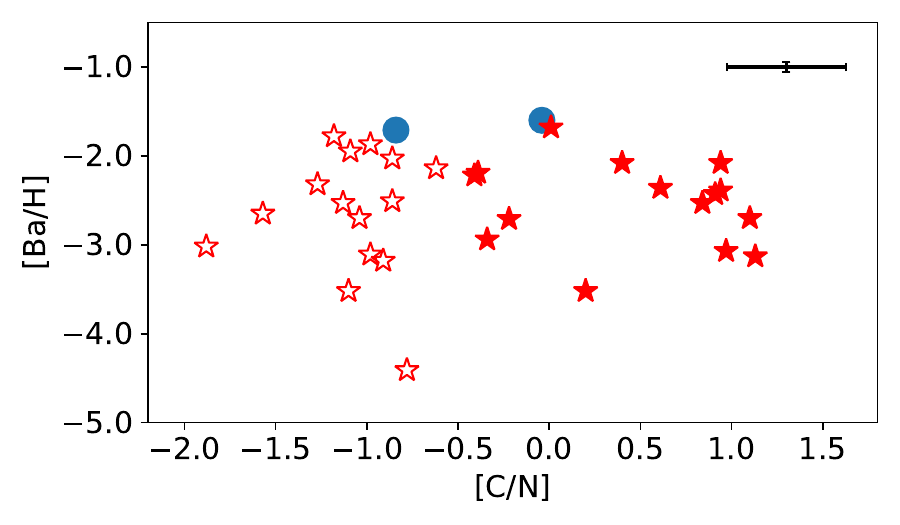}
   \includegraphics[width=0.49\hsize]{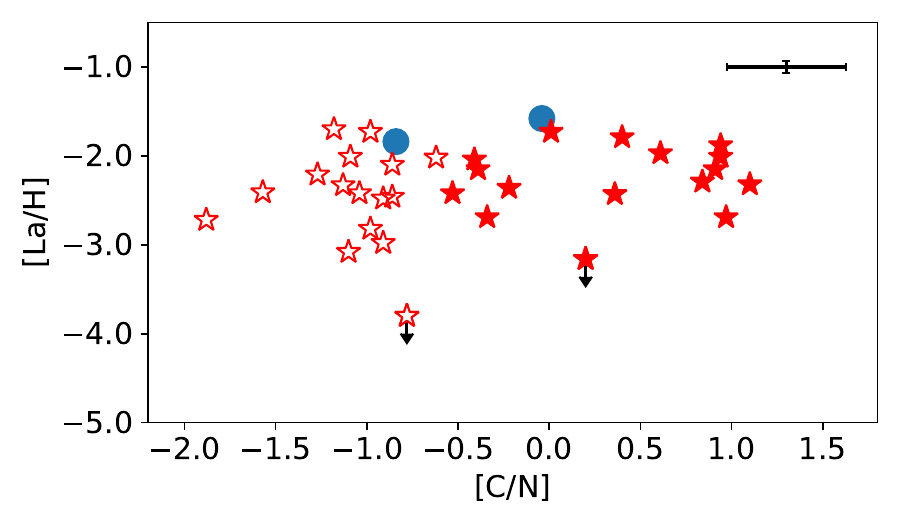}
   \includegraphics[width=0.49\hsize]{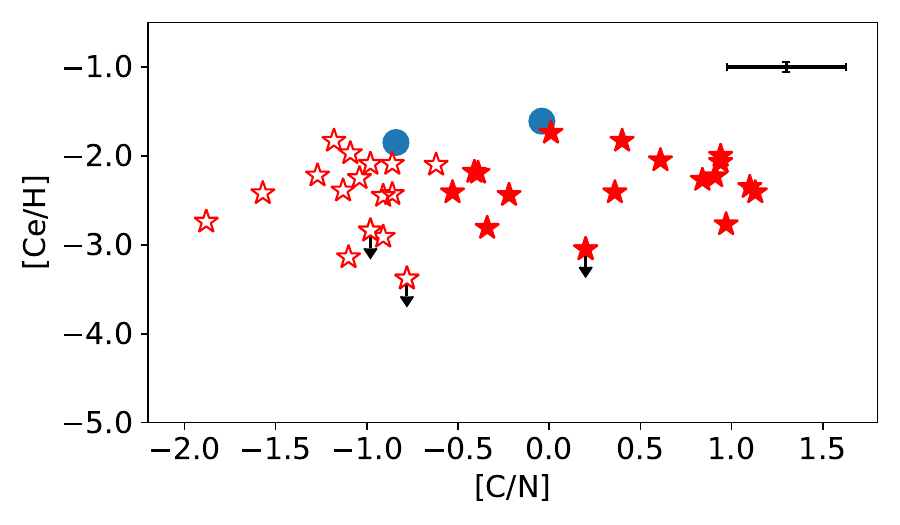}
   \includegraphics[width=0.49\hsize]{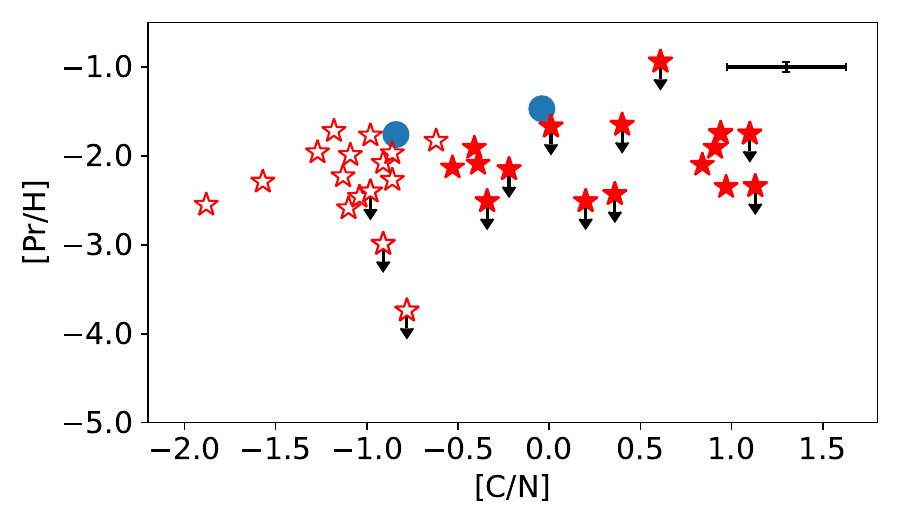}
   \includegraphics[width=0.49\hsize]{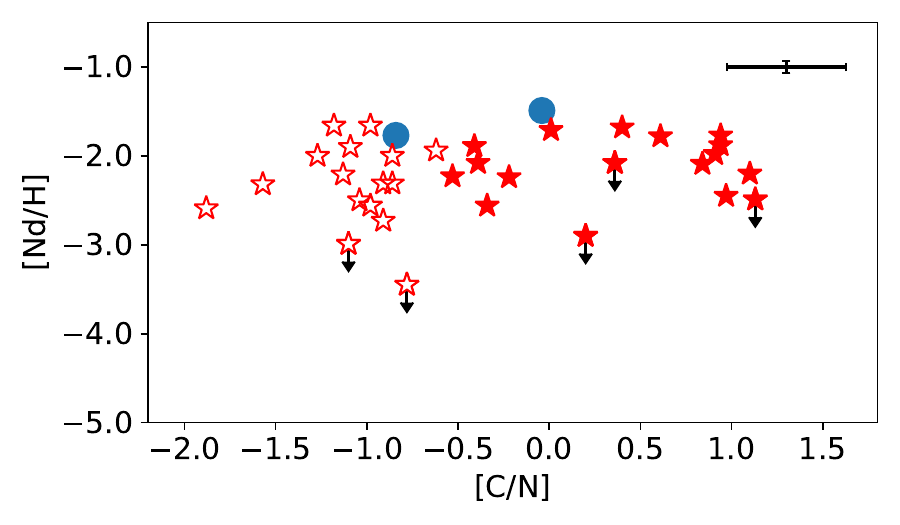}
   \includegraphics[width=0.49\hsize]{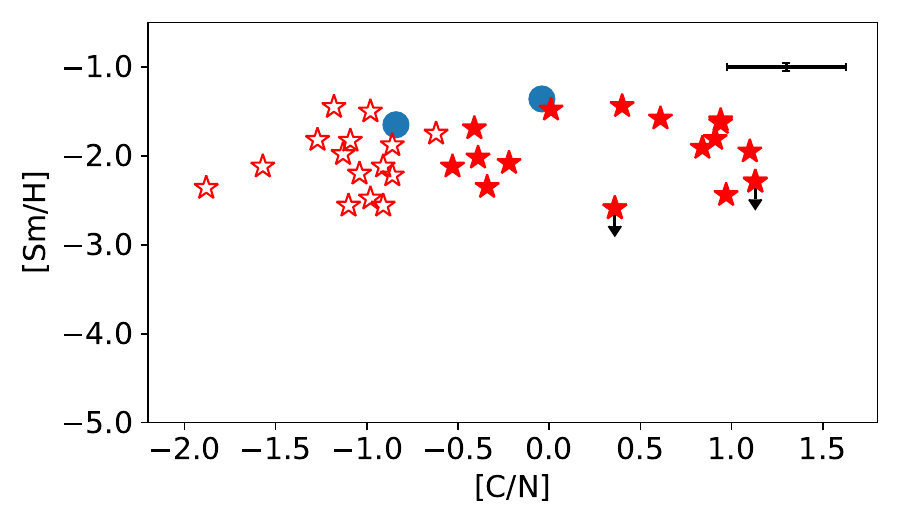}
   \includegraphics[width=0.49\hsize]{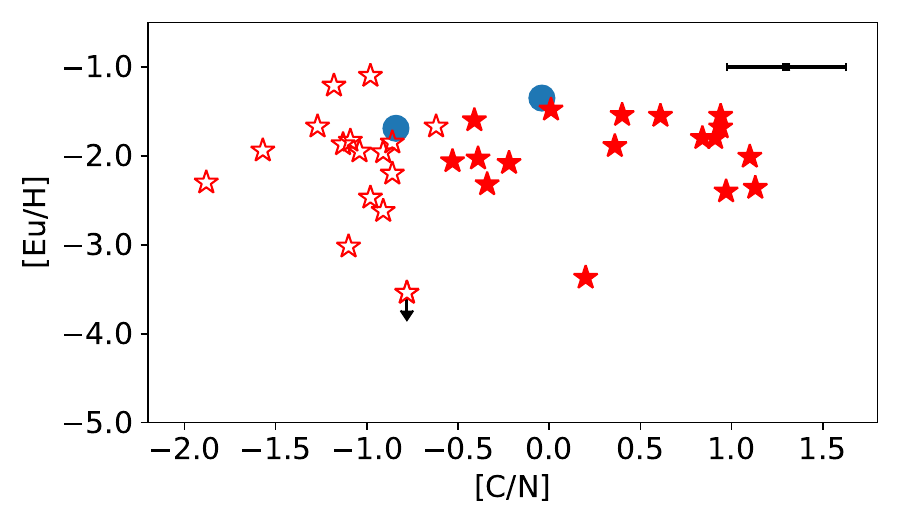}
     \caption{
     [Ba/H], [La/H], [Ce/H], [Pr/H], [Nd/H], [Sm/H], [Eu/H] abundances as a function of [C/N] for our sample of stars. The open/filled star symbols represent mixed/unmixed stars according to the classification in Paper~II (mixed: [N/Fe]$>$0.5 and [C/N]$<$$-0.6$, unmixed: [N/Fe]$<$0.5 and [C/N]$>$$-0.6$). The filled circles represent stars that have values outside the ranges defined for mixed and unmixed stars. 
              }
         \label{Fig:XH_CN}
   \end{figure*}

   \begin{figure*}[h!]
   \centering
   \includegraphics[width=0.49\hsize]{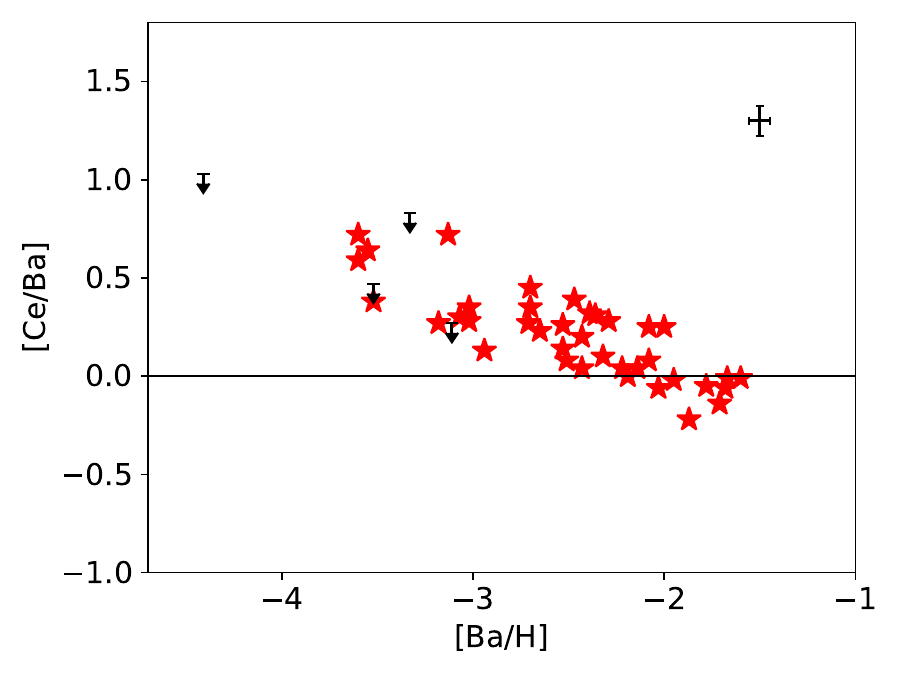}
   \includegraphics[width=0.49\hsize]{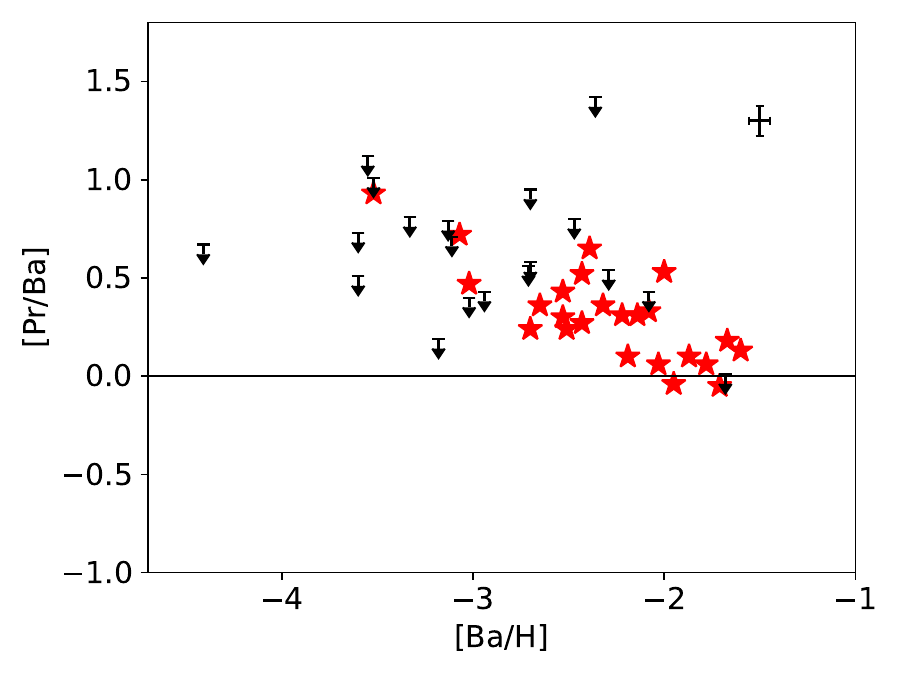}
      \caption{[Ce/Ba] and [Pr/Ba] abundance ratios as a function of [Ba/H]. The black arrows represent upper limits of abundance ratios. 
      A representative error bar is shown in the upper-right corner of each panel. 
              }
         \label{Fig:abu_ratio_ba}
   \end{figure*}

   \begin{figure*}[h!]
   \centering
   \includegraphics[width=0.49\hsize]{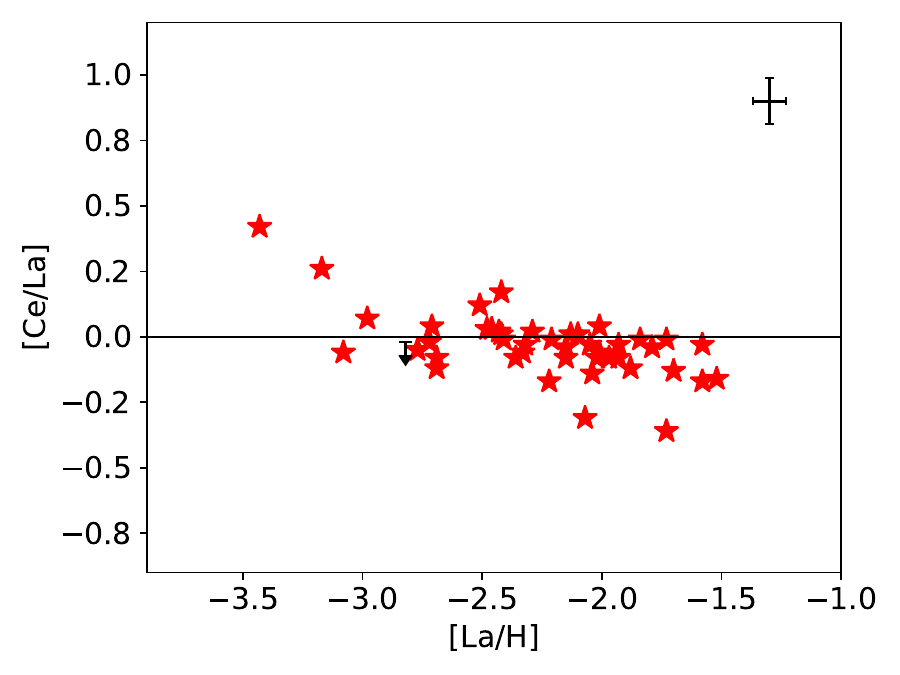}
   \includegraphics[width=0.49\hsize]{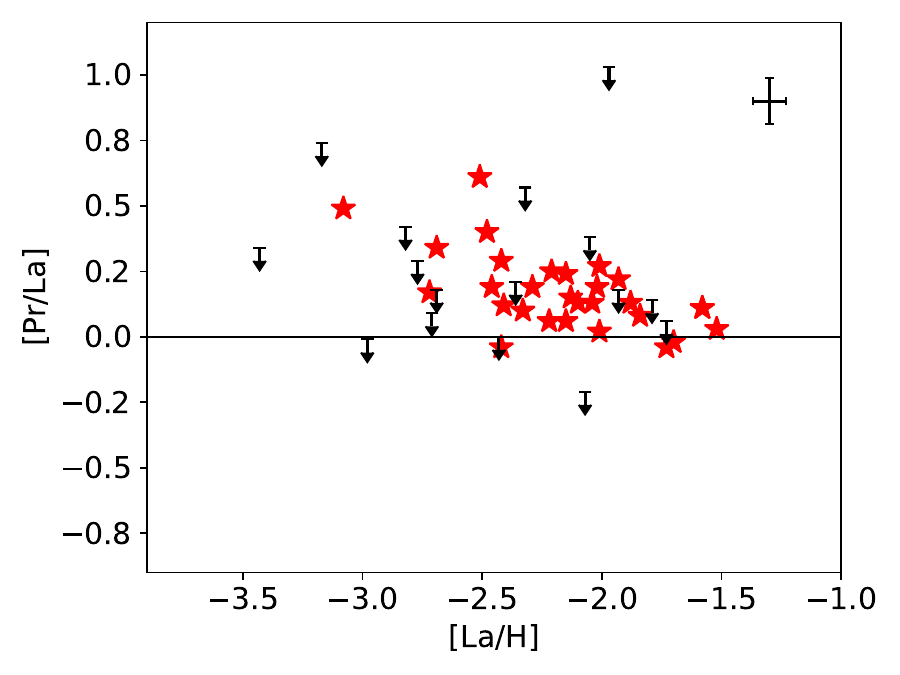}
   \includegraphics[width=0.49\hsize]{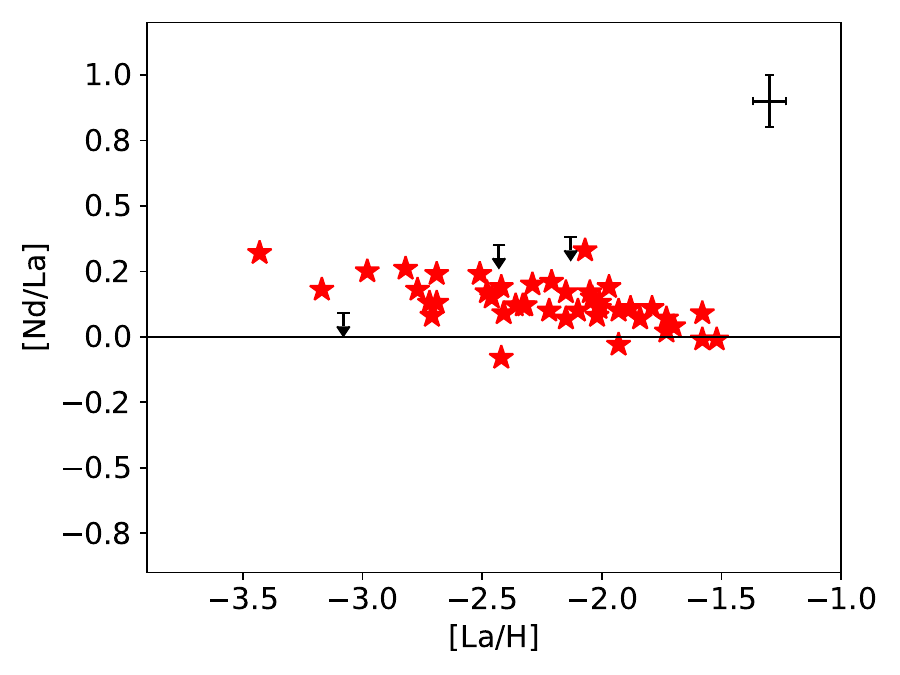}
   \includegraphics[width=0.49\hsize]{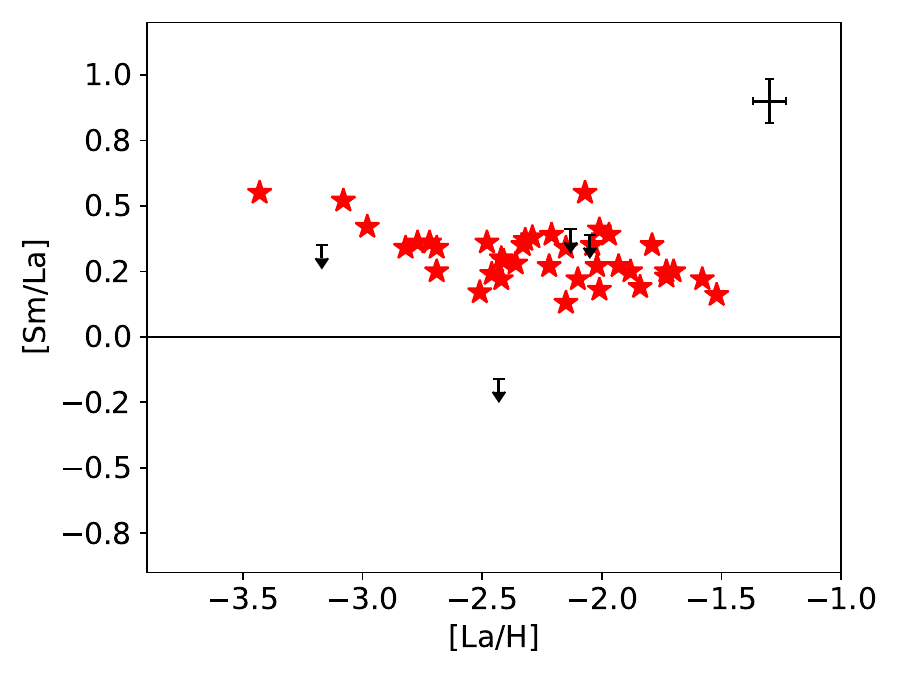}
      \caption{[Ce/La], [Pr/La], [Nd/La], [Sm/La] abundance ratios as a function of [La/H]. Coloured symbols as in Fig.~\ref{Fig:abu_ratio_ba}.
              }
         \label{Fig:abu_ratio_la}
   \end{figure*}
   
   \begin{figure*}[h!]
   \centering
   \includegraphics[width=0.49\hsize]{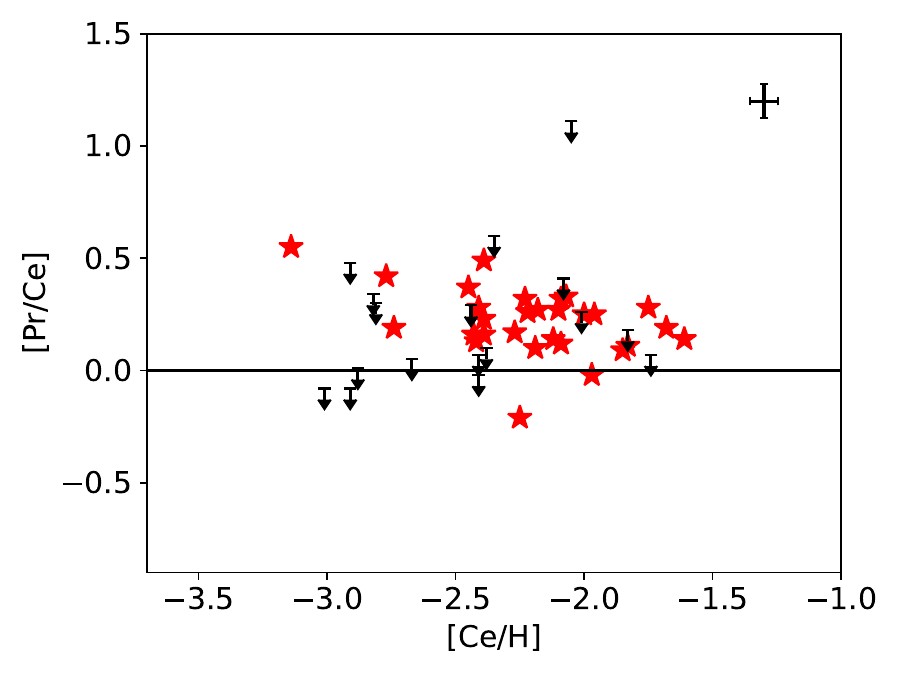}
   \includegraphics[width=0.49\hsize]{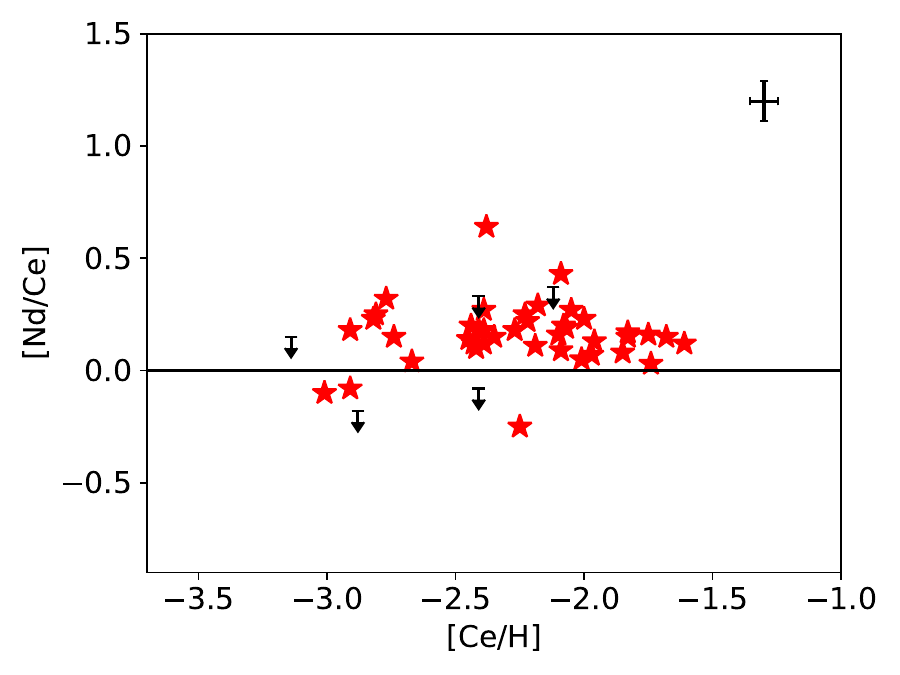}
   \includegraphics[width=0.49\hsize]{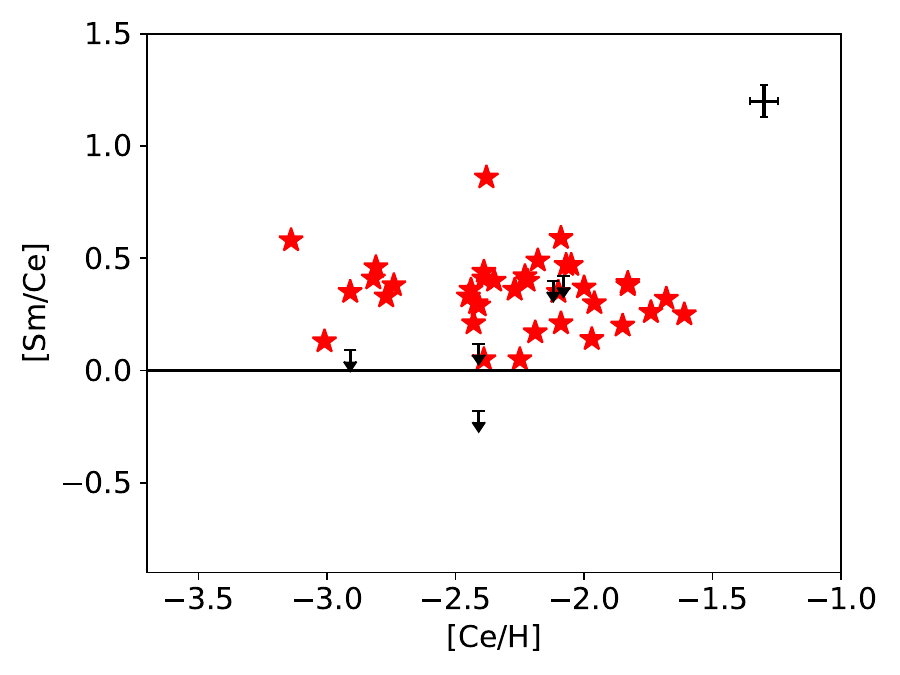}
      \caption{[Pr/Ce], [Nd/Ce], [Sm/Ce] abundance ratios as a function of [Ce/H]. Coloured symbols as in Fig.~\ref{Fig:abu_ratio_ba}.
              }
         \label{Fig:abu_ratio_ce}
   \end{figure*}

   \begin{figure*}[h!]
   \centering
   \includegraphics[width=0.49\hsize]{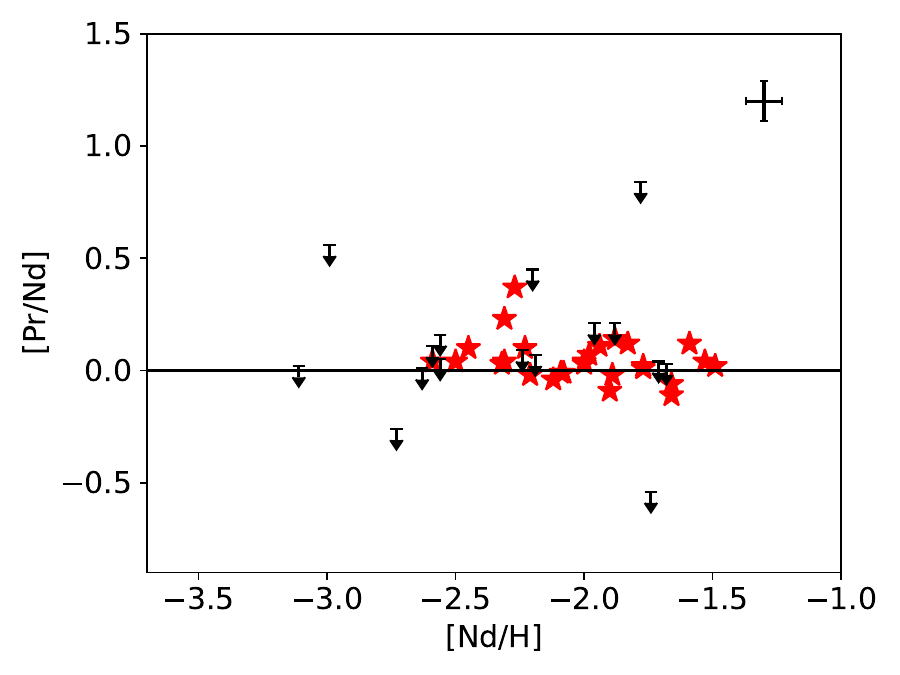}
   \includegraphics[width=0.49\hsize]{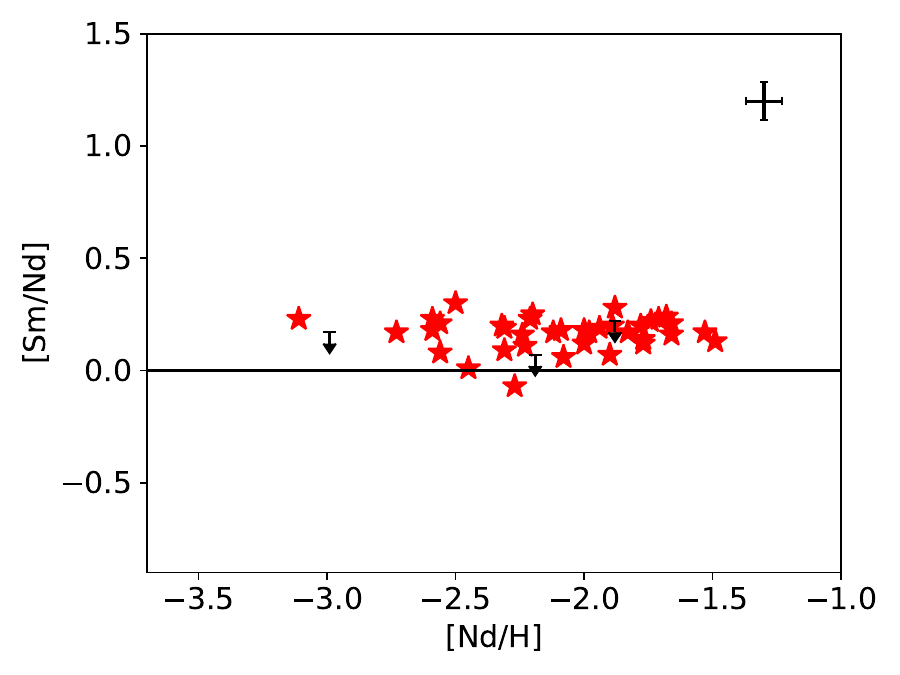}
   \includegraphics[width=0.49\hsize]{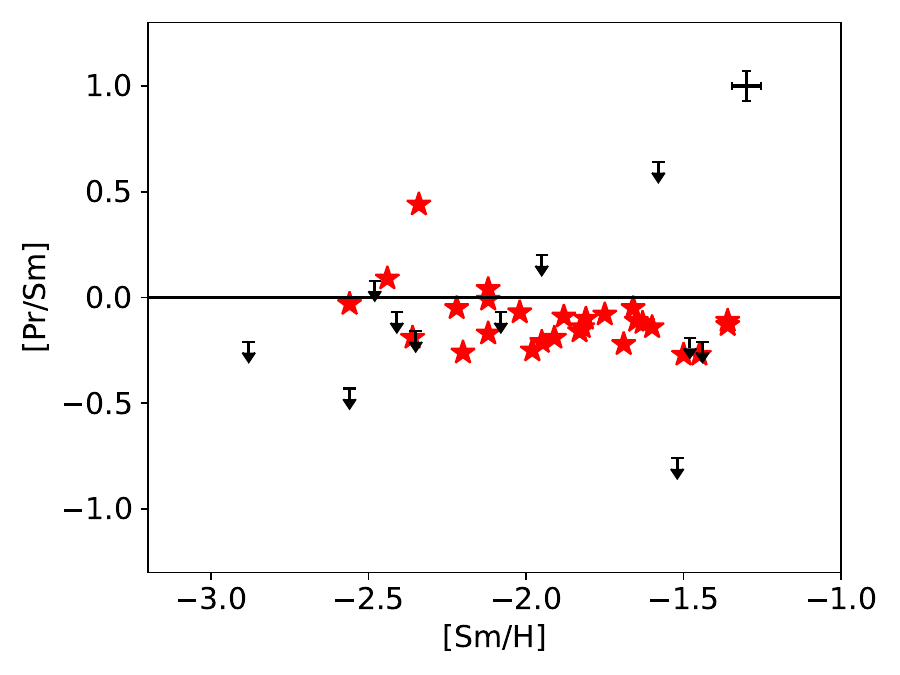}
      \caption{[Pr/Nd] and [Sm/Nd] abundance ratios as a function of [Nd/H], and [Pr/Sm] abundance ratio as a function of [Sm/H]. Coloured symbols as in Fig.~\ref{Fig:abu_ratio_ba}.
              }
         \label{Fig:abu_ratio_nd_sm}
   \end{figure*}

   \begin{figure*}[h!]
   \centering
   \includegraphics[width=0.49\hsize]{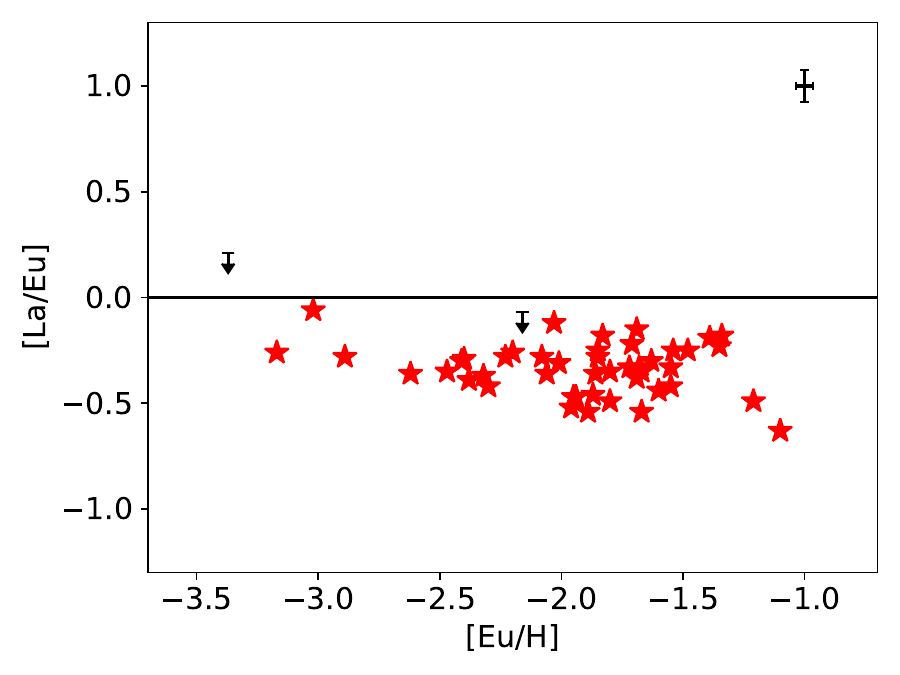}
   \includegraphics[width=0.49\hsize]{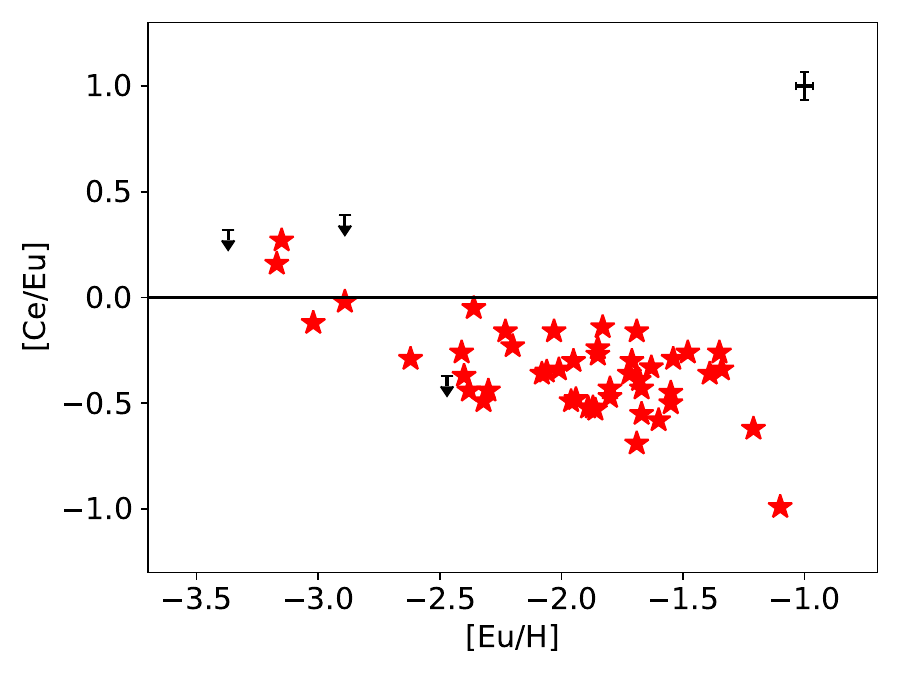}
   \includegraphics[width=0.49\hsize]{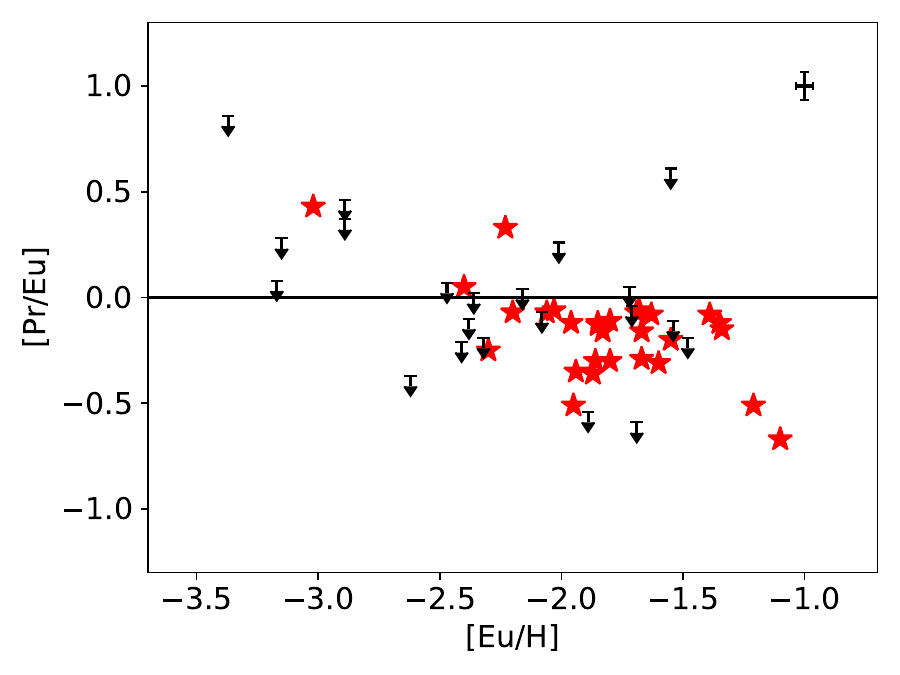}
   \includegraphics[width=0.49\hsize]{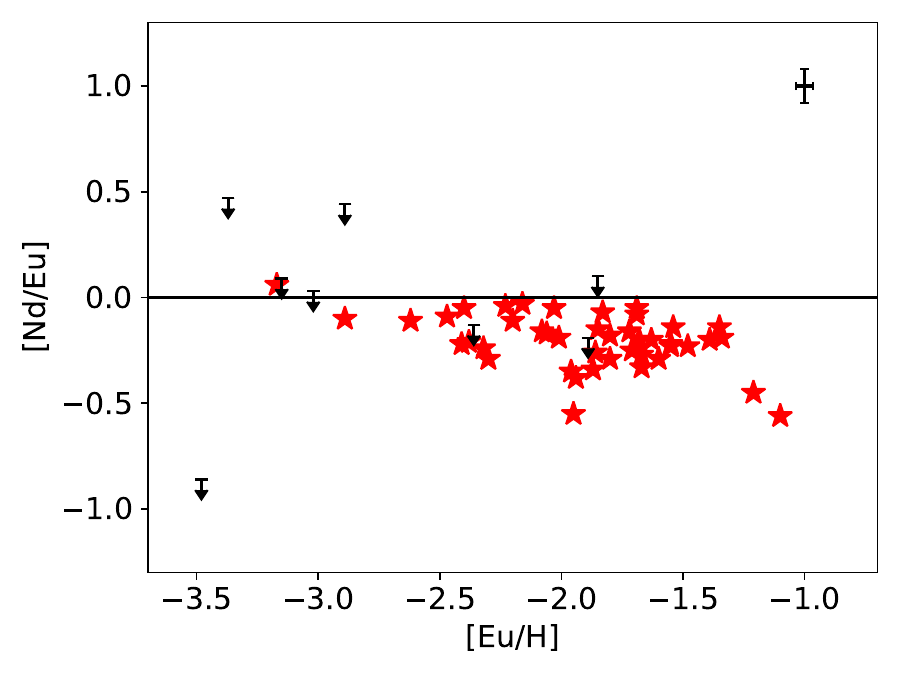}
   \includegraphics[width=0.49\hsize]{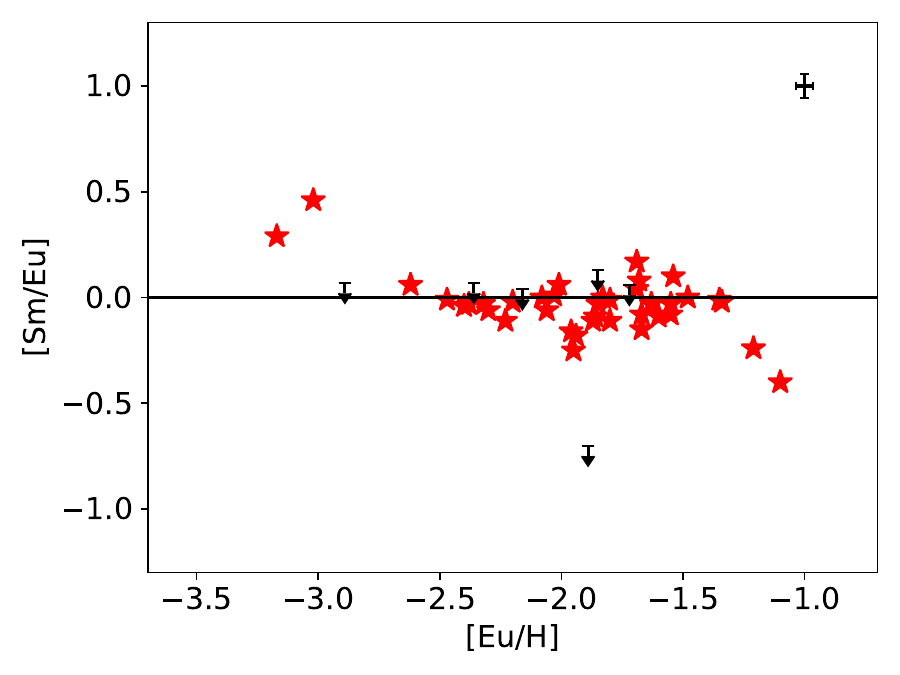}
      \caption{[La/Eu], [Ce/Eu], [Pr/Eu], [Nd/Eu] and [Sm/Eu] abundance ratios as a function of [Eu/H]. Coloured symbols as in Fig.~\ref{Fig:abu_ratio_ba}.
             }
         \label{Fig:abu_ratio_eu}
   \end{figure*}

   \begin{figure*}[h!]
   \centering
   \includegraphics[width=\hsize]{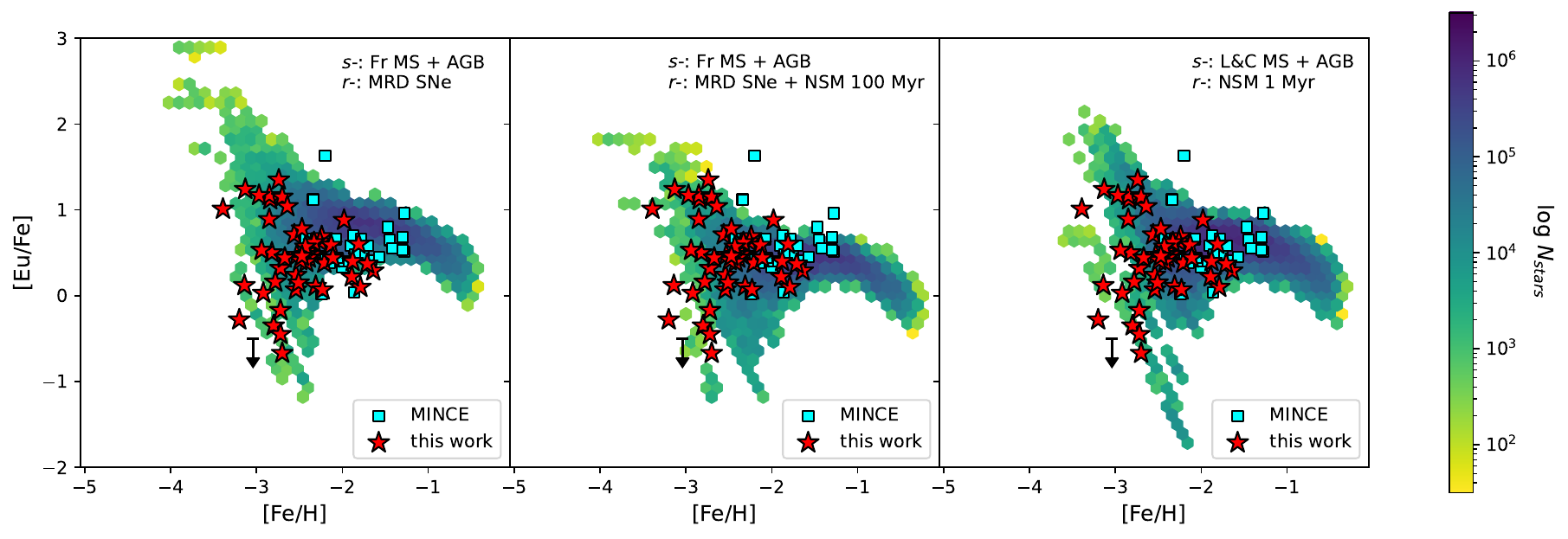}
      \caption{[Eu/Fe] abundance ratios as a function of [Fe/H] for our sample of stars (red stars) and MINCE data \citep[][cyan squares]{2024A&A...686A.295F} compared to the GCE models  of \cite{Cescutti2014A&A...565A..51C} (left), \cite{Cescutti2015A&A...577A.139C} (centre), and \cite{Rizzuti2021MNRAS.502.2495R} (right, see text for more details).
              }
         \label{Fig:EuFe_FeH_GCE}
   \end{figure*}

\end{appendix}

\end{document}